\documentclass[preprint,10pt]{elsarticle}

\usepackage{amsmath}
\usepackage{amssymb}
\usepackage{commath}
\usepackage{array}
\usepackage{calc}
\usepackage{longtable}
\usepackage{multirow}
\usepackage{physics}
\usepackage[squaren]{SIunits}
\usepackage{tensor}
\usepackage{upgreek}

\usepackage[capitalise]{cleveref}
\usepackage{enumitem}
\usepackage{comment}

\usepackage{pstricks}
\usepackage{graphicx}
\graphicspath{{figures/}}
\usepackage{xspace}
\usepackage{listings}
\usepackage[section]{placeins}

\usepackage{booktabs}
\usepackage{todonotes}

\usepackage[left=2.5cm,right=2.5cm]{geometry}

\newcommand{\Ariadne}{A\protect\scalebox{0.8}{RIADNE}\xspace}

\newcommand{\MadGraph}{M\protect\scalebox{0.8}{AD}G\protect\scalebox{0.8}{RAPH}5\_aM\protect\scalebox{0.8}{C}@N\protect\scalebox{0.8}{LO}\xspace}
\newcommand{\Pythia}{P\protect\scalebox{0.8}{YTHIA}\xspace}
\newcommand{\Rivet}{R\protect\scalebox{0.8}{IVET}\xspace}

\newcommand{\Vincia}{V\protect\scalebox{0.8}{INCIA}\xspace}

\newcommand{\D}{\ensuremath{\,\text{d}}}

\newcommand{\alphaS}{\ensuremath{\alpha_\text{s}}}
\newcommand{\sIK}{\ensuremath{s_{IK}}}
\newcommand{\sik}{\ensuremath{s_{ik}}}
\newcommand{\sij}{\ensuremath{s_{ij}}}

\newcommand{\Order}[1]{\ensuremath{\mathcal{O}\left(#1\right)}}
\newcommand{\tms}{\ensuremath{t_\text{MS}}}
\newcommand{\weightCKKWL}{\ensuremath{w_\text{\tiny{CKKW-L}} }}

\journal{Computer Physics Communications}

\begin{document}

\begin{frontmatter}

\title{Efficient Multi-Jet Merging with the \Vincia Sector Shower}
\author{Helen Brooks\fnref{move1}}
\fntext[move1]{Now at CCFE, Oxfordshire, UK.}
\author{Christian T Preuss\corref{corAuth}}
\cortext[corAuth]{Corresponding author.}
\ead{christian.preuss@monash.edu}
\address{School of Physics and Astronomy, Monash University, Wellington Road, Clayton, VIC-3800, Australia}

\begin{abstract}
We here present an extension of the CKKW-L multi-jet merging technique to so-called sector showers as implemented in the \textsc{Vincia} antenna shower. The bijective nature of sector showers allows for efficient multi-jet merging at high multiplicities, as any given configuration possesses only a single \textquotedblleft history\textquotedblright, while retaining the accuracy of the CKKW-L technique.
Our method reduces the factorial scaling of the number of parton shower histories to a constant of a single history per colour-ordered final state.
We show that the complexity of constructing shower histories is reduced to an effective linear scaling with the number of final-state particles. Moreover, we demonstrate that the overall event generation time and the memory footprint of our implementation remain approximately constant when including additional jets. We compare both to the conventional CKKW-L implementation in \textsc{Pythia} and gain a first estimate of renormalisation scale uncertainties at high merged multiplicities.
As a proof of concept, we show parton-level predictions for vector boson production in proton-proton collisions with up to nine hard jets using the new implementation.
Despite its much simpler nature, we dub the new technique MESS, in analogy to the conventional MEPS nomenclature.
\end{abstract}


\end{frontmatter}


\clearpage
\tableofcontents

\section{Introduction}
While fixed-order calculations accurately describe observables in regions of phase space where hard, well-separated jets dominate, they are insufficient in the resummation region, where additional particles are emitted at low energies or angles. In these collinear- and soft-enhanced phase space regions, parton shower Monte Carlo generators provide a reliable and versatile tool to resum the leading logarithms (LL) arising from QCD matrix elements to all orders in the perturbative expansion in the strong coupling.
In order to achieve an accurate description over all of phase space, these two approaches need to be combined, by techniques known as matching or merging. With experimental analyses becoming available for high jet multiplicities, cf.\ e.g.~\cite{Aad:2014qxa,Aaboud:2017hbk}, and in the advent of the high-luminosity LHC, the demand for calculations with both an accurate description of many hard jets as well as QCD bremsstrahlung is ever increasing.

To date, a vast amount of matching and merging schemes has been developed, with matching to leading-order (LO) \cite{Bengtsson:1986hr,Giele:2007di,Giele:2011cb,Fischer:2017yja} or next-to-leading-order (NLO) \cite{Frixione:2002ik,Nason:2004rx,Frixione:2007vw} matrix elements on the one hand and merging with LO \cite{Mangano:2001xp,Mangano:2006rw,Catani:2001cc,Lonnblad:2001iq,Hoeche:2009rj,Hamilton:2009ne,Lonnblad:2011xx,Lonnblad:2012ng,Platzer:2012bs} and NLO \cite{Lavesson:2008ah,Hoche:2010kg,Lonnblad:2012ix,Gehrmann:2012yg,Hoeche:2012yf,Hamilton:2012np,Frederix:2012ps,Bellm:2017ktr} calculations on the other. First steps towards the inclusion of NNLO calculations have also been taken \cite{Hamilton:2012rf,Alioli:2015toa,Hoeche:2014aia,Hoche:2014dla,Hoche:2018gti,Alioli:2019qzz,Monni:2019whf}.
Together with the automation of tree-level matrix-element generation \cite{Papadopoulos:2000tt,Kanaki:2000ey,Krauss:2001iv,Moretti:2001zz,Mangano:2002ea,Kilian:2007gr,Gleisberg:2008fv,Alwall:2014hca} the path has been paved for tree-level matching and merging at high multiplicities. However, the computational overhead of such calculations grows at least factorially with the number of particles on both the fixed-order and the resummation side, quickly rendering such computations intractable.

Although the most restrictive bottlenecks in merged calculations arise in the context of generating high-multiplicity matrix elements \cite{Hoeche:2019rti}, especially the phase-space integration, these parton-level samples can be generated \textquotedblleft once and for all\textquotedblright, meaning they can be re-utilised for many different particle-level analyses, given the generation is sufficiently factorised. A novel framework for factorised fixed-order and parton-shower calculations in the high-multiplicity regime has been presented in \cite{Hoeche:2019rti}. The parton-level events of vector boson production with up to 9 additional jets generated there have been made publicly available\cite{z9jets,wm9jets,wp9jets}.

Leaving the difficulties and pitfalls of high-multiplicity matrix-element generation aside, the main bottleneck in merged calculations then arises from the fact that merging schemes usually rely on the construction of parton shower histories, i.e., the sequences of states the parton shower would have produced to arrive at a given configuration.
The purpose of constructing parton shower histories is to obtain Sudakov factors to reweight \textit{inclusive} event samples to make them \textit{exclusive}, so that double-counting of emissions is avoided.
Hence, this procedure has to be undertaken for every parton-shower merged calculation and therefore particle-level event generation run.

In conventional dipole- or DGLAP-based parton showers, the number of histories grows at least factorially with the number of final-state particles. 
Naturally, one can resort to a deterministic (sometimes referred to as \textquotedblleft winner-takes-all\textquotedblright) scheme, where a simple jet-clustering algorithm is employed to construct the shower history. Such a scheme can, however, in principle lead to under- or over-counted phase space regions and consequently may not correctly reflect the Sudakov factors generated by the shower.
Constructing and weighting all possible histories therefore becomes highly time- and resource-intensive for large final-state multiplicities. Moreover, the memory required to store all possible histories until the most probable is picked may exceed the available memory, cf.~\cite{Hoeche:2019rti}.
In that study, the construction of all possible histories was therefore limited to up to 6 additional jets, after which a deterministic (\textquotedblleft WTA\textquotedblright) scheme was employed. Although the effect was not found to be large, in a precision calculation it is desirable that the Sudakov factors exactly match the ones in the shower evolution.

We here present a new implementation to combine tree-level matrix elements with so-called sector showers \cite{LopezVillarejo:2011ap,Brooks:2020upa}, based on the CKKW-L merging prescription \cite{Catani:2001cc,Lonnblad:2001iq,Krauss:2002up,Lonnblad:2011xx}. In the sector shower framework, only a single splitting kernel contributes at any point in phase space, making the shower operator effectively bijective, i.e., uniquely invertible, while retaining its leading-logarithmic accuracy. Hence, for any given multi-parton configuration, there exists only a single path to every previous shower state and the factorially growing history tree is replaced by a single, linear history branch.
The method presented here, which, in analogy to the MEPS nomenclature, we dub MESS as a shorthand for matrix elements + sector shower, alleviates the scaling of the memory footprint as well as computation time on the parton shower side of multi-jet merging.
The MESS presented here was made publicly available in the \Pythia 8.304 release.

This paper is structured as follows. We review the CKKW-L merging scheme for the \Vincia\ sector shower with a particular focus on the construction of shower histories in \cref{sec:SectorMerging} and validate the new implementation in \cref{sec:Validation}.
Our central results -- the run time scaling and memory footprint of our implementation -- are presented in \cref{sec:Results} before concluding in \cref{sec:Conclusions}.

\section{CKKW-L Merging with Sector Showers}\label{sec:SectorMerging}
Generally, we shall be performing tree-level CKKW-L merging \cite{Lonnblad:2001iq}, following the prescription in \cite{Lonnblad:2011xx} with slight modifications to adapt it to sector showers. 
Although this method is in principle more generally applicable, we will limit our discussion to the \Vincia\ antenna shower, for which the extension to sector showers was discussed in detail in \cite{Brooks:2020upa}. We will briefly comment on the sectorisation of other shower approaches in \cref{sec:SectorShowers}.
For a review of the CKKW and the CKKW-L approach, we refer to \cite{Lavesson:2007uu,Lonnblad:2011xx}; here it shall suffice to present a brief review only.

\subsection{CKKW-L Merging in a Nutshell}\label{sec:CKKWL}
To safely combine multiple event samples, including the effect of shower simulations, without over-counting emissions, the initially inclusive events have to be made exclusive. In order to do so, first a \textquotedblleft merging scale\textquotedblright, $\tms$, has to be defined. It is used to separate the fixed-order and resummation regions, so that each shower-generated jet falls below the merging scale and each matrix-element generated jet falls above it.
Starting from a given $\text{Born}+n$-jet event that passes this constraint, a typical merging algorithm can then be separated into three steps:
\begin{enumerate}
    \item Construct the most likely \textquotedblleft shower history\textquotedblright\ consisting of sequential clustering \textquotedblleft nodes\textquotedblright\
    \item Reweight the event with Sudakov factors to account for unresolved radiation
    \item Reweight the event with $\alphaS$ factors, evaluated at appropriate \textquotedblleft node scales\textquotedblright\ 
\end{enumerate}

The idea of the CKKW-L merging prescription as presented in \cite{Lonnblad:2001iq,Lonnblad:2011xx} is to generate the Sudakov factors in the second step dynamically and in the same way as the parton shower at hand would have done while reaching the given $\text{Born}+n$ configuration, as described below.

For a configuration with $n$ additional jets with respect to the Born configuration, which has been generated according to a tree-level matrix element with a regularisation cutoff $k_\text{cut}$, the most probable shower history is reconstructed in the first step. Denoting the hard event by $\mathcal{H}_{\mathrm{Born}+n}$ and shower states by $\mathcal{S}_{\mathrm{Born}+i}$, this generates a sequence of nodes
\begin{equation}
    \left\{\mathcal{S}_{\mathrm{Born}}, \mathcal{S}_{\mathrm{Born}+1}, \ldots, \mathcal{S}_{\mathrm{Born}+n-1}, \mathcal{H}_{\mathrm{Born}+n} \right\} \, ,
\end{equation}
with a corresponding sequence of node scales, $\left\{\rho_{0}, \rho_{1}, \ldots, \rho_{n-1}, \rho_{n} \right\}$, typically given by the shower evolution variable.

Subsequently, Sudakov form factors are generated by trial showers between history nodes, during which an event is vetoed if a branching between two nodes is produced. This generates no-branching probabilities $\Pi_{\mathcal{S}_{\mathrm{Born}+i}}(\rho_i, \rho_{i+1})$ in the same way the shower had if it would have been started off the reconstructed nodes. As the no-branching probability $\Pi_{\mathcal{S}_i}$ generated by the shower generally differs from Sudakov factors by PDF ratios, the events are additionally weighted by
\begin{equation}
    w_i^\mathrm{PDF} = \frac{f_i(x_i,\rho_i)}{f_i(x_i,\rho_{i+1})} \, .
\end{equation} 
To account for the running of the strong coupling and other higher-order corrections included in the shower evolution, events are weighted with ratios 
\begin{equation}
    w^{\alphaS}_i = \frac{\alpha_\text{s,PS}(\rho_i)}{\alpha_{\text{s,ME}}}
\end{equation}
for each intermediate node, where  $\alpha_\text{s,PS}$ and $\alpha_\text{s,ME}$ reflect the scale and scheme choice of the shower and fixed-order calculation, respectively. 

In the last trial shower step, the treatment differs between intermediate and highest-multiplicity nodes.
The event is vetoed if the trial shower off the hard $\mathcal{H}_{\text{Born}+n}$ configuration generates an emission above the merging scale, $t_n(\mathcal{H}_{\text{Born}+n}) > \tms$ and $n$ is below the maximal number of additional jets $N$. Hard emissions off configurations with the highest jet multiplicity $n\equiv N$ are, however, retained. Here, the notation $t_n(\mathcal{S})$ denotes the evaluation of the state $\mathcal{S}$ with respect to the same metric as used for the merging scale $\tms$. This can be a simple jet-$p_\perp$ cut, the shower evolution variable, or more complicated  definitions including the use of jet clustering algorithms.

A hard $\text{Born}+n$ parton configuration is therefore accepted if and only if the trial showers did not generate additional hard emissions. Accepted events are thus weighted by
\begin{equation}
    \weightCKKWL = \frac{f_n(x_n,\rho_n)}{f_n(x_n,\mu^2_\text{F})} \prod\limits_{i=0}^{n-1} \frac{\alpha_\text{s,PS}(\rho_{i+1})}{\alpha_{\text{s,ME}}} \frac{f_i(x_i,\rho_i)}{f_i(x_i,\rho_{i+1})} \Pi_{\mathcal{S}_{\mathrm{Born}+i}}(\rho_i, \rho_{i+1}) \, ,
\end{equation}
where $\mu^2_\text{F}$ denotes the factorisation scale of the hard process.

\begin{figure}[t]
    \centering
    \includegraphics[width=0.55\textwidth]{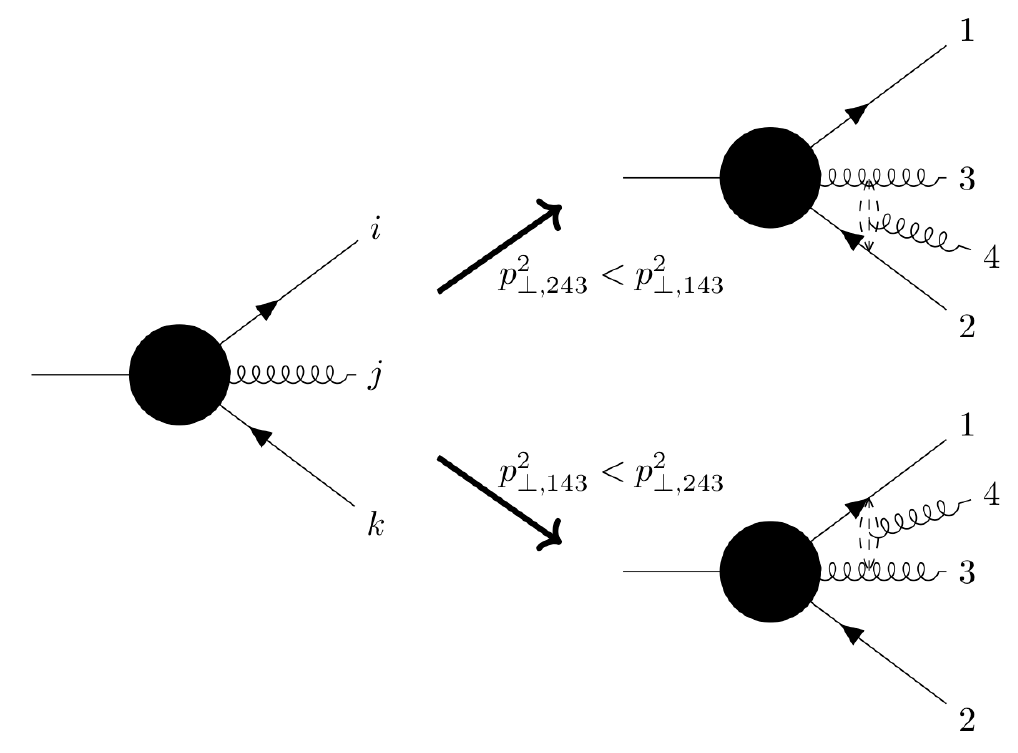}
    \caption{Illustration of the sector-shower evolution off a colour-ordered $Z \to q g\bar q$ configuration. The emission in the $i$-$j$ antenna is accepted if and only if its post-branching $p_\perp$ is the smallest in the tentative post-branching state.}
    \label{fig:sector3to4step}
\end{figure}

\subsection{Sectorised Shower Evolution}\label{sec:SectorShowers}
The \Vincia\ antenna showers are evolved in a generalised \Ariadne $p_\perp$,
\begin{equation}
	p^2_{\perp} = \frac{\bar q^2_{ij} \bar q^2_{jk}}{s_\text{max}} \, ,
    \quad
    \bar q^2_{ij} = \pm [(p_i \pm p_j)^2 - p_I^2] = \begin{cases}
    \displaystyle \sij + m_i^2 + m_j^2 - m_I^2 & i \text{ is final} \\[4mm]
    \displaystyle \sij - m_i^2 - m_j^2 + m_I^2 & i \text{ is initial}
    \end{cases} \, ,
\label{eq:EvolutionVar}
\end{equation}
where capital indices denote pre-branching partons and $\sij = 2p_i \cdot p_j$ with $s_\text{max}$ the maximal invariant of the current antenna,
\begin{equation}
    s_\text{max} = \begin{cases}
    \displaystyle \sIK & \text{final-final} \\
    \displaystyle \sij + \sik & \text{initial-final} \\
    \displaystyle \sik & \text{initial-initial}
    \end{cases} \, .
\end{equation}
In the sector shower formalism, only a single antenna contributes at each point in phase space. In order to nevertheless capture the correct leading-logarithmic behaviour, a single antenna function incorporates both the full soft and the full collinear singularity. The exact form of sector antenna functions is, however, ambiguous and only limited by the constraint that the correct single-unresolved limits are entirely contained within a single function. We refer to \cite{Brooks:2020upa} for a full set of helicity- and mass-dependent sector antenna functions, which, in its colour factor- and coupling-stripped variant, we denote by $\bar a^\text{sct}$ here.
The shower operator is made bijective by rejecting any branching that does not correspond to the most singular configuration in the tentative post-branching state, cf.~\cref{fig:sector3to4step}, defined in terms of the sector resolution variable
\begin{equation}
	Q^2_{\text{res}_j} = \begin{cases}
	\displaystyle p_\perp^2 & \text{if } j \text{ is a gluon} \\[4mm]
 	\displaystyle \bar q^2_{ij} \sqrt{\frac{\bar q^2_{jk}}{s_\text{max}}} & \text{if } j \text{ is a(n) (anti)quark}
	\end{cases}\, . \label{eq:SectorResolutionVar}
\end{equation}
Here, the asymmetric choice for gluon splittings accounts for the fact that the $g_I X_K \mapsto q_i \bar q_j X_k$ branching is not singular in the $j$-$k$ collinear limit, cf.~\cite{LopezVillarejo:2011ap}.
Thus, the sector shower produces no-branching probabilities of the form
\begin{multline}
	\Pi_n (p^2_{\perp,n}, p^2_{\perp, n+1}) = \\ \exp\left(-  4\uppi \sum\limits_{j \in \{n \mapsto n+1\}} \,
	\int \limits_{p^2_{\perp,n+1}}^{p^2_{\perp,n}} \frac{f_i(x_i,p^2_\perp)f_k(x_k,p^2_\perp)}{f_I(x_I,p^2_\perp)f_K(x_K,p^2_\perp)} \, \alphaS(p^2_\perp) \mathcal C_{j/IK}\bar a_{j/IK}(p^2_\perp, \zeta) \, \Theta^\text{sct}_j(p^2_\perp,\zeta) \D \Phi^{\text{ant}}_j\right)\, , \label{eq:noBranching}
\end{multline}
where the Heaviside function $\Theta^\text{sct}$ enforces the constraint that only a single antenna radiates per phase-space point. In general, it depends non-trivially on the post-branching kinematics.

Consequently, any given configuration produced by the sector shower can be uniquely inverted by iteratively minimising \cref{eq:SectorResolutionVar}, effectively yielding a $p^2_\perp$-based jet-clustering algorithm, which, however, still \emph{exactly} represents the (leading-colour) shower history.

Before moving on to the treatment of shower histories, it is worthwhile to discuss how other shower models may be extended to sector showers.
Compared to conventional shower algorithms, sector showers differ mainly in the choice of branching kernels and their phase-space coverage. These are constructed based on the following two requirements \cite{Kosower:1997zr,Kosower:2003bh}:
\begin{enumerate}
    \item a single branching kernel contains all single-unresolved limits of the respective colour dipole
    \item the branching phase space is decomposed into non-overlapping sectors, where only a single branching kernel contributes per sector and the set of sectors provides a decomposition of unity
\end{enumerate}
Other aspects, which would typically be left unchanged when sectorising a shower model, but which will of course affect the resummation of leading logarithms include a judicious choice of the ordering variable \cite{Dokshitzer:2008ia,Nagy:2009re,Skands:2009tb,Bewick:2019rbu,Dasgupta:2020fwr}, correct assignment of colour factors \cite{Friberg:1996xc,Gustafson:1992uh,Dasgupta:2018nvj,Hamilton:2020rcu,Holguin:2020joq,Forshaw:2020wrq}, and the treatment of branching recoils \cite{Dasgupta:2020fwr,Forshaw:2020wrq}. The choices made in the \Vincia sector shower are discussed in detail in \cite{Brooks:2020upa}.

As a second specific example, consider a shower based on Catani-Seymour dipoles \cite{Catani:1996vz,Catani:2002hc}. As opposed to (global) antenna showers, in which the gluon-gluon collinear limits are partitioned between two neighbouring antennae, in Catani-Seymour dipoles it is the soft limit which is partitioned between the two dipole legs. Sector splitting kernels can therefore be constructed by combining two Catani-Seymour dipoles, e.g.\ for gluon emission from a quark-gluon dipole:
\begin{equation}
    V^\text{sct}_{q_ig_j,g_k}(z_i,z_k,y_{ij,k}) = \frac{V_{q_ig_j,k}(z_i,y_{ij,k})}{s_{ij}} + \frac{V_{g_kg_j,i}(z_k,y_{jk,i})}{s_{jk}} \, ,
\end{equation}
where $V_{ij,k}$ denotes the colour-stripped versions of the spin-averaged splitting functions $\langle\boldsymbol{V}_{ij,k}\rangle$ in \cite{Catani:1996vz} and
\begin{equation}
    z_i = \frac{s_{ik}}{s_{ik}+s_{jk}} \, , \quad y_{ij,k} = \frac{s_{ij}}{s_{ij}+s_{jk}+s_{ik}} \, .
\end{equation}
Splitting functions constructed this way fulfil the first requirement above, as they approach the soft eikonal and the correct DGLAP kernels in the respective soft and collinear limits.
Supplemented with a sensible measure of singularity, for instance taken to be $k_\perp^2 = z_i(1-z_i)s_{ij}$, they can be restricted to the appropriate phase space sectors in the same way as in \cref{eq:noBranching},
\begin{equation}
    1 = \sum_j \theta(k^2_{\perp,\text{min}} - k^2_{\perp,j}) \, ,
\end{equation}
therefore fulfilling the second requirement above.

\subsection{Shower Histories}\label{sec:SectorHistories}
The advantage of sector showers is that, at least for gluon emissions at leading colour, there is just a single history, because only one antenna is active at each point in phase space. Thus, the history for gluon emissions may be constructed deterministically by minimising the resolution criterion \cref{eq:SectorResolutionVar}.

\begin{figure}[t]
    \centering
    \includegraphics[width=0.8\textwidth]{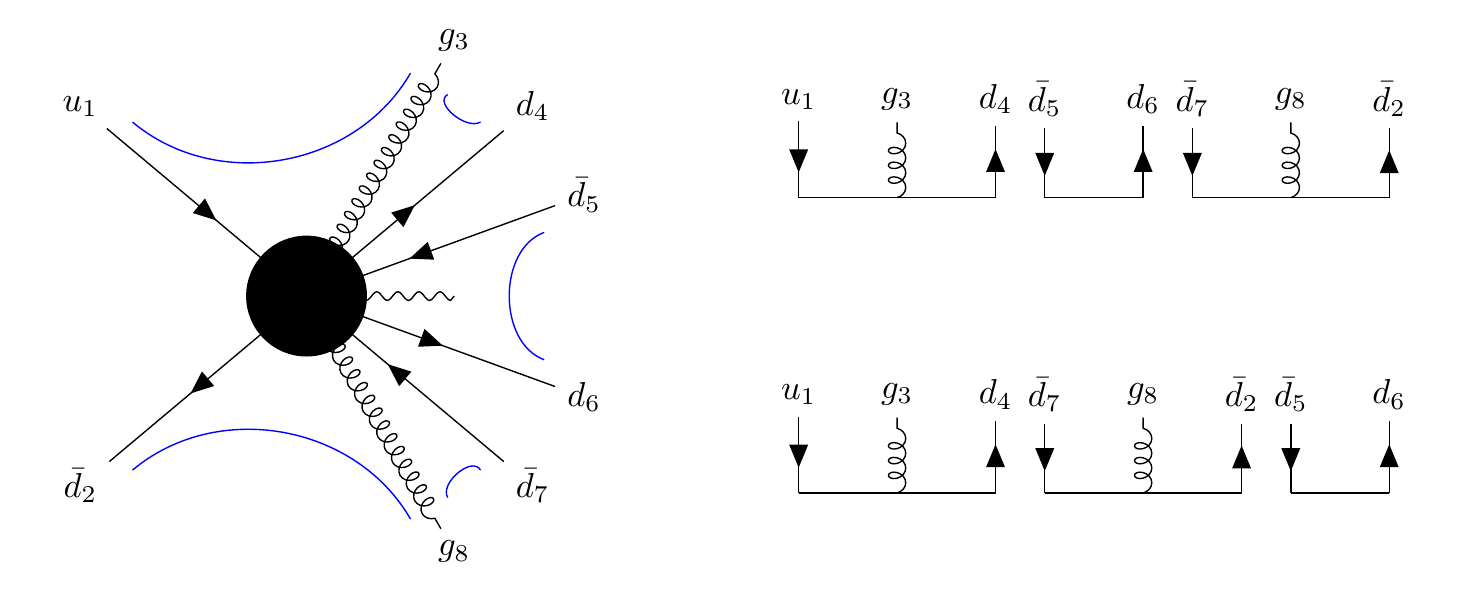}
    \caption{Illustration of colour chains used for history construction. For the colour-ordered configuration (\textit{left}), two different permutations of the same three colour chains contribute (\textit{right}). Both constitute one sector shower history.}
    \label{fig:colchains}
\end{figure}

For quark pairs, however, the situation is less clear-cut. While for gluons, we can use their colour-connected neighbours to determine their possible parents (and the sector criterion determines the order in which they were emitted), for quark-antiquark pairs, there is no colour information to determine which pairs should be clustered. Therefore, in principle we must consider all clusterings of opposite-sign same-flavour pairs\footnote{Assuming that there are no flavour-changing (i.e.\ electroweak) emissions from the shower.}.
We can do this by taking all possible orderings of the colour-ordered chains of gluons, each of which starts on a quark and ends on an antiquark, such that quarks which are juxtaposed are of the same flavour, and clustering all pairs of quark-antiquark pairs results in a viable Born-level topology, cf.~\cref{fig:colchains}.
Nevertheless, this results in a significant gain in efficiency, since the number of orderings only grows as $\Order{\prod_i (n_{q_i} !)}$, where $n_{q_i}$ is the number of pairs of quark flavour $i$, which is at worst $n/2$, but typically much smaller. We emphasise that once a colour ordering of the quark pairs is picked, the shower history is again deterministic.

As we are only trying to capture the leading singular behaviour of the matrix element, we calculate
\begin{equation}
	\vert \mathcal{M}_{\text{Born}} \vert^2 \prod\limits_{i=1}^n \bar a^\text{sct}_i(\{p\}_i) \propto \vert \mathcal{M}_{\text{Born}+n} \vert^2 \, ,
	\label{eq:qqcondition}
\end{equation}
for each viable colour ordering and then maximise over this quantity. Therefore we only need to save maximally two histories concurrently: the current one plus the \textquotedblleft best-so-far\textquotedblright, i.e., that which maximises \cref{eq:qqcondition}.

In summary, the sector shower history is constructed as follows:
\begin{enumerate}
    \item Find all colour-connected chains of gluons.
    \item Find all possible orderings of colour chains compatible with the Born-level process.
    \item For each available permutation:
    \begin{itemize}
         \item Sequentially perform the clustering which corresponds to the minimal value of the resolution criterion $Q^2_\mathrm{res}$, cf.\ \cref{eq:SectorResolutionVar}. For each state $\mathcal{S}_n$:
         \begin{itemize}
            \item For all gluons and internal quark pairs, calculate the sector resolution variable. Note that for quarks, there is an ambiguity in the recoiler, so there are two antennae per quark pair.
            \item Cluster the partons which correspond to the minimal value of the sector resolution variable \cref{eq:SectorResolutionVar}. 
            \item Reconstruct the $(n-1)$-parton kinematics using the exact inverse kinematics map, cf.~\cite{Brooks:2020upa}.
            \item Store mother/daughter information.
            \item Update the colour chain information.  
            \item Calculate the evolution variable \cref{eq:EvolutionVar} for the branching (in general not the same as the sector resolution variable).
            \item Calculate the sector antenna function from the invariants of the pre-clustering partons.
        \end{itemize}
        \item Retain the history for the current permutation only if it corresponds to the maximal value of \cref{eq:qqcondition} so far.
    \end{itemize} 
\end{enumerate}
Below, we address some subtleties connected to the construction of sector shower histories.

\subsubsection*{Commensurate Resolution Scales}
We construct the sector shower history by minimising the sector resolution variable \cref{eq:SectorResolutionVar}. In principle, however, two (or more) different clusterings may have very similar sector resolutions and it may seem unreasonable to choose one over the other. In those cases, one could consider to pick one of the commensurate-resolution clusterings randomly. However, given that this would destroy the merit of the sector shower being uniquely invertible, as the same would have to be done in the shower evolution, we refrain from this procedure and always pick the one with the (slightly) smaller scale. Given that this precisely mimicks the sector shower behaviour, this is a well-motivated choice. Nevertheless, the effect of randomising the choice of such clusterings is an interesting subject for a later study and could be used for uncertainty estimates.

\begin{figure}[ht]
    \centering
    \includegraphics[width=0.35\textwidth]{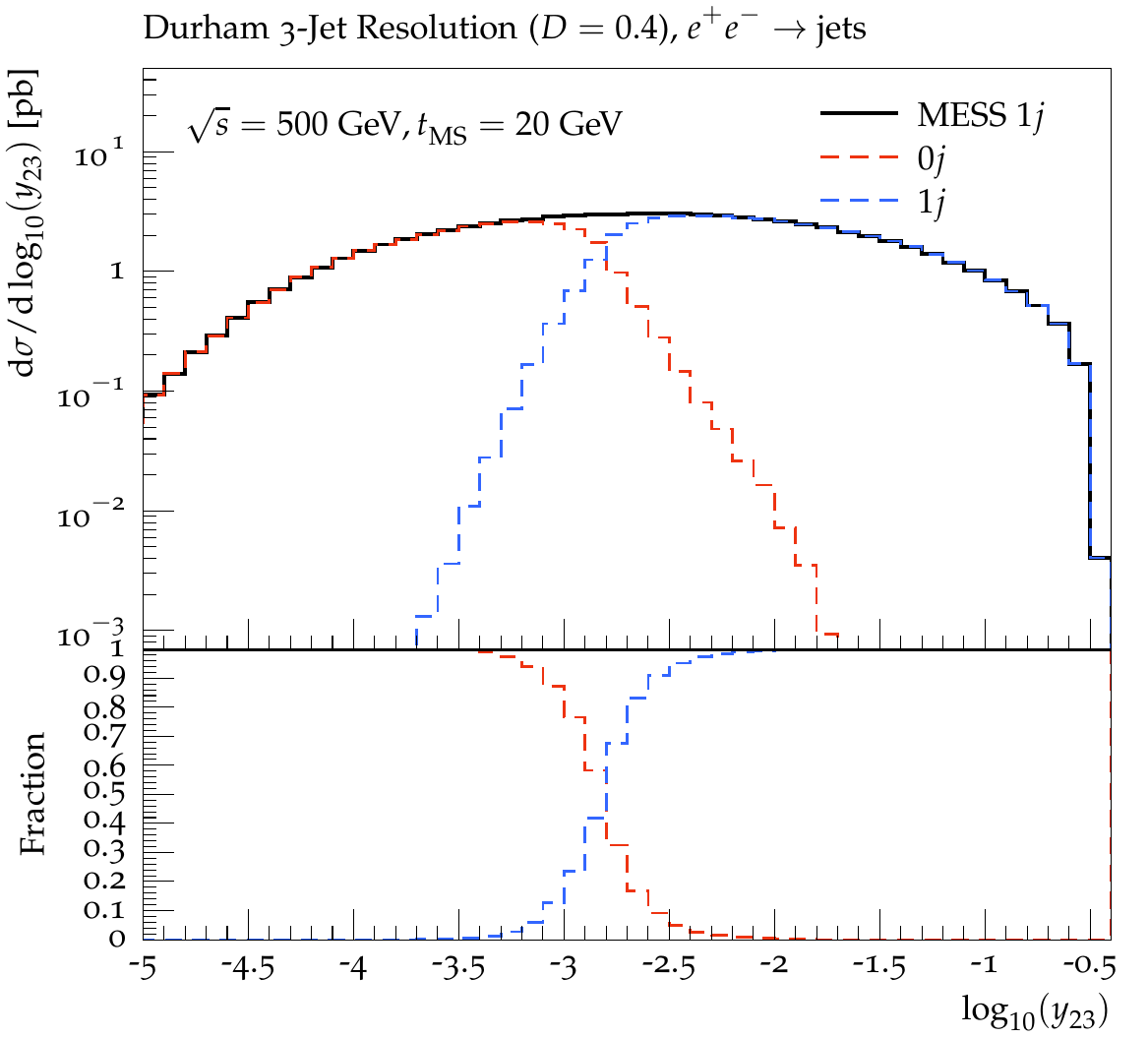}
    \includegraphics[width=0.35\textwidth]{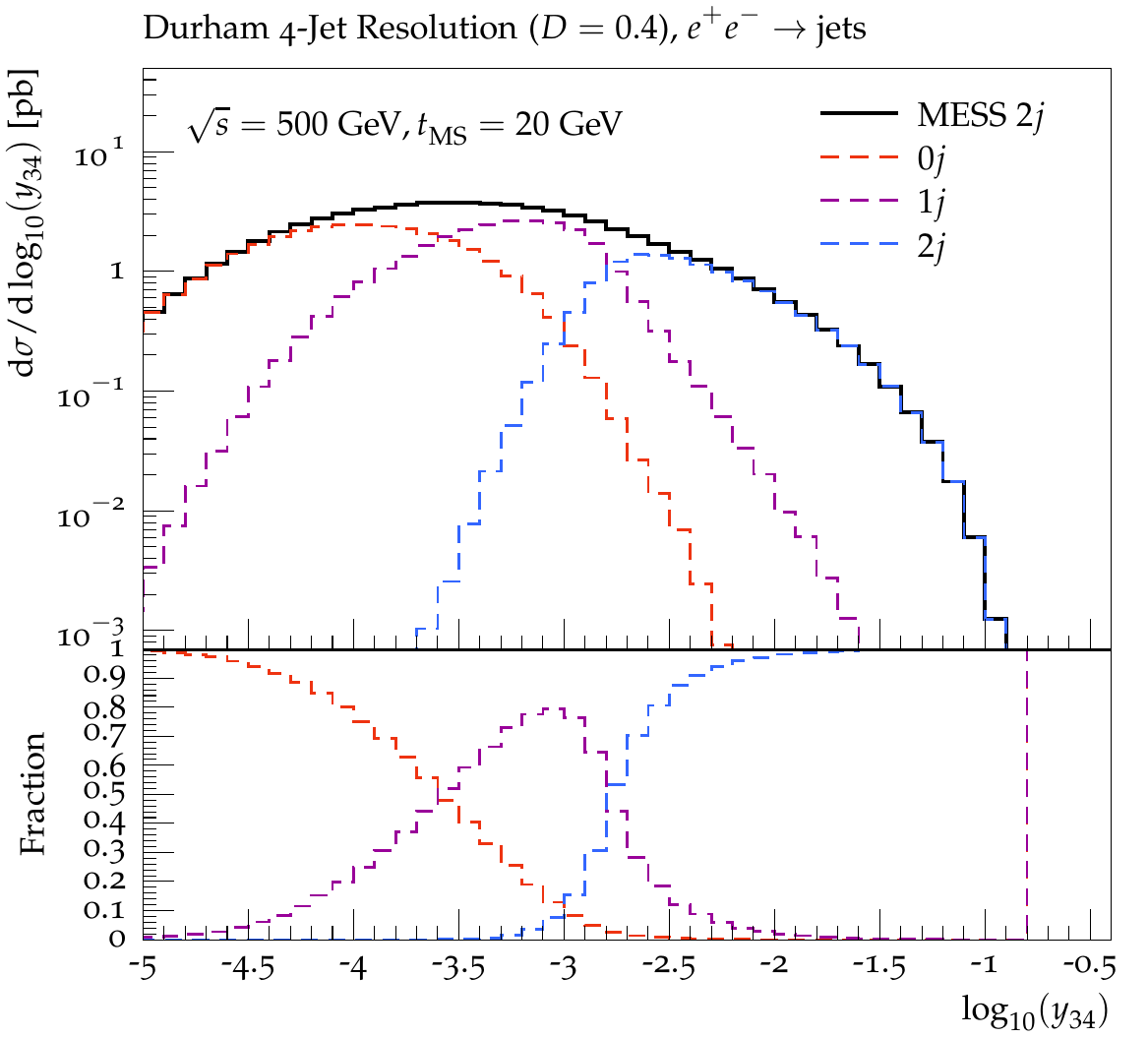}
    \caption{Contributions of the individual hard-event samples in Durham 3-jet (\textit{left}) and 4-jet (\textit{right}) resolution scales in $e^+e^- \to \mathrm{jets}$ at $\sqrt{s} = 500~\giga e\volt$.}
    \label{fig:eeContributions}
	\bigbreak
    \includegraphics[width=0.35\textwidth]{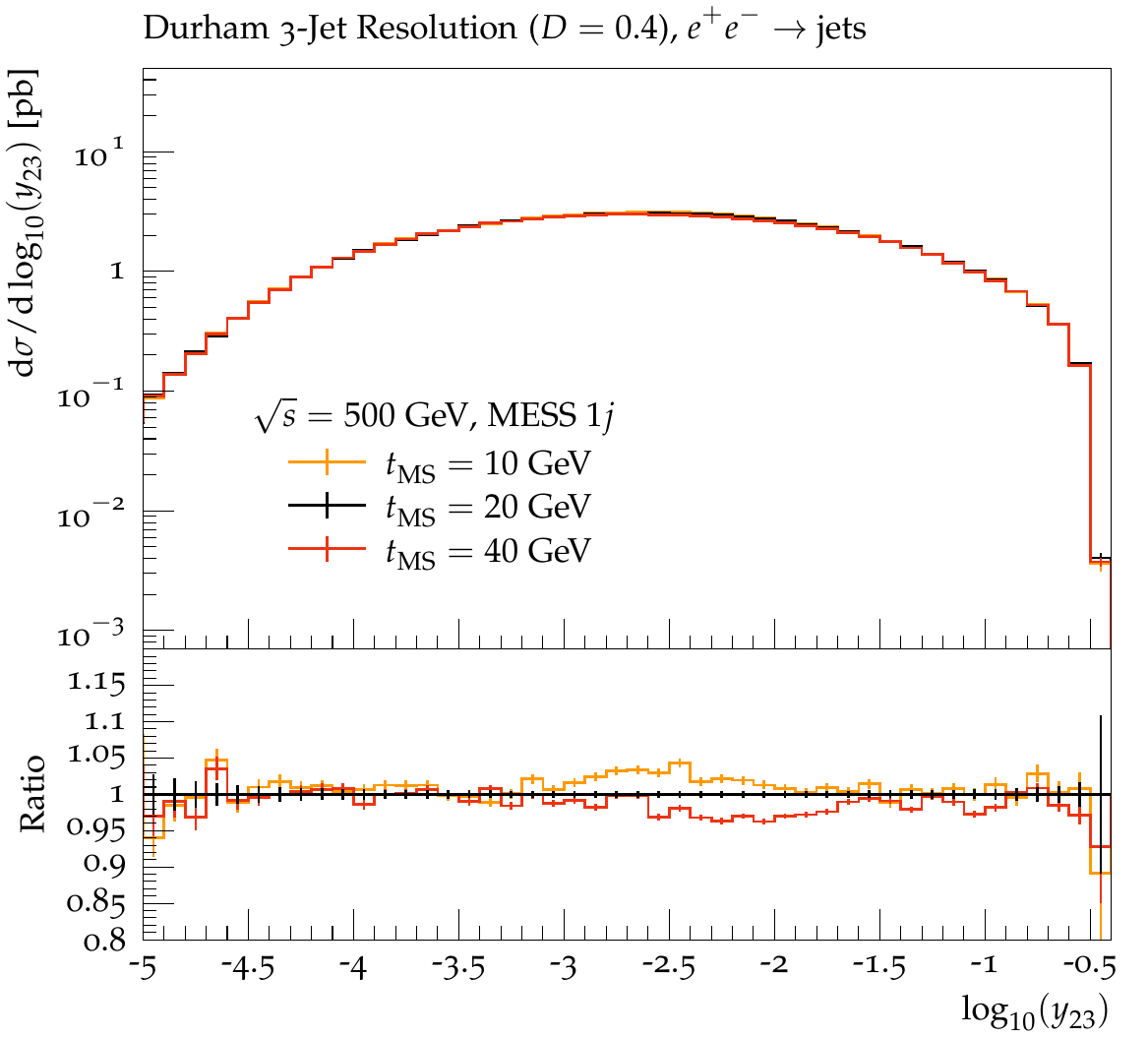}
    \includegraphics[width=0.35\textwidth]{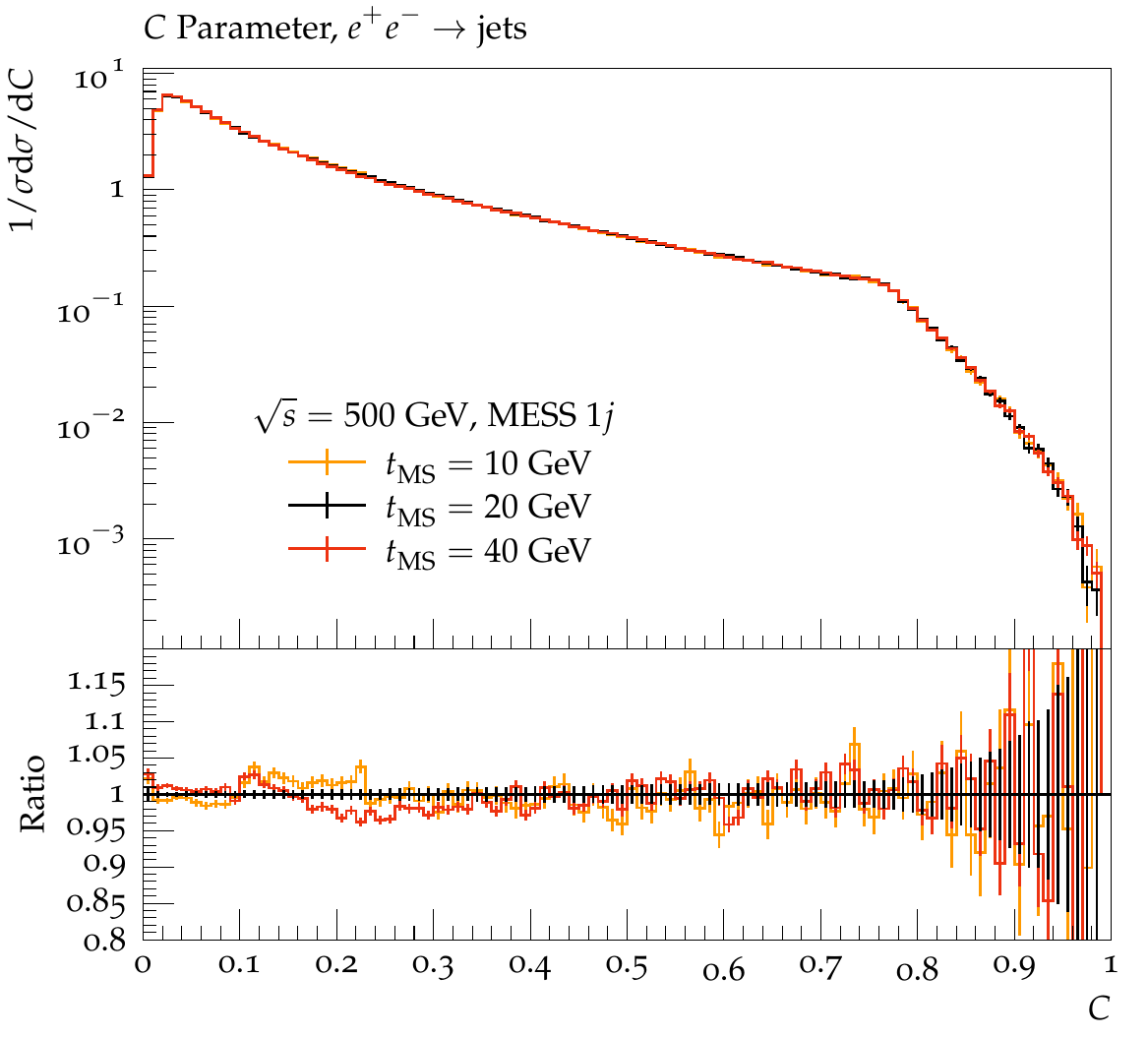}
    \caption{Influence of the merging scale choice on sector shower plus one-jet merged predictions of the Durham 3-jet (\textit{left}) and the $C$ parameter (\textit{right}) distribution in $e^+e^- \to \mathrm{jets}$ at $\sqrt{s} = 500~\giga e\volt$.}
    \label{fig:eeMergingScaleVar}
\end{figure}

\subsubsection*{Unordered Histories}
Although the history is constructed based on minimising the sector resolution, \cref{eq:SectorResolutionVar}, there is no guarantee that this produces a clustering sequence ordered in either the resolution or the evolution variable, \cref{eq:EvolutionVar}.
As the sector shower is based purely on $2\mapsto 3$ branchings, it will never populate regions of phase space with branchings unordered in the evolution variable.
Hence, no Sudakov factors must be included for unordered (sub-)sequences.
This is similar to the treatment in \cite{Lonnblad:2011xx}; we note also that an unordered history is only selected if no ordered one exists.

\subsubsection*{Incomplete Histories}
Occasionally it may occur that it is impossible to perform any (further) parton shower clusterings. Physically, these topologies correspond to states that cannot be reached by the parton shower from any lower multiplicity state. Therefore, there is no danger of double-counting with the lower-multiplicity states, and these states are treated as coming from separate Born configurations.
In the event that there are multiple colour histories, we must calculate a modification to 
\cref{eq:qqcondition} as our criterion to maximise, namely:
\begin{equation}
	\vert \mathcal{M}_{\text{Born+m}} \vert^2 \prod\limits_{i=m}^{n} \bar a^\text{sct}_i(\{p\}_i) \propto \vert \mathcal{M}_{\text{Born}+n} \vert^2,
	\label{eq:incomplete}
\end{equation}
where $m$ is the number of additional emissions relative to the Born in the maximally clustered node of the incomplete history. This still allows to select the most singular path, since in effect this compares $\vert \mathcal{M}_{\text{Born}+n} \vert^2 $ with $\vert \mathcal{M}_{\text{Born}+m} \vert^2 \prod_{i=m}^n \bar a^\text{sct}_i(\{p\}_i)$; since the latter captures the singularity structure of the former, it is a fair comparison.
We follow the procedure of \cite{Lonnblad:2011xx} and accept an incomplete history only if no colour permutation with a complete one exists. 

\begin{figure}[t]
    \centering
    \includegraphics[width=0.35\textwidth]{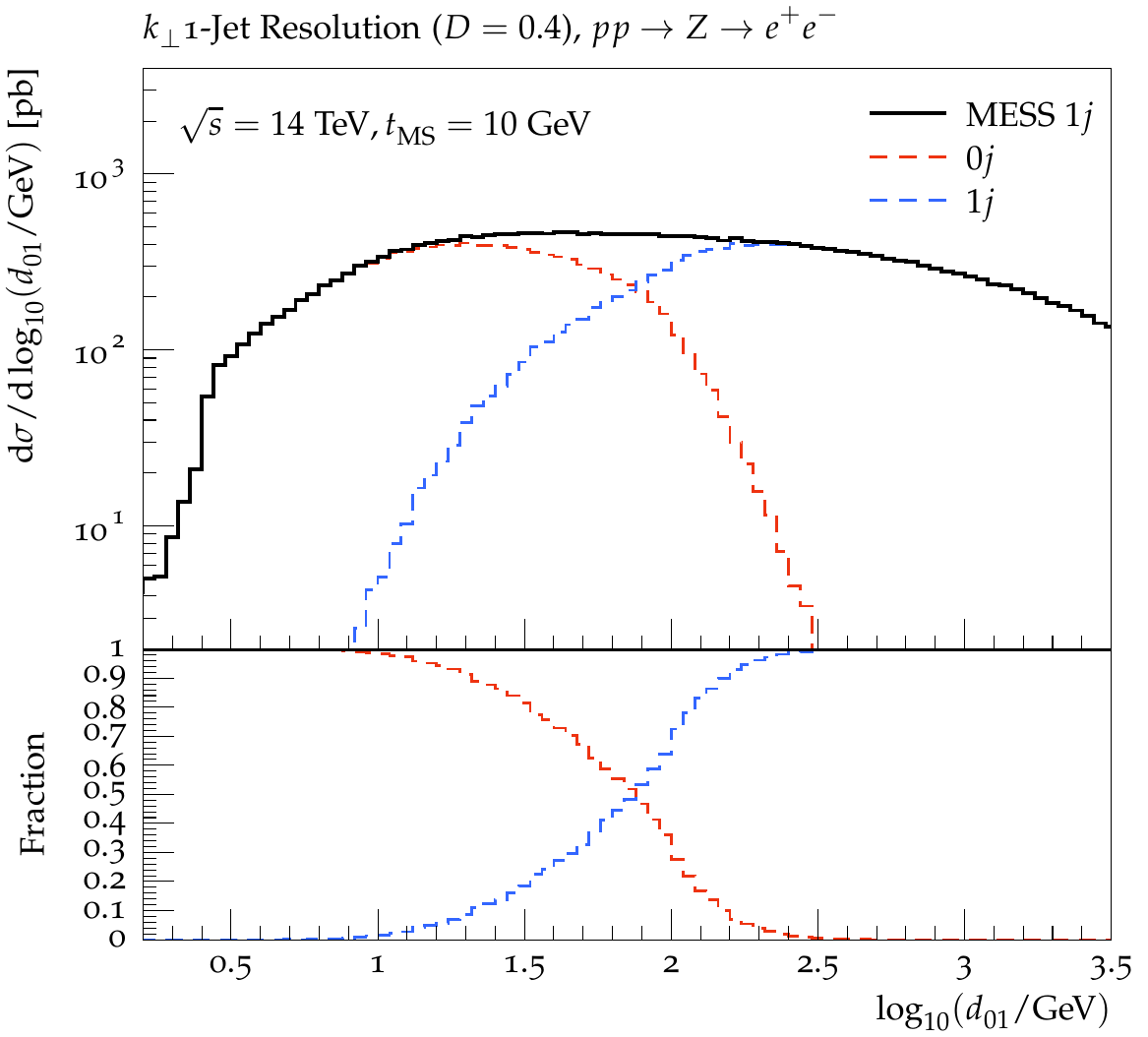}
    \includegraphics[width=0.35\textwidth]{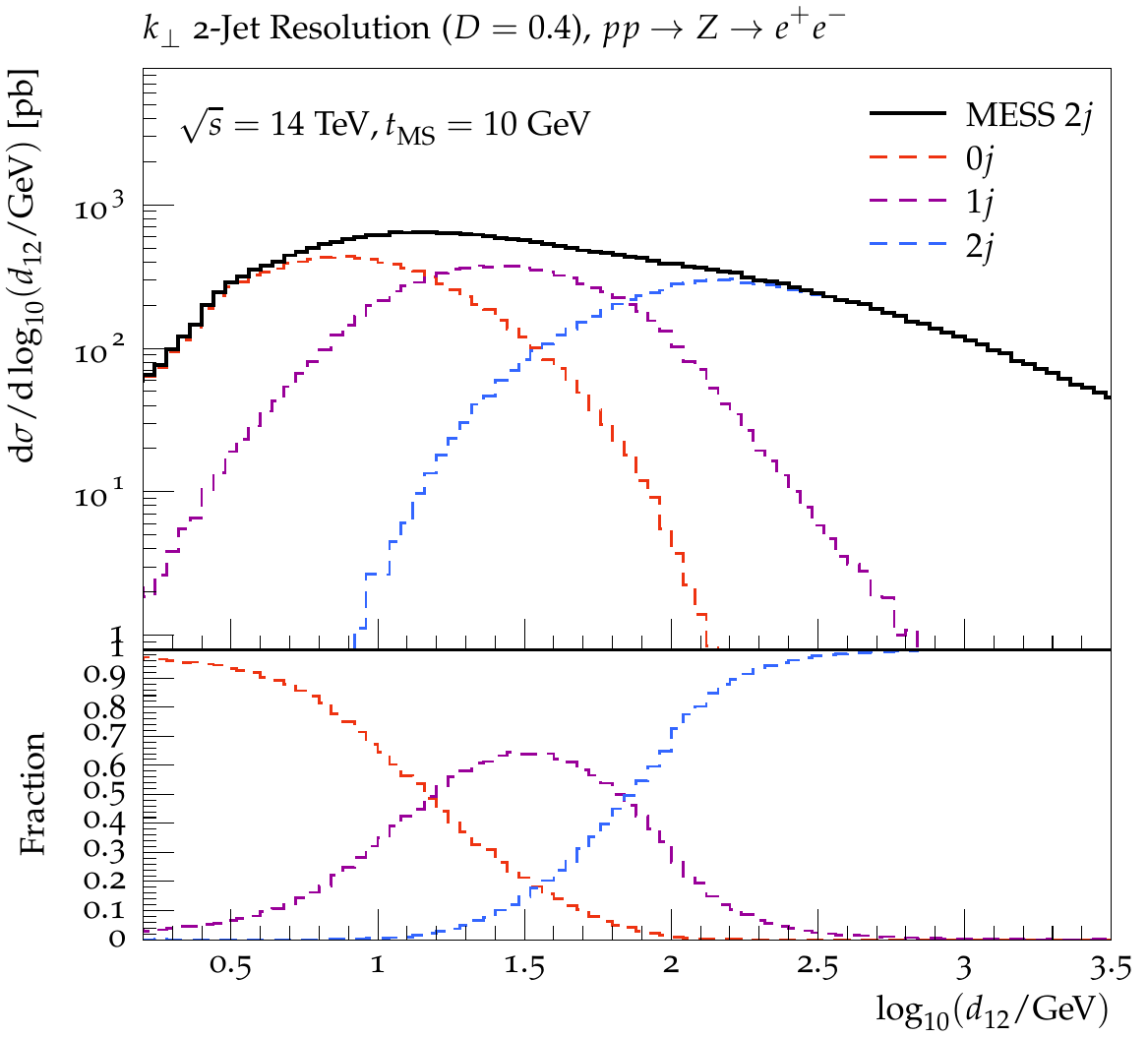}
    \caption{Contributions of the individual hard-event samples in $k_\perp$ 1-jet (\textit{left}) and 2-jet (\textit{right}) resolution scales in the electron channel of $Z$ production $pp \to Z + \mathrm{jets}$ at $\sqrt{s} = 14~\tera e\volt$.}
    \label{fig:zContributions}
	\bigbreak
    \includegraphics[width=0.35\textwidth]{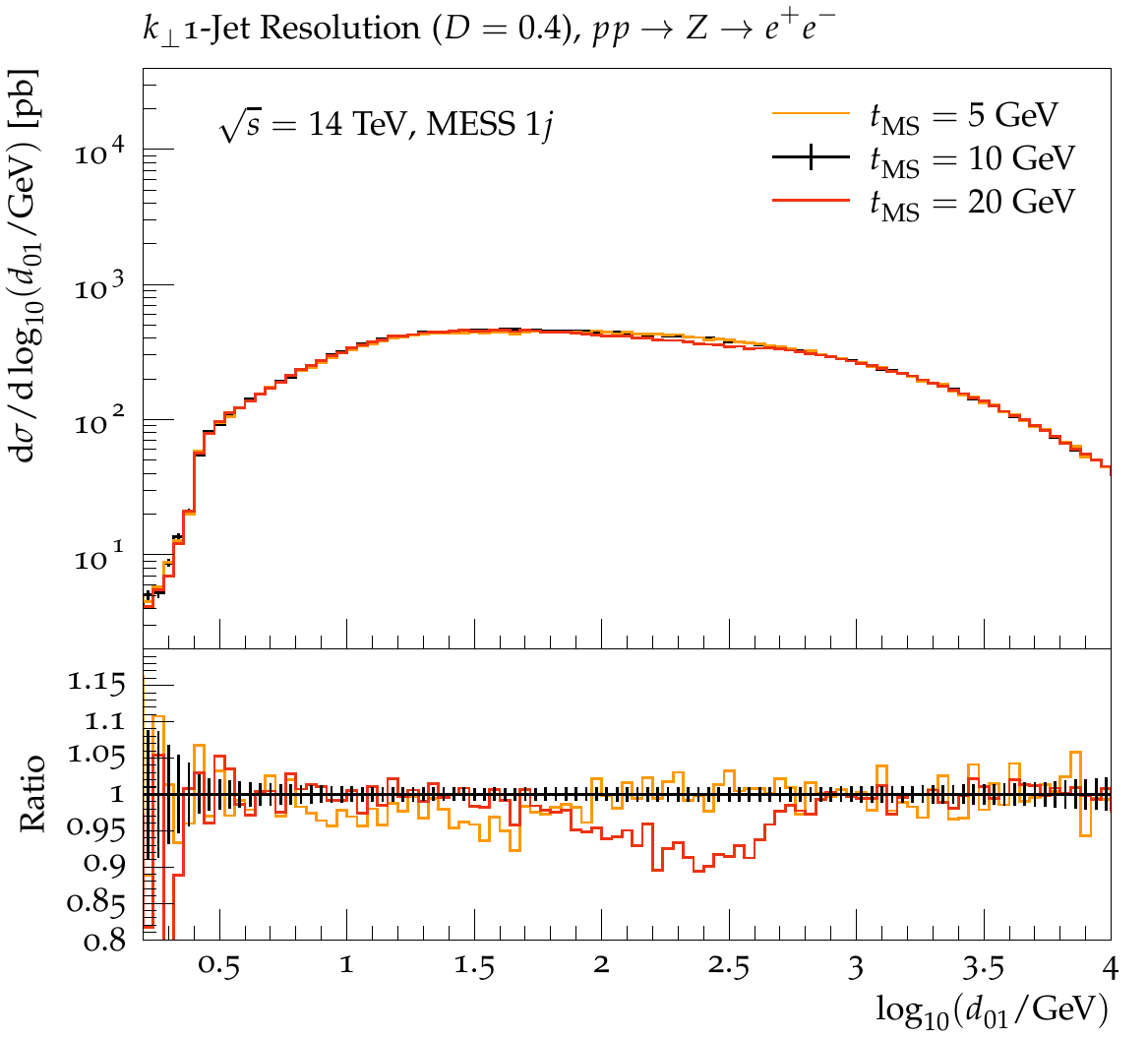}
    \includegraphics[width=0.35\textwidth]{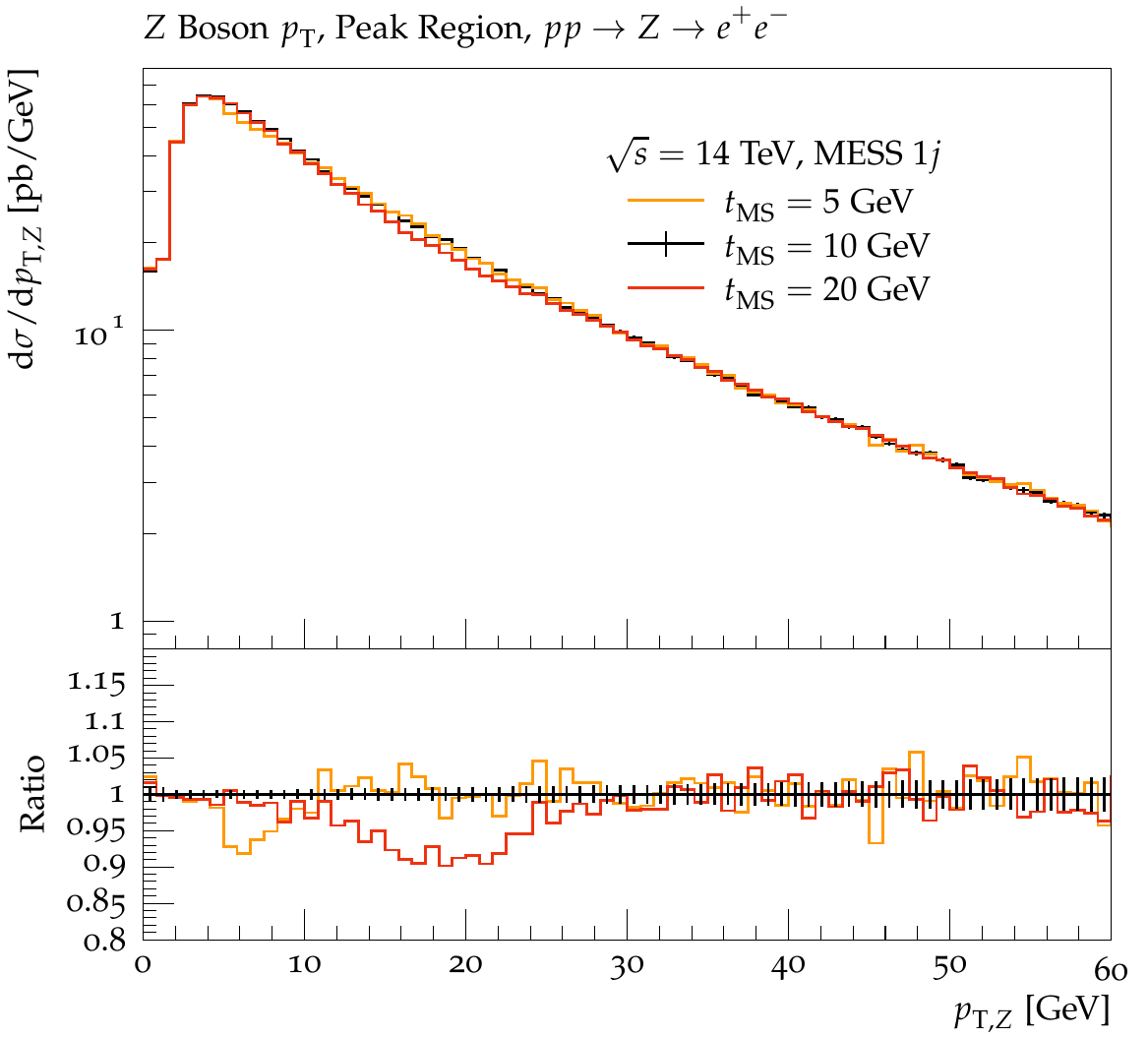}
    \caption{Influence of the merging scale choice on sector shower plus one-jet merged predictions of the $k_\perp$ 1-jet (\textit{left}) and the $Z$ boson transverse momentum (\textit{right}) distribution in $pp \to Z + \mathrm{jets}$ at $\sqrt{s} = 14~\tera e\volt$.}
    \label{fig:zMergingScaleVar}
\end{figure}

\subsubsection*{Interleaved Multi-Parton Interactions }
In the context of interleaved showers for hadronic initial states, it is possible
that the trial shower may generate a multi-parton interaction (MPI) \textquotedblleft emission\textquotedblright\ from an intermediate clustered 
state in the history. Since such topologies are not reachable by the matrix-element
and it would not be physical to limit the scale of MPI to below the merging scale,
such \textquotedblleft new\textquotedblright\ topologies are taken to replace the original event, and showering continues from the scale at which the MPI was generated. This is precisely the same treatment as \cite{Lonnblad:2011xx}.

\begin{figure}[t]
    \centering
    \includegraphics[width=0.35\textwidth]{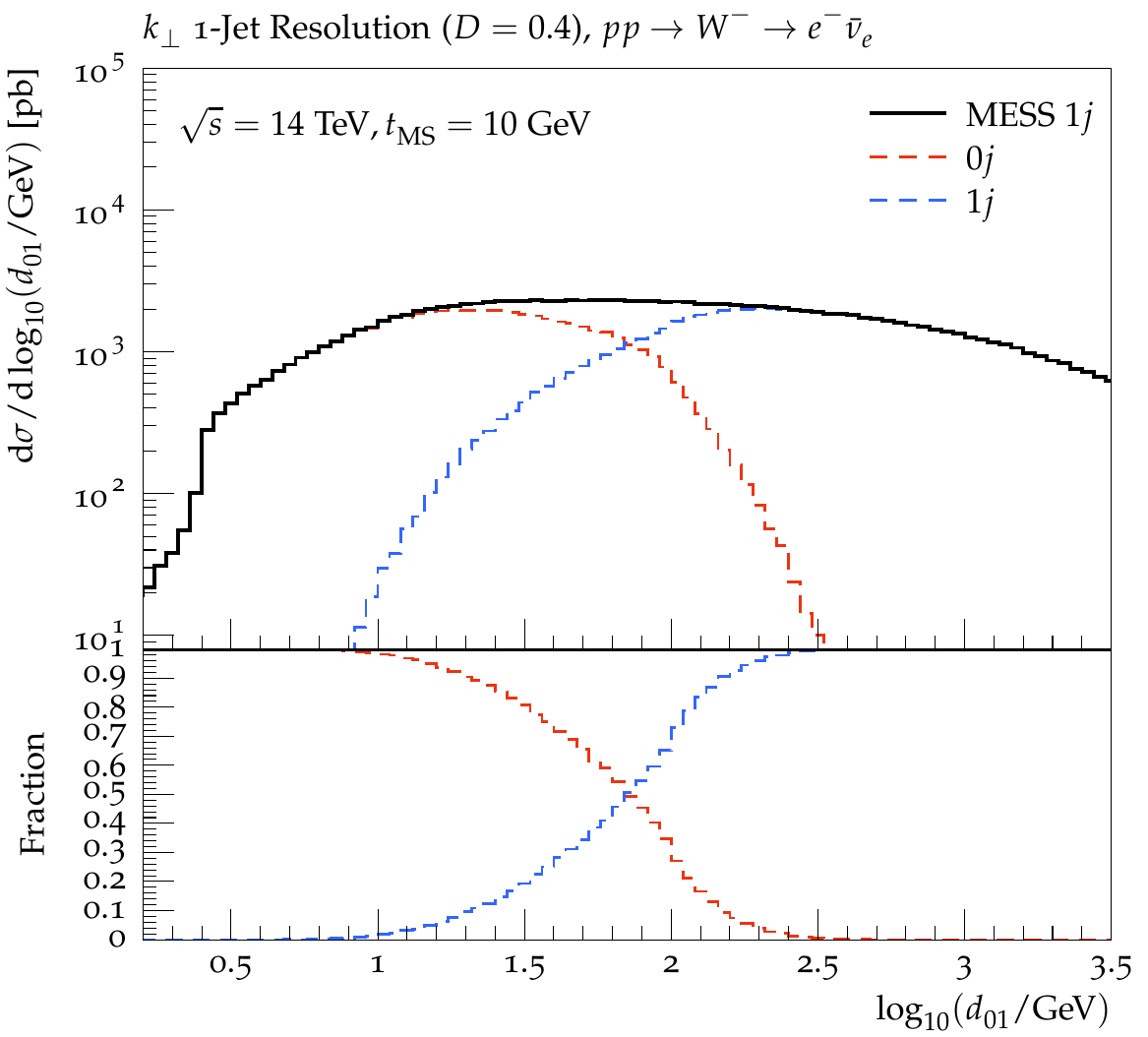}
    \includegraphics[width=0.35\textwidth]{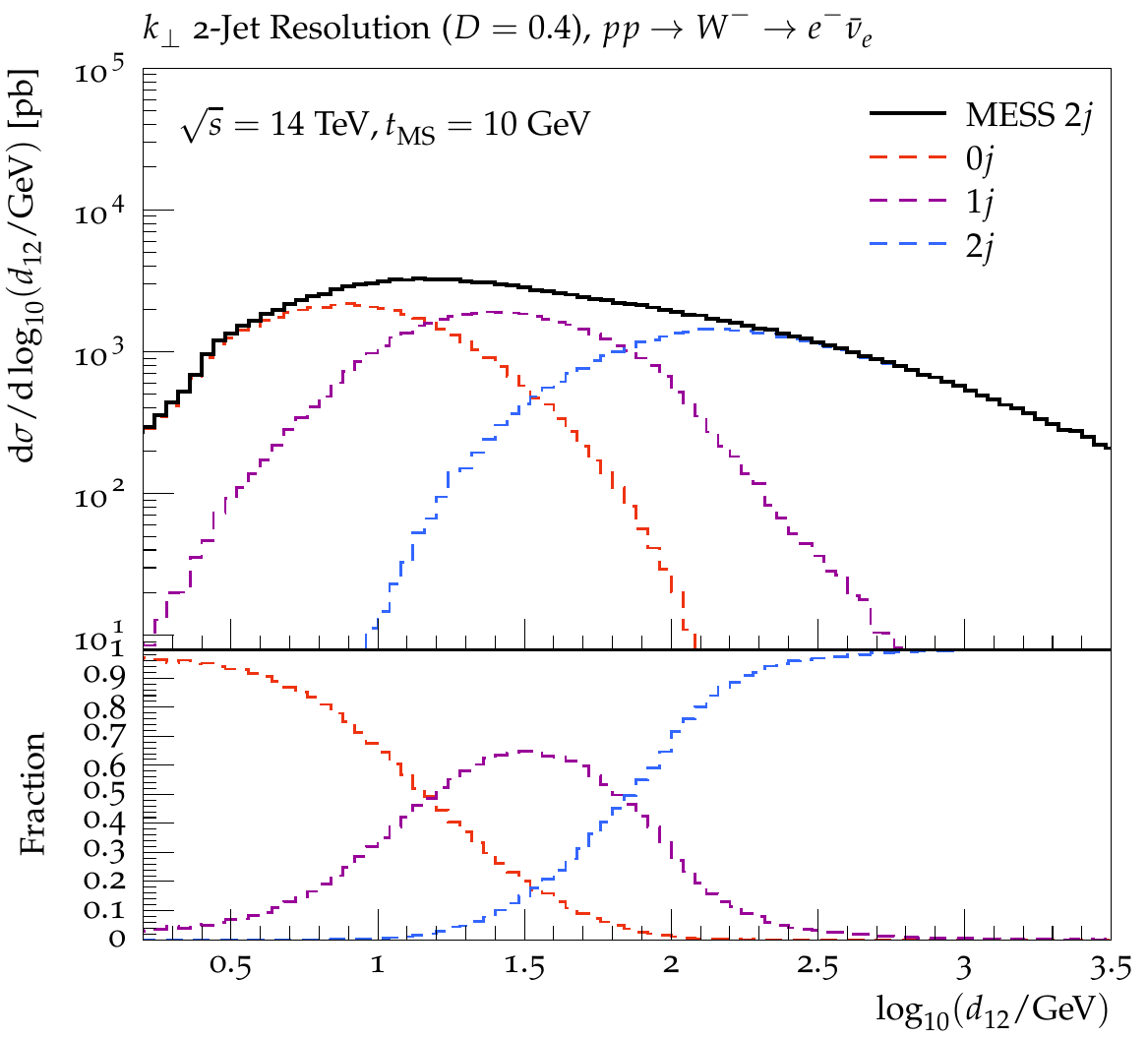}
    \caption{Contributions of the individual hard-event samples in $k_\perp$ 1-jet (\textit{left}) and 2-jet (\textit{right}) resolution scales in the electron channel of $W^-$ production $pp \to W^- + \mathrm{jets}$ at $\sqrt{s} = 14~\tera e\volt$.}
    \label{fig:wmContributions}
	\bigbreak
    \includegraphics[width=0.35\textwidth]{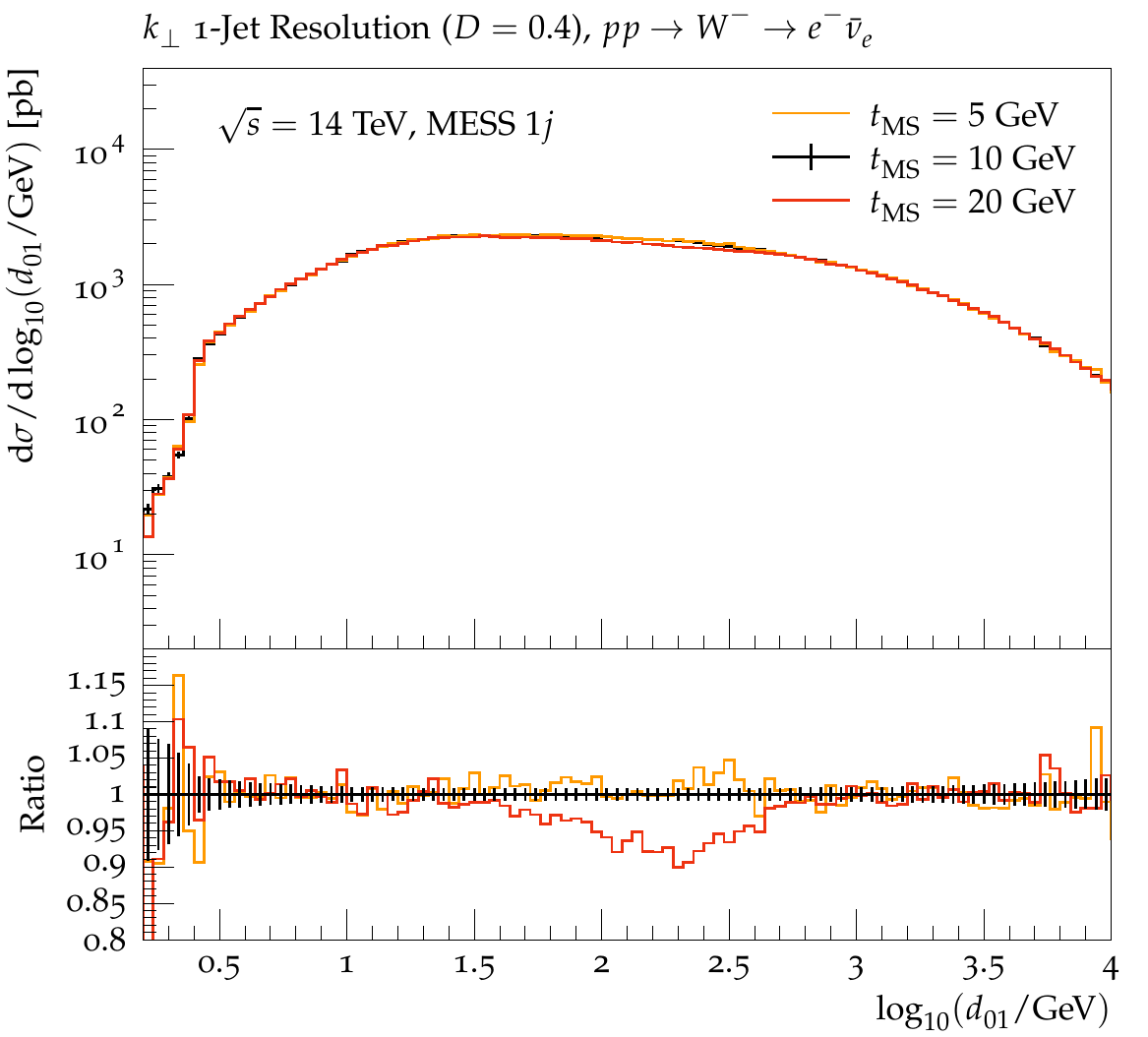}
    \includegraphics[width=0.35\textwidth]{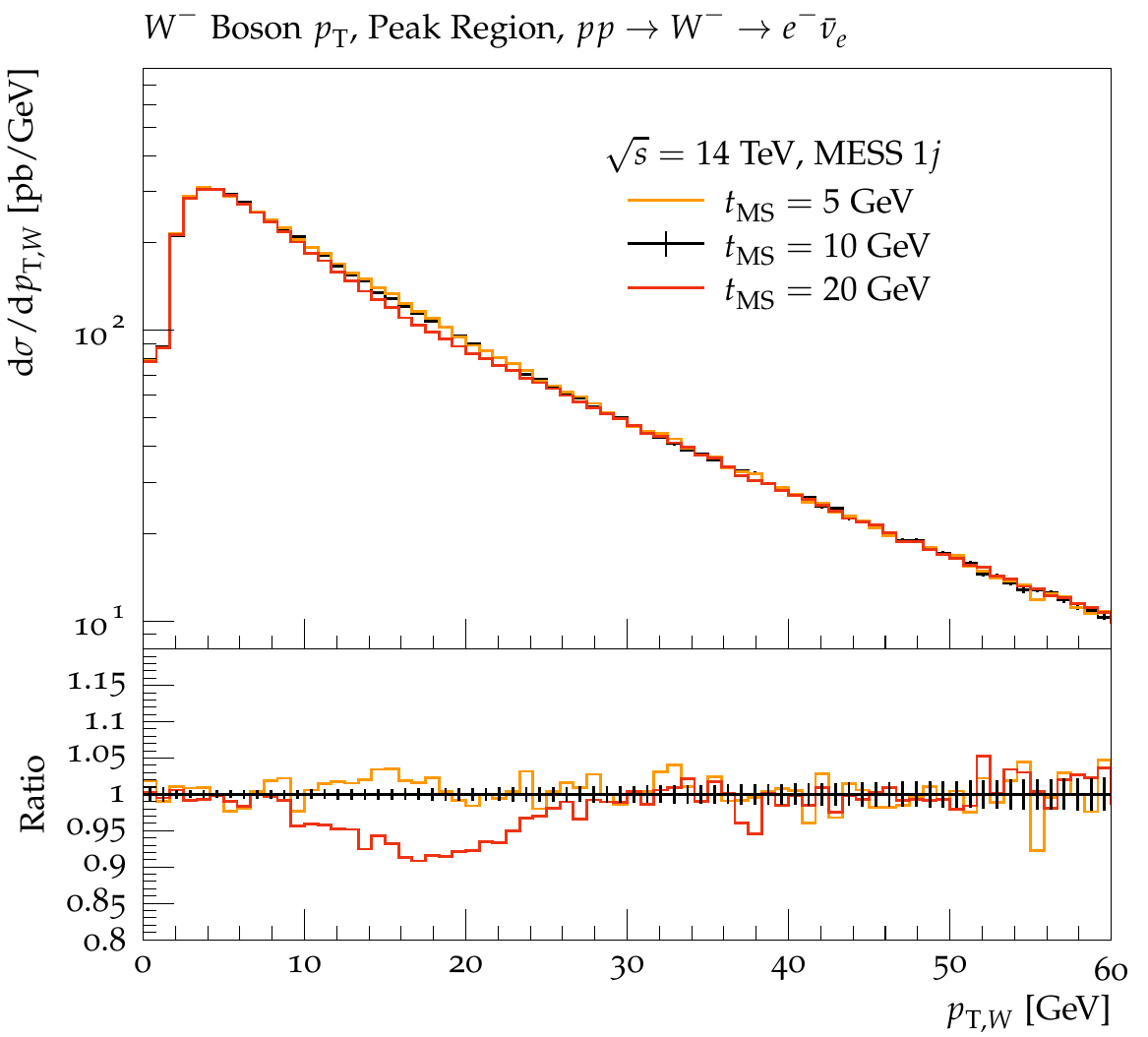}
    \caption{Influence of the merging scale choice on sector shower plus one-jet merged predictions of the $k_T$ 1-jet (\textit{left}) and the $W$ boson transverse momentum (\textit{right}) distribution in $pp \to W^- + \mathrm{jets}$ at $\sqrt{s} = 14~\tera e\volt$.}
    \label{fig:wmMergingScaleVar}
\end{figure}

\subsubsection*{Scale definitions}
In the sector shower merging algorithm, up to four different scale definitions may be present:
\begin{enumerate}
    \item the shower evolution variable $p_\perp$
    \item the sector resolution variable $Q_\text{res}$
    \item the merging scale $\tms$
    \item the matrix element cutoff $k_\text{cut}$
\end{enumerate}
The sector resolution variable $Q_\text{res}$ is only used to construct the shower history and does not play a role in the merging algorithm beyond that. If the other three scales are not chosen to coincide, care has to be taken to neither over- nor undercount phase space regions.

To ensure a smooth transition between the shower evolution variable and the merging scale, we reject hard configurations if any of the intermediate states violates the merging scale cut, i.e. if $t_i(\mathcal S_{\mathrm{Born}+i}) < \tms$. This is different to the implementation in \Pythia, where intermediate nodes are not required to be above the merging scale, as multiple shower paths contribute to the same phase space point. We refrain from this treatment, as in our case, given the same hard configuration multiple times, the sector shower history will always be the same.
This, however, is not sufficient to ensure that the hard phase space is saturated when the matrix element cutoff is chosen with respect to a different scale definition than the merging scale. This will only be the case when the available phase space with the merging scale cut is a subset of the phase space with the matrix element cut, $\Phi_{>t_\text{MS}} \subset \Phi_{>k_\text{cut}}$. 

\subsection{The Full Algorithm}\label{sec:FullAlgo}
For completeness, we here summarise the full CKKW-L merging algorithm for sector showers, closely following \cite{Lonnblad:2011xx}:
\begin{enumerate}[label=(\arabic*)]
 \item Pick a hard event $\mathcal H_{\text{Born}+n}$ containing $n$ additional partons relative to the Born topology:
 \begin{itemize}
    \item If the hard configuration does not satisfy the merging scale cut, i.e., $t_n(\mathcal H_{\text{Born}+n}) < \tms$, veto the event and start from (1). 
    \item For each viable colour-ordering, reconstruct the (deterministic) sequence of shower states $\mathcal{S}_{\text{Born}+i}$,
    \begin{equation}
        \left \{\mathcal{S}_{\text{Born}}, \mathcal{S}_{\text{Born}+1}, \ldots , \mathcal{S}_{\text{Born}+n-1}, \mathcal{H}_{\text{Born}+n} \right\}
    \end{equation}
    with a corresponding sequence of evolution variables
     \begin{equation}
        \left \{p^2_{\perp,0}, p^2_{\perp,1}, \ldots , p^2_{\perp,n-1}, p^2_{\perp,n} \right\} \, .
    \end{equation}
    Here, $p^2_{\perp,i}$ denotes the sector shower evolution scale of the branching to produce each state and $p^2_{\perp,0}$ is the kinematic limit of the Born process, i.e., $p^2_{\perp,1}$ is the scale to produce the first branching, $p^2_{\perp,2}$ is the scale to produce the second branching and so on.
    \item If any state does not satisfy the merging scale cut, i.e., if $t_i(\mathcal S_{\text{Born}+i}) < \tms$ for any $0 \leq i < n$, veto the event and start from (1).
 \end{itemize}
 \item For each pair of scales $(p^2_{\perp,i}, p^2_{\perp,i+1})$, where $m \leq i < n$ ($m=0$ for complete histories):
 \begin{itemize}
    \item If the pair is unordered, $p^2_{\perp,i+1} > p^2_{\perp,i}$, move to the next pair.
    \item Else, perform trial shower between the two scales:
    \begin{itemize}
     \item If the generated state $\mathcal{S}_{\text{Born}+j}$ has a MPI, accept event and move to step (3).
     \item Else, if $\mathcal{S}_{\text{Born}+j}$ has $p^2_{\perp,j} > p^2_{\perp,i+1}$, veto the event and start from (1).
    \item Else, calculate the weights
    \begin{equation}
         w^{\alphaS}_i = \frac{\alpha_\text{s,PS}(p^2_{\perp,i+1})}{\alpha_\text{s,ME}} \, , \quad
         w^{\text{PDF}}_i = \frac{f^A_i(x^A_i, p^2_{\perp,i})}{f^A_i(x^A_i, p^2_{\perp,i+1})}\frac{f^B_i(x^B_i, p^2_{\perp,i})}{f^B_i(x^B_i, p^2_{\perp,i+1})}
    \end{equation}
    \end{itemize}
 \end{itemize}
 \item If the event was not vetoed:
 \begin{itemize}
     \item Multiply the event weight by
     \begin{equation}
        \weightCKKWL = \frac{f^A_n(x^A_n, p^2_{\perp,n})}{f^A_n(x^A_n, \mu^2_\text{F})}\frac{f^B_n(x^B_n, p^2_{\perp,n})}{f^B_n(x^B_n, \mu^2_\text{F})} \prod\limits_{i=0}^{n-1} w^{\alphaS}_i w^{\text{PDF}}_i \, .
    \end{equation}
    This is a probabilistic way to generate the factor $\prod\limits_{i=0}^{n-1} w_i^{\alphaS} \Delta(p^2_{\perp,i},p^2_{\perp,i+1})$.
    \item Start the regular parton shower from the state $\mathcal{H}_{\text{Born}+n}$ at scale $p^2_{\perp,n}$. If $n+1 \leq N$ veto the event and start from (1) if $t_{n+1}(\mathcal S_{(\text{Born}+n)+1}) > \tms$.
 \end{itemize}
 \item Start over from (1).
\end{enumerate}
The algorithm outlined above has been implemented for the \Vincia\ parton shower within the \Pythia~8.3 event generator. Notwithstanding the use of some existing data structures in \Pythia, our implementation is largely
independent from that of the original CKKW-L merging algorithm.

\begin{figure}[t]
	\centering
	\includegraphics[width=0.35\textwidth]{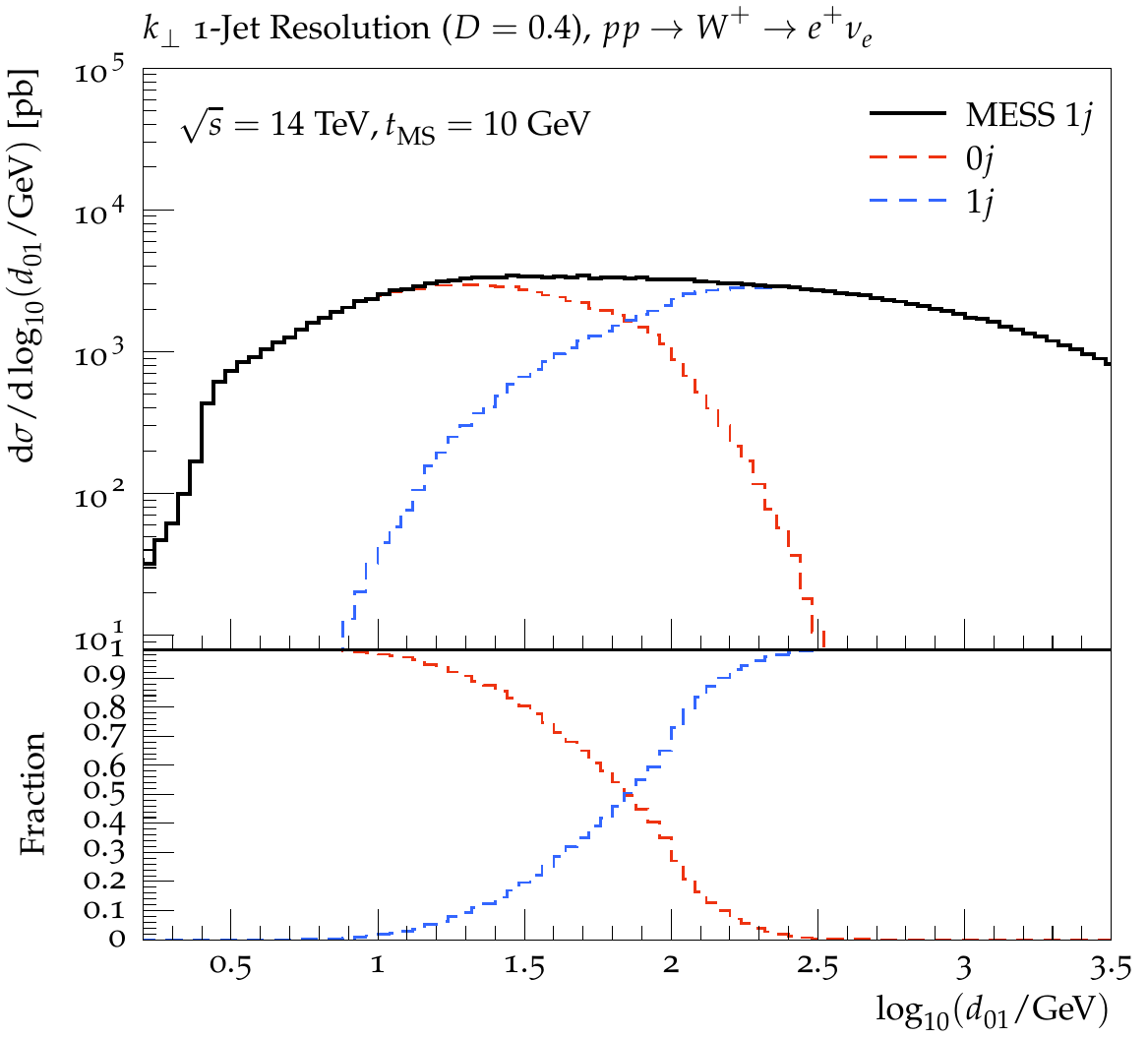}
	\includegraphics[width=0.35\textwidth]{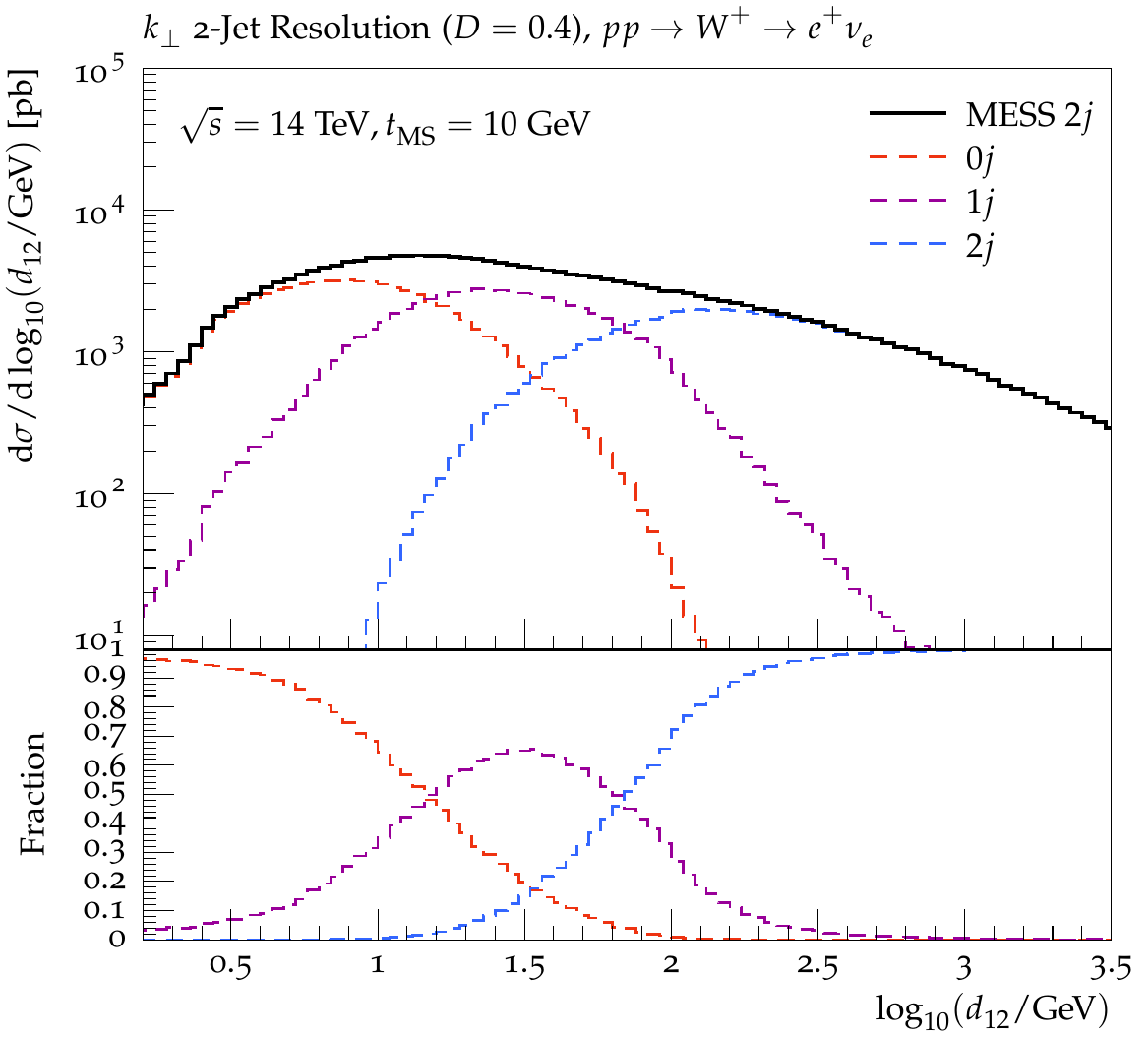}
	\caption{Contributions of the individual hard-event samples in $k_\perp$ 1-jet (\textit{left}) and 2-jet (\textit{right}) resolution scales in the electron channel of $W^+$ production $pp \to W^+ + \mathrm{jets}$ at $\sqrt{s} = 14~\tera e\volt$.}
	\label{fig:wpContributions}
	\bigbreak
	\includegraphics[width=0.35\textwidth]{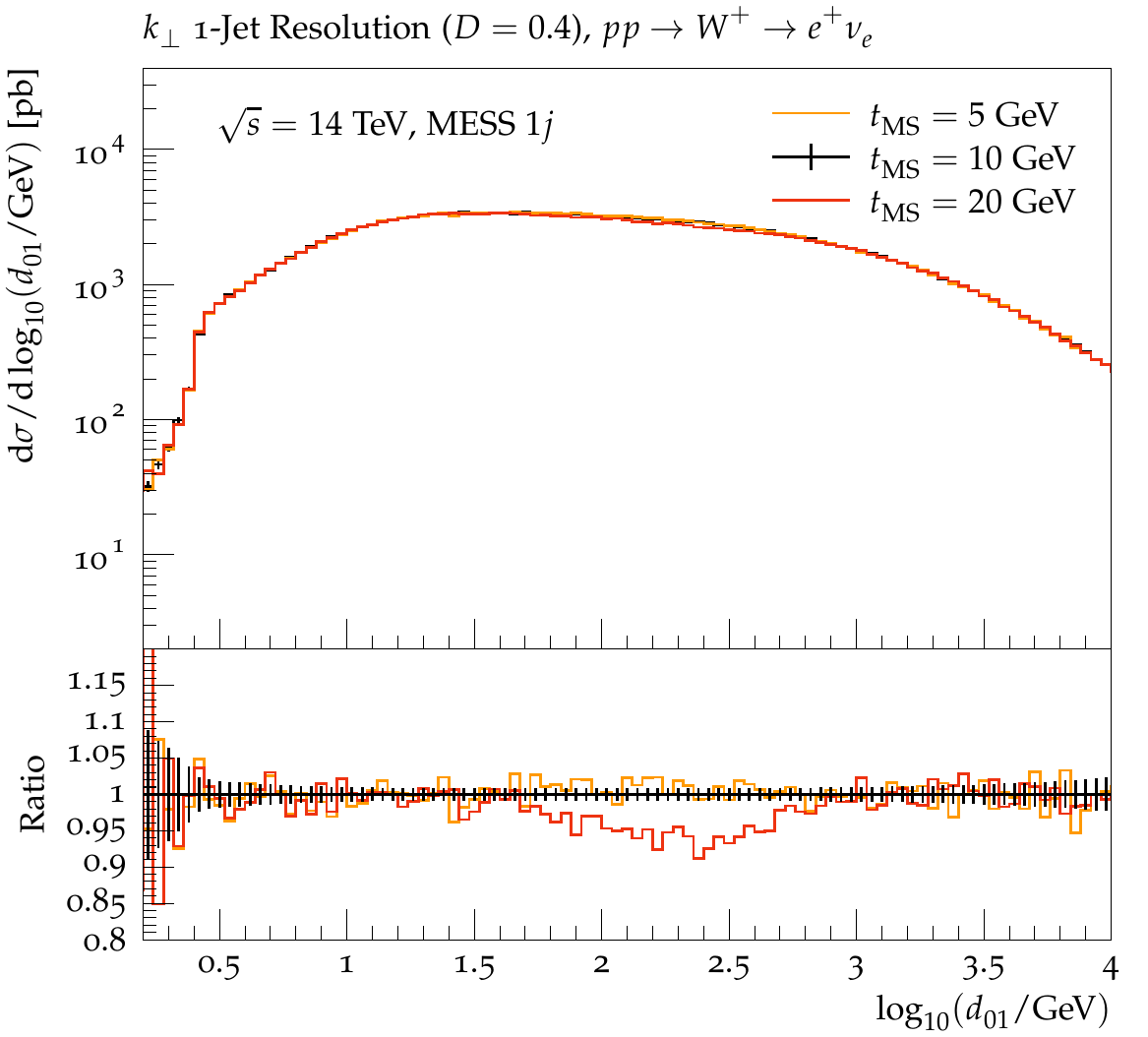}
	\includegraphics[width=0.35\textwidth]{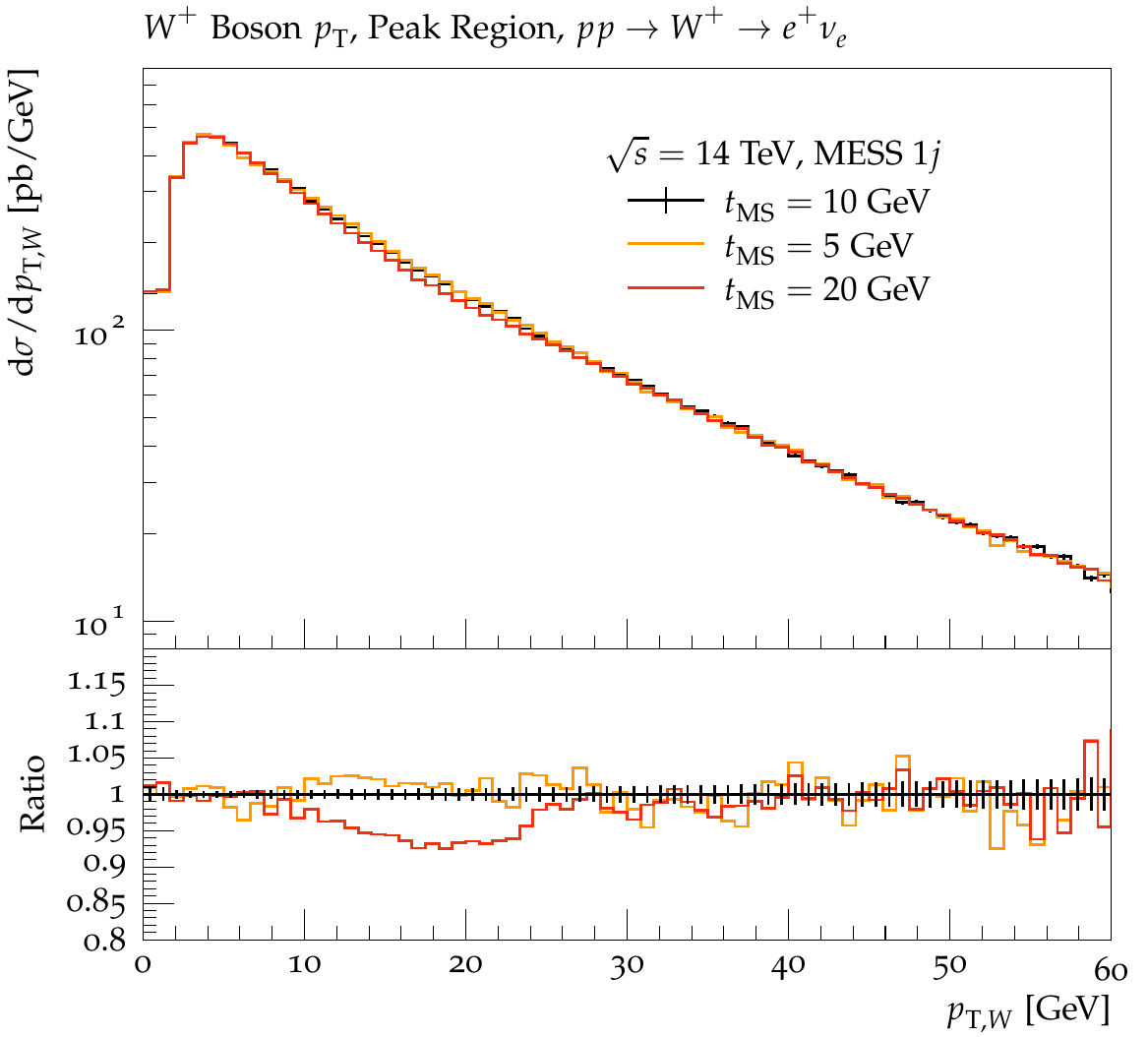}
	\caption{Influence of the merging scale choice on sector shower plus one-jet merged predictions of the $k_T$ 1-jet (\textit{left}) and the $W$ boson transverse momentum (\textit{right}) distribution in $pp \to W^+ + \mathrm{jets}$ at $\sqrt{s} = 14~\tera e\volt$.}
	\label{fig:wpMergingScaleVar}
\end{figure}

\section{Validation}\label{sec:Validation}
We validate our implementation in electron-positron annihilation processes and vector boson production in proton-proton collisions.
Event samples are generated with the \MadGraph event generator \cite{Alwall:2014hca} using the NNPDF23\_lo\_as\_0130\_qed PDF set with fixed renormalisation and factorisation scale corresponding to the mass of the $Z$, $\mu_\text{R} = \mu_\text{F} = M_Z$. Jets are defined using the $k_\perp$ jet clustering algorithm with a radius parameter of $D = 0.4$ and matrix elements are regularised by a $k_\perp$ cut.
To not obscure the effect of the merging, we consider parton-level results only and do not include MPIs. The merging scale is defined by the matrix element $k_\perp$ cut in all cases and we use the default \Vincia\ tune of the perturbative parameters with $\alphaS^{\overline{\mathrm{MS}}}(M_Z) = 0.118$, summarised in \ref{sec:tuneParms}. For electron-positron annihilation we choose a default merging scale of $\tms = 20~\giga e\volt$, while for vector boson production, we choose a lower default merging scale of $\tms = 10~\giga e\volt$, as the sector shower is currently not corrected to matrix elements. Analyses are performed using \Rivet~\cite{Buckley:2010ar,Bierlich:2019rhm}.

\cref{fig:eeContributions} shows the contribution of hard 3-jet and 4-jet events in Durham jet resolution scales in $e^+e^- \to \mathrm{jets}$ at a centre-of-mass energy of $\sqrt{s}=500~\giga e \volt$. The effect of varying the merging scale by a factor of two is shown in \cref{fig:eeMergingScaleVar} for the Durham 3-jet resolution and the $C$ parameter (for a definition see e.g.~\cite{Parisi:1978eg}).

In \cref{fig:zContributions}, the influence of merging the sector-shower predictions with up to two hard matrix elements on $k_\perp$ 1- and 2-jet resolution scales in Drell-Yan processes in the electron channel, $pp \to Z+ \text{jets}$ at $\sqrt{s}=14~\tera e \volt$ is studied. The effect of varying the merging scale by a factor of two is presented in \cref{fig:zMergingScaleVar}.

The individual contributions of the Born, 1-, and 2-jet event samples to $k_\perp$ 1- and 2-jet resolution scales in $W$ boson production in proton-proton collisions at $\sqrt{s} = 14~\tera e\volt$ are shown in \cref{fig:wmContributions,fig:wpContributions}. \cref{fig:wmMergingScaleVar,fig:wpMergingScaleVar} shows the influence of different merging scale choices on the merged predictions of the $k_\perp$ 1-jet splitting scale and the $W$ boson transverse momentum spectrum.

The jet-resolution scale distributions clearly show the expected behaviour that showers off Born configurations dominate the low-energy region on the left-hand side, while showers off higher-multiplicity states dominate in the hard region towards the right-hand side of the plots. The transition region smoothly interpolates between these two phase-space regions. A variation of the merging scale by factors of two has negligible effects on the distributions in $e^+e^-$ annihilation. For vector boson production in proton-proton collisions, the effect of choosing a higher merging scale results in a more pronounced effect on the distributions. Given that the underlying sector shower starts at the factorisation scale and is not corrected to matrix elements in a process that is subject to sizeable corrections from 1- and 2-jet matrix elements, we deem this a reasonable effect. This argument is supported by the fact that choosing a smaller merging scale has a far less-pronounced effect.

We have verified that the above statements remain true for a significantly larger set of observables than presented here.

\begin{figure}[t]
    \centering
    \includegraphics[width=0.46\textwidth]{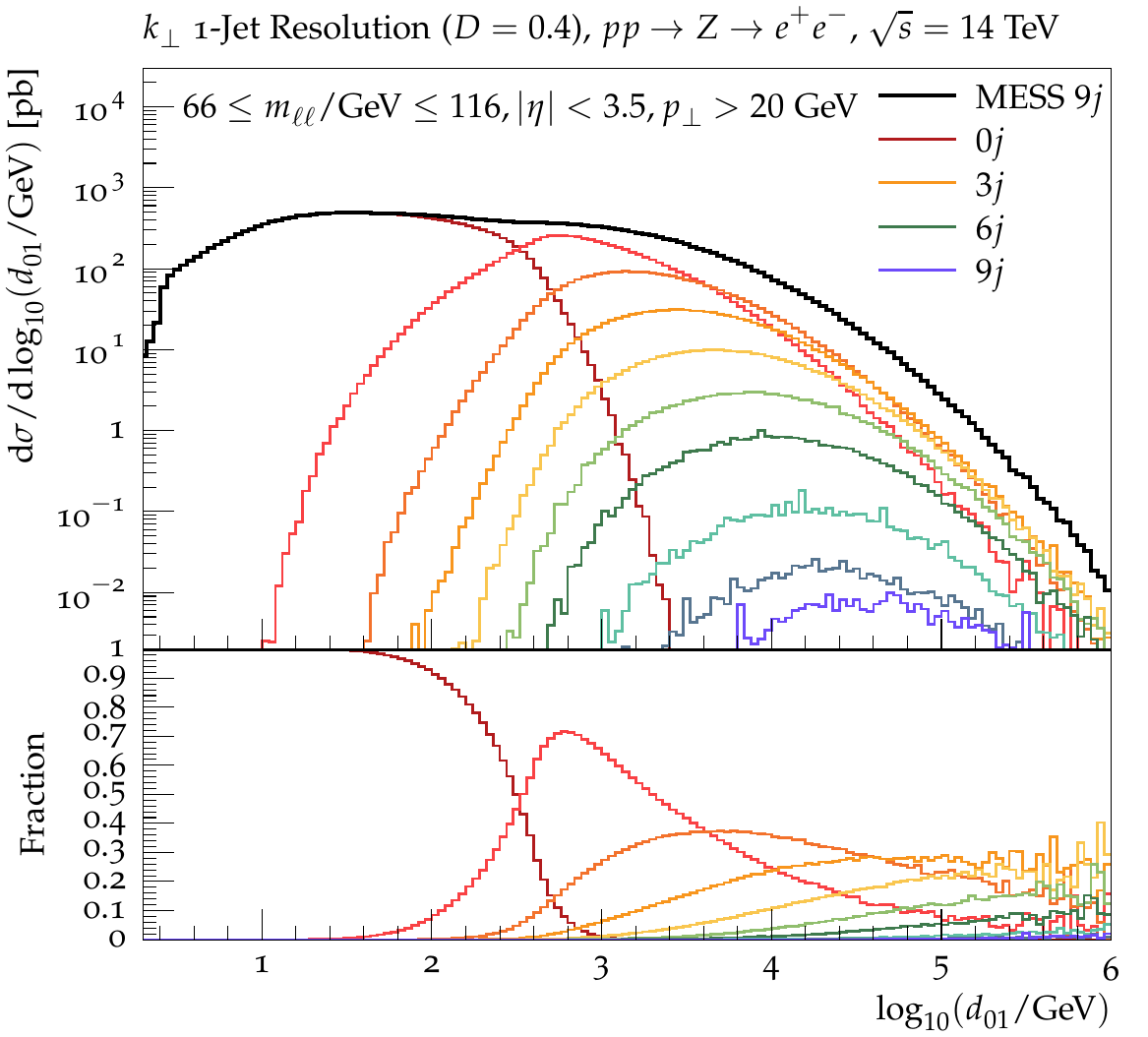}
    \includegraphics[width=0.46\textwidth]{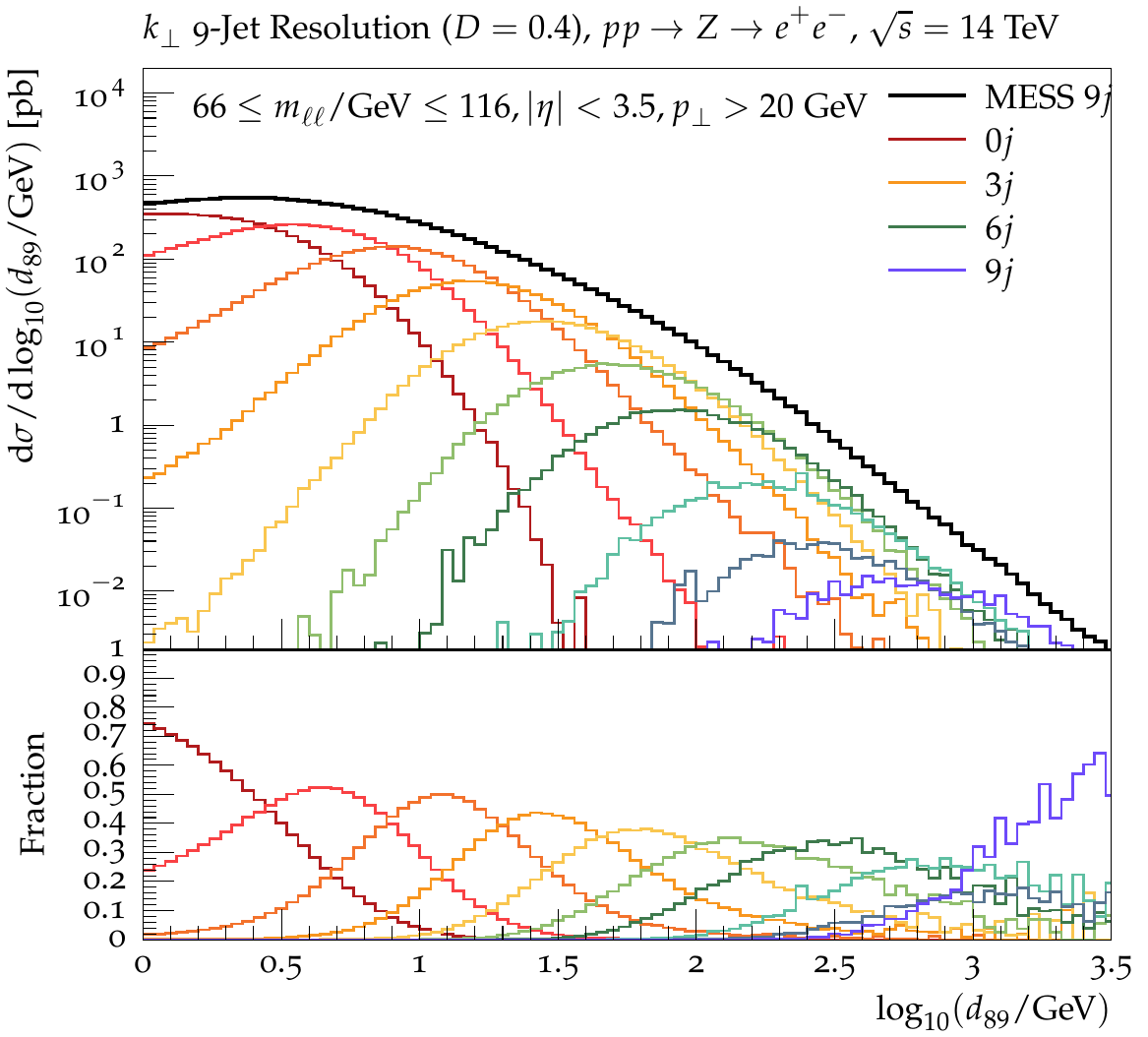}
    \caption{\Vincia\ merged parton-level predictions for the $k_\perp$ $1$-jet (\textit{left}) and $9$-jet (\textit{right}) resolution scales in Drell-Yan processes in $pp$ collisions at $14~\tera e\volt$.}
    \label{fig:zj9VinciaMess}
\end{figure}
\section{Results}\label{sec:Results}
To study the scaling behaviour of our implementation in the high-multiplicity regime, we use the parton-level event files \cite{z9jets,wm9jets,wp9jets} for vector boson production with up to 9 jets used in \cite{Hoeche:2019rti}. The merging scale is chosen to coincide with the $k_\perp$ cut of $20~\giga e\volt$ used in the event samples.

\begin{figure}[t]
    \centering
    \includegraphics[width=0.46\textwidth]{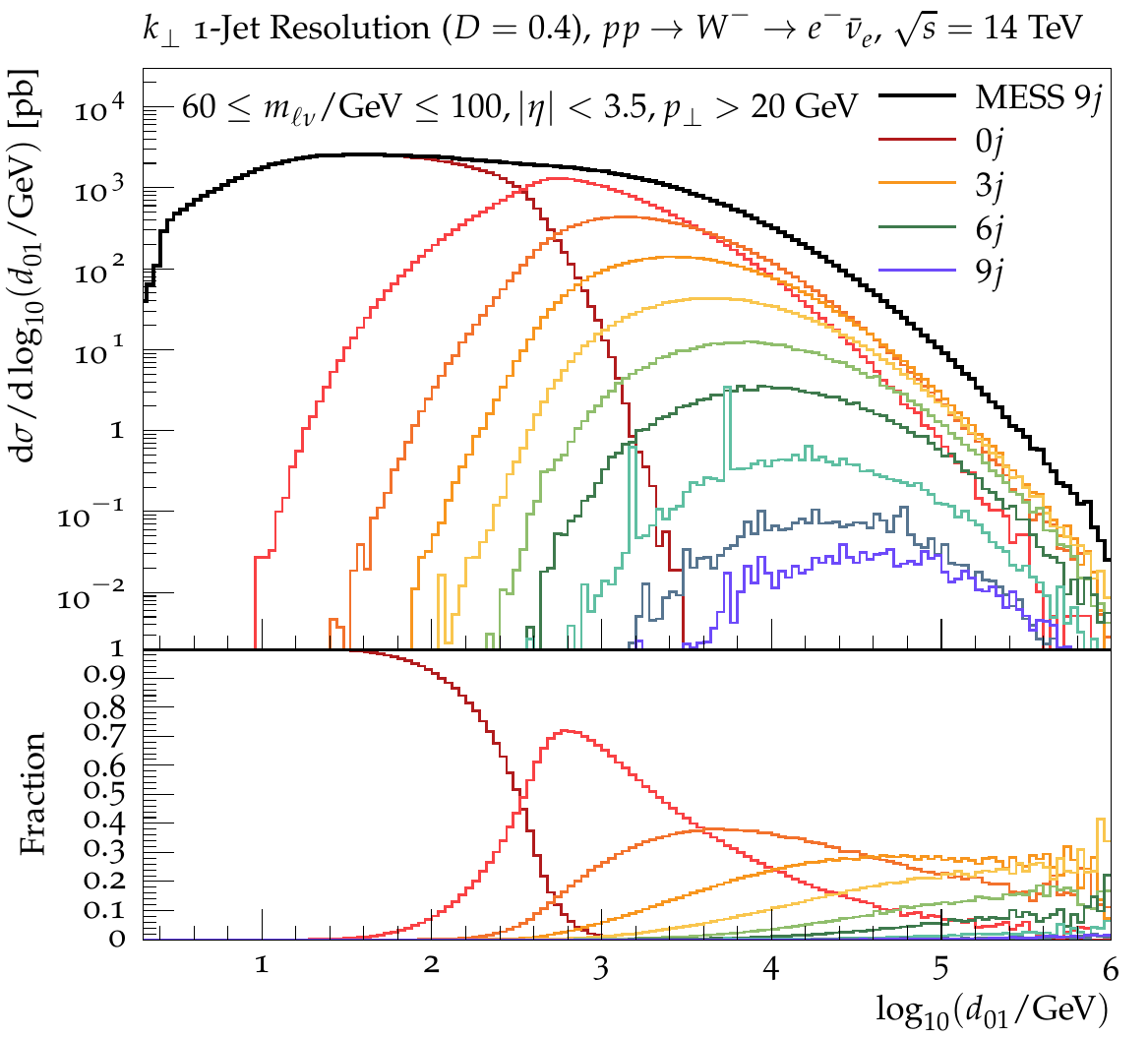}
    \includegraphics[width=0.46\textwidth]{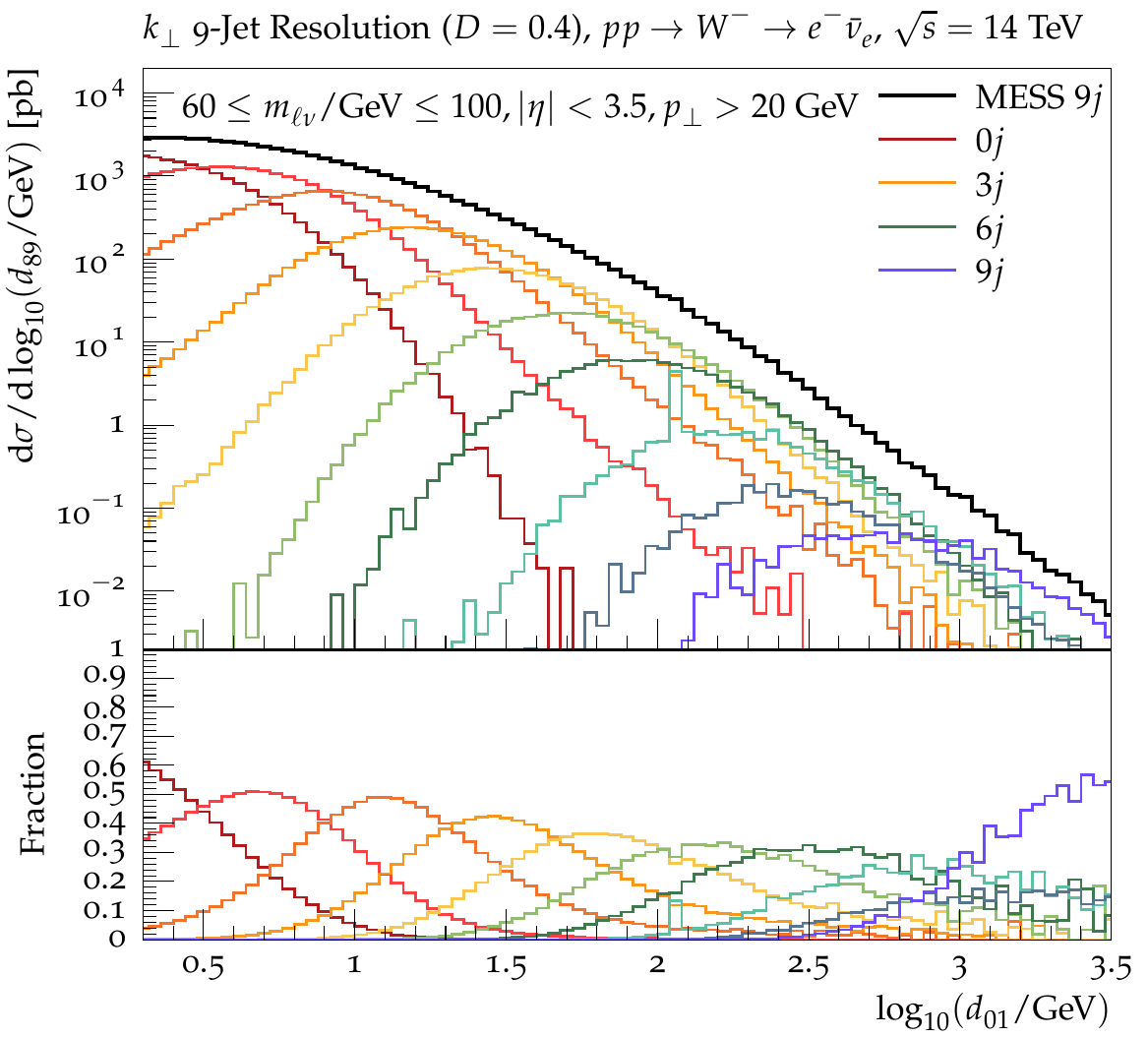}
    \caption{\Vincia\ merged parton-level predictions for the $k_\perp$ $1$-jet (\textit{left}) and $9$-jet (\textit{right}) resolution scales in $W^-$ production in $pp$ collisions at $14~\tera e\volt$.}
    \label{fig:wmj9VinciaMess}
\end{figure}

\begin{figure}[t]
    \centering
    \includegraphics[width=0.46\textwidth]{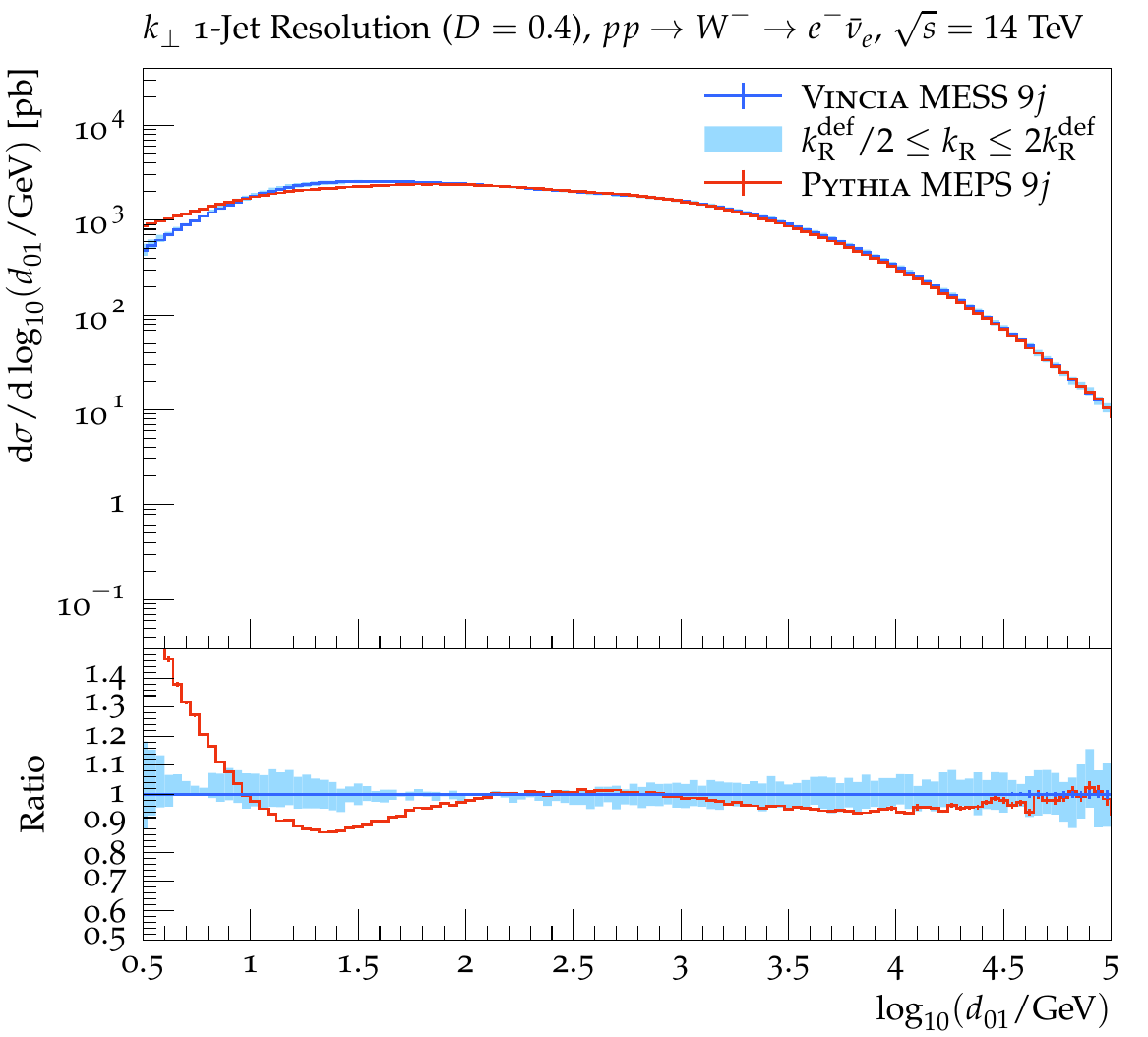}
    \includegraphics[width=0.46\textwidth]{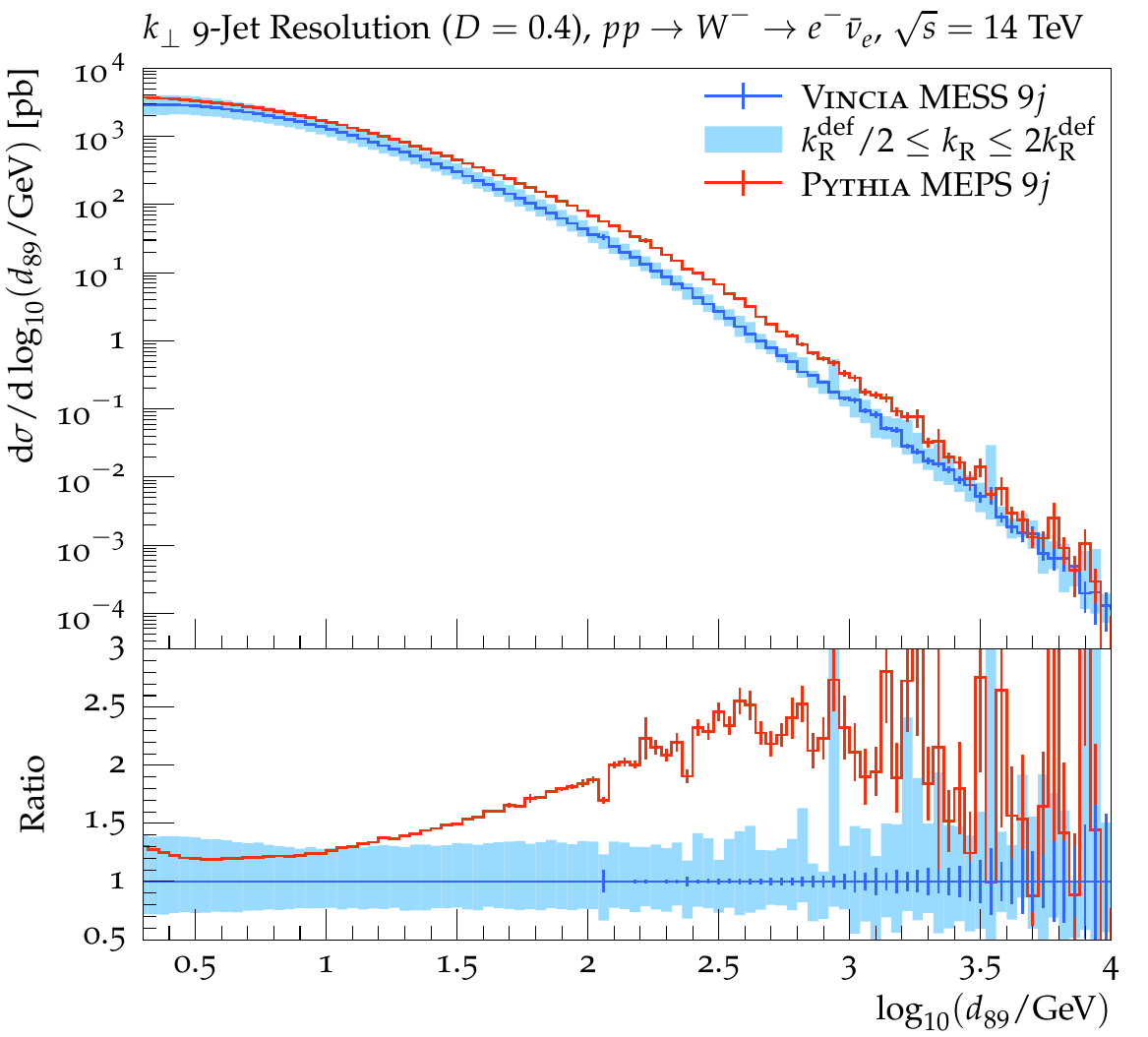}
    \caption{Comparison of \Vincia\ and \Pythia\ merged predictions for $k_\perp$ 1-jet (\textit{left}) and 9-jet (\textit{right}) resolution scales in $pp\to W^- + \mathrm{jets}$ at $\sqrt{s} = 14~\tera e \volt$.}
    \label{fig:wmjMessMeps}
\end{figure}

As a proof of concept, we show merged parton-level predictions for $k_\perp$ $1$- and $9$-jet resolution scales in $pp \to Z$ and $pp\to W^-$ with up to 9 hard jets in \cref{fig:zj9VinciaMess,fig:wmj9VinciaMess}. These results are obtained with the preliminary default \Vincia\ tune with two-loop running $\alphaS$ in the CMW scheme and $\alphaS^{\overline{\mathrm{MS}}}(M_Z) = 0.118$ as summarised in \ref{sec:tuneParms}.
Despite obtaining the accuracy of the additional tree-level matrix elements, the merged predictions retain the LL+LO precision of the shower, including unitarity violations due to the CKKW-L method. As events are weighted by $\alphaS$ ratios, the non-unitarity of the method becomes manifest with increasing jet multiplicity, leading to larger scale dependencies at higher orders.
To gain a first estimate of this effect, we vary \Vincia's renormalisation scale factors $k_\mathrm{R}$ used to evaluate the strong coupling  (cf.~\ref{sec:tuneParms}) by a factor of 2 with respect to the default values, cf.~\cref{fig:wmjMessMeps}, where we also compare merged parton-level predictions for $k_\perp$ resolution scales in $pp\to W^-$ from \Vincia's and \Pythia's CKKW-L implementation. For \Pythia, we use the default tune of the strong coupling, i.e., a one-loop running coupling in the $\overline{\mathrm{MS}}$ scheme with a numerical value of $\alphaS^{\overline{\mathrm{MS}}}(M_Z)=0.1365$.

Except for the region near the hadronisation cutoff, the two implementations agree well for the $1$-jet resolution scale, cf.\ the left-hand pane in \cref{fig:wmjMessMeps}, while there is a larger discrepancy in the distributions of the $9$-jet resolution scale. As expected, the scale variations only have a small effect on the former, while for the 9-jet clustering scale, these are far more significant. There remains, however, a shape difference between \Vincia\ and \Pythia\ for the 9-jet scale, which may at least partly be traced back to the rather high merging scale for the uncorrected sector shower, cf.~\cref{sec:Validation}. Given that neither of the showers are tuned for merging at these high multiplicities, the observed difference provides an interesting subject for further studies.

As the central objective of our improved merging scheme, we study the run time and memory usage of our implementation and compare it to the CKKW-L implementation in \Pythia.

\subsection{Run Time}
As a first gauge of the scaling behaviour of the default \Pythia\ and our \Vincia\ sector shower CKKW-L implementations, we measure the CPU time to find the (most probable) shower history in both. 
We consider the processes $pp\to Z$ and $pp\to W^-$ with up to 9 additional hard jets and run each multiplicity individually on a single core of a 2.3 GHz Intel Core i5 processor and only count complete histories, i.e., ones for which at least one reconstructed shower sequence to the Born exists.

\begin{figure}[t]
    \centering
    \includegraphics[width=0.4\textwidth]{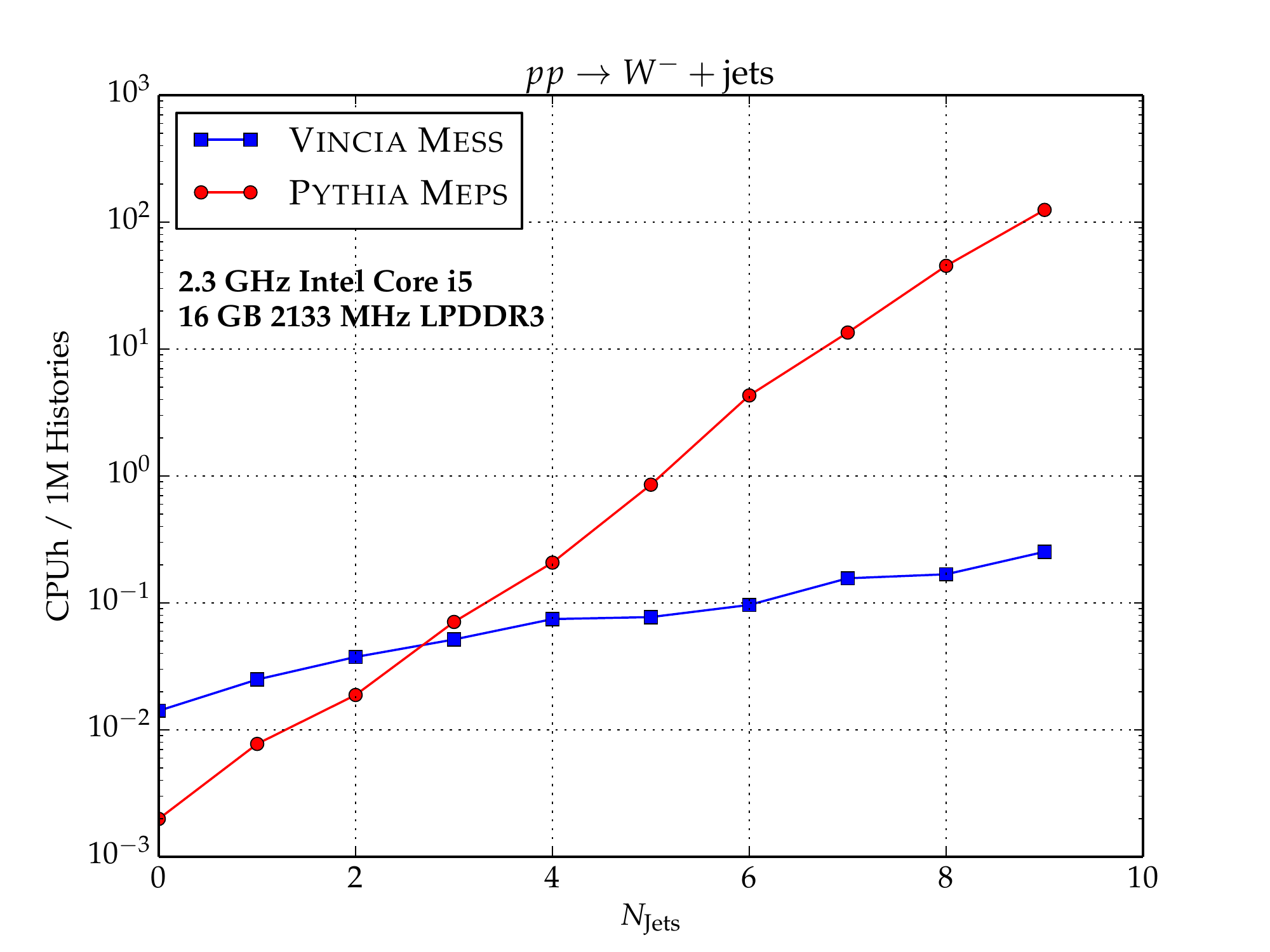}
    \includegraphics[width=0.4\textwidth]{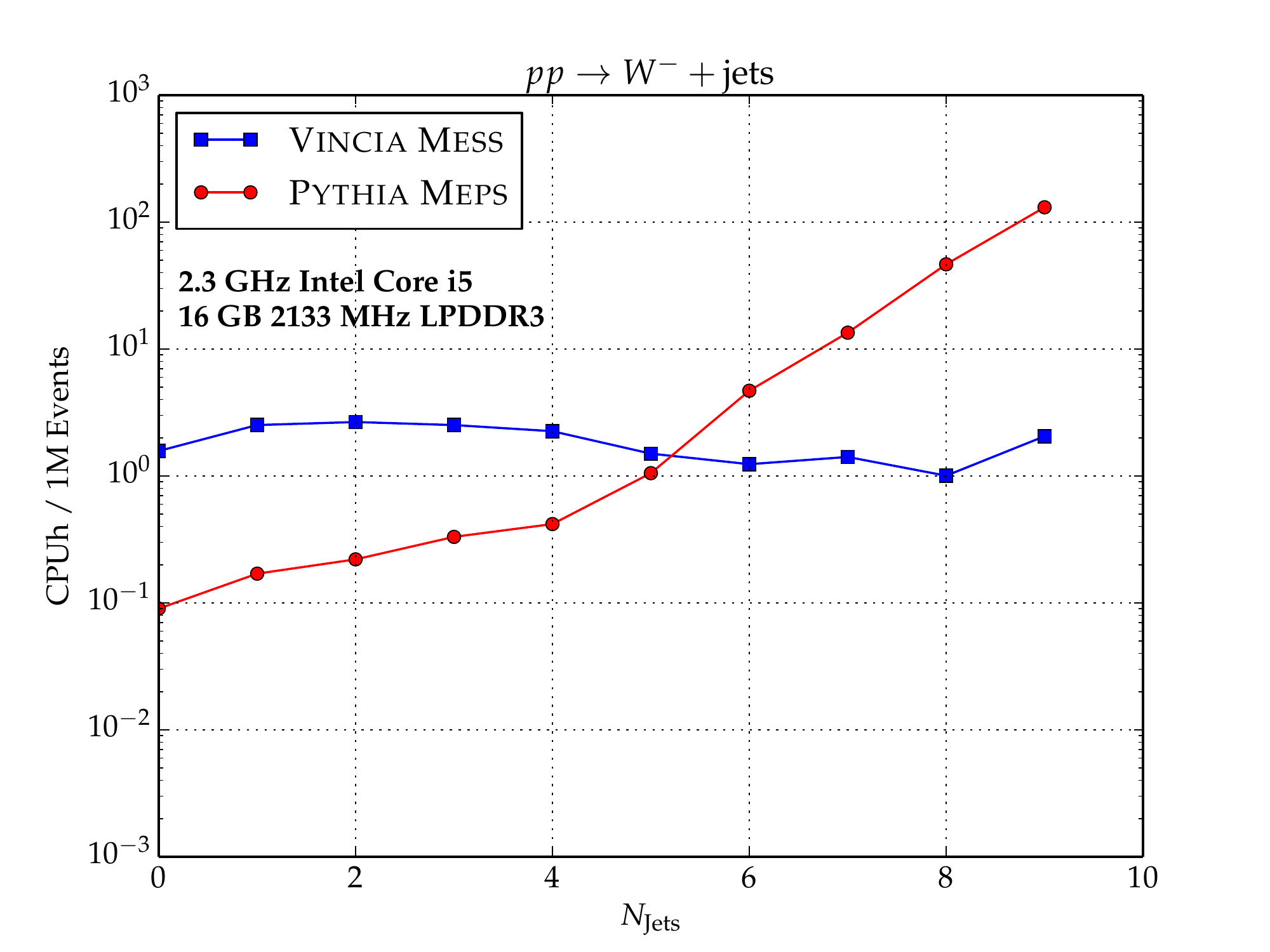}
    \caption{\Pythia\ and \Vincia\ CPU time scaling in history construction (\textit{left}) and parton-level event generation (\textit{right}) for $pp \to W^- + \mathrm{jets}$ merging at $\sqrt{s} = 14~\tera e\volt$.}
    \label{fig:cputimeScalingWmj}
\end{figure}

\begin{figure}[ht]
    \centering
    \includegraphics[width=0.4\textwidth]{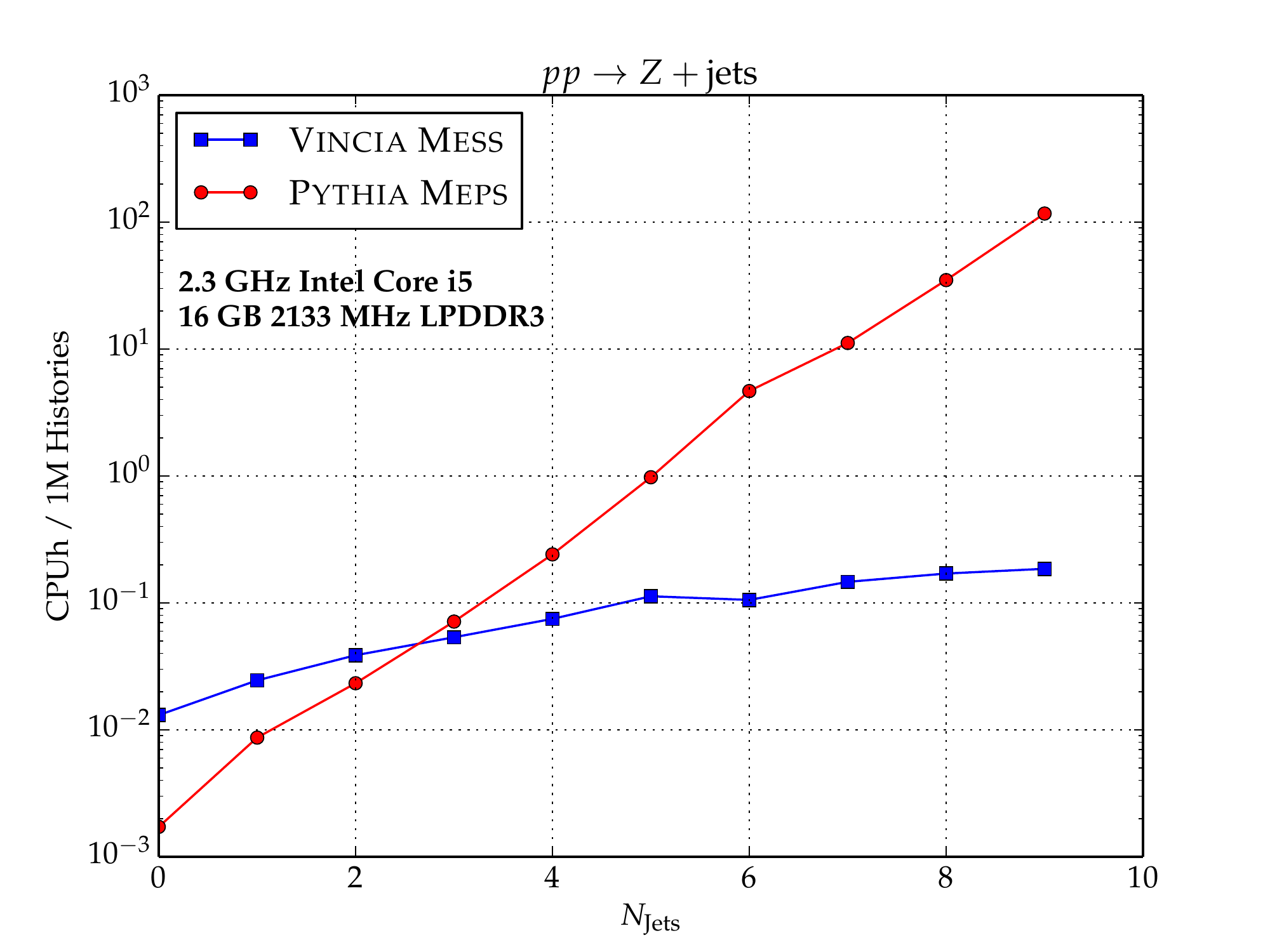}
    \includegraphics[width=0.4\textwidth]{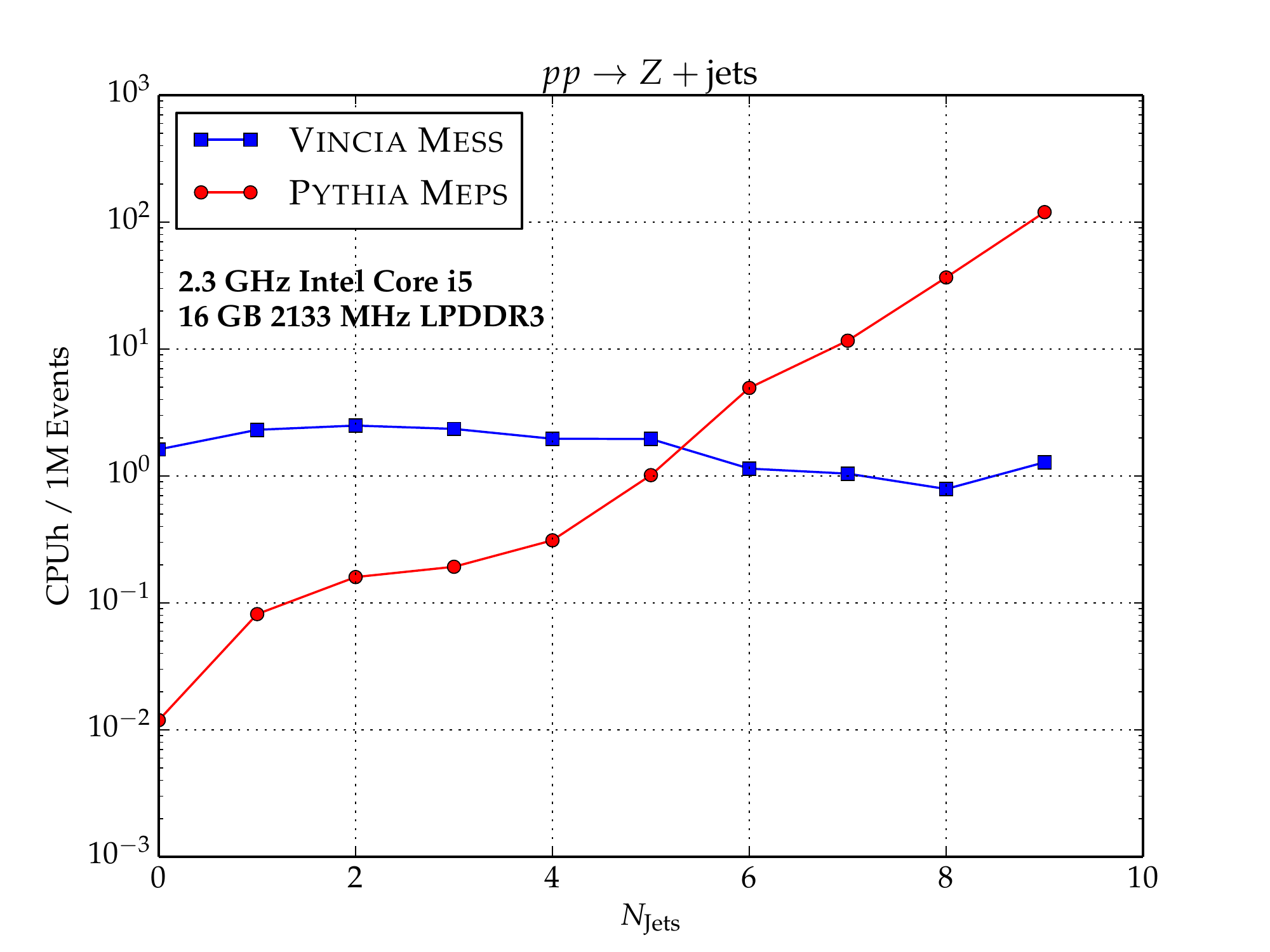}
    \caption{\Pythia\ and \Vincia\ CPU time scaling in history construction (\textit{left}) and parton-level event generation (\textit{right}) for $pp \to Z + \mathrm{jets}$ merging at $\sqrt{s} = 14~\tera e\volt$.}
    \label{fig:cputimeScalingZj}
\end{figure}

In the left-hand panes of \cref{fig:cputimeScalingZj,fig:cputimeScalingWmj}, we show the scaling of the CPU time for shower history construction in $pp \to Z + \mathrm{jets}$ and $pp\to W^- +\mathrm{jets}$, respectively. We find that the recursive strategy of the default \Pythia\ history construction is faster for low multiplicities, but develops a steep exponential scaling for higher multiplicities. The iterated \Vincia\ sector shower history construction, on the other hand, scales linearly with the number of jets. 
Starting from the four-jet sample, it becomes notably faster than the \Pythia\ history construction. At the extreme of 9 jets, \Pythia\ spends about half a second per event to construct all shower histories, while \Vincia\ does not even need a millisecond per history.

As being of practical importance, we study the overall CPU time per generated event in the right-hand panes of \cref{fig:cputimeScalingZj,fig:cputimeScalingWmj}. We consider the time for \Pythia\ to generate a new parton-level event, either using the default merging and shower implementation or the \Vincia\ one. Again, the default implementation is notably faster for low multiplicities but develops a steep exponential scaling. By comparing with the time needed for history construction, it can be seen that, starting from the four-jet sample, the default \Pythia\ implementation spends most of the event generation time on constructing the history of the input event. 
Because of the more complex shower algorithm, the picture is completely different for the \Vincia\ MESS implementation, where most of the time is spent on the showering of an accepted hard event.
As the number of accepted events decreases with the multiplicity, and therefore less often a full sector shower has to be performed, the overall event generation time stays approximately constant when adding more jets, with a slight decrease towards high multiplicities. 
The increase for the 9-jet sample is explained by the inclusive treatment of the last node in CKKW-L merging, due to which a full sector shower is performed for more events. When adding further exclusive multiplicities beyond 8 jets, the total event generation time of the \Vincia\ MESS implementation will approach the pure history construction time.

It should be noted that the baseline sector shower used in this study does not utilise optimised sampling strategies to deal with competing sectors, cf.\ e.g.~\cite{Kleiss:2016esx,Kleiss:2017iir,Skands:2020lkd}, which can improve the performance relative to the results shown here. Such optimisation studies are currently ongoing.

\subsection{Memory Usage}
As the even more prohibiting bottleneck of conventional CKKW-L merging schemes at high multiplicities, we study the memory usage.
We use \texttt{Valgrind}'s \texttt{Massif} tool to monitor the heap usage of the default \Pythia\ CKKW-L merging and our \Vincia\ sector shower merging implementations. In particular, this means that neither the stack nor the memory at the page level is recorded. For comparability and reproducibility, we use the \texttt{--time-unit=B} option in \texttt{Valgrind} to measure the runtime of the program in terms of the number of allocated and deallocated bytes. We use the same main program and event samples for both runs and consider a fictitious $Z+10~\mathrm{jet}$ merging run, so that every event multiplicity, including the 9-jet sample, is processed as an intermediate node. We run each multiplicity independently with the maximal possible number of snapshots available, which may be at most (but is not necessarily identical to) 1000. To gain the most
detailed possible picture of the memory allocations, we choose a relatively small number of 1000 events per run. For higher statistical significance, we perform up to ten independent runs for each multiplicity. On the technical level, the 7-, 8-, and 9-jet event samples in \cite{wp9jets,wm9jets,z9jets} are separated into multiple files, corresponding to irreducible groups of processes with similar diagrammatic structure, cf.~\cite{Hoeche:2019rti}. For these multiplicities, at least one run is performed per group.

\begin{figure}[t]
    \centering
    \includegraphics[width=0.3\textwidth]{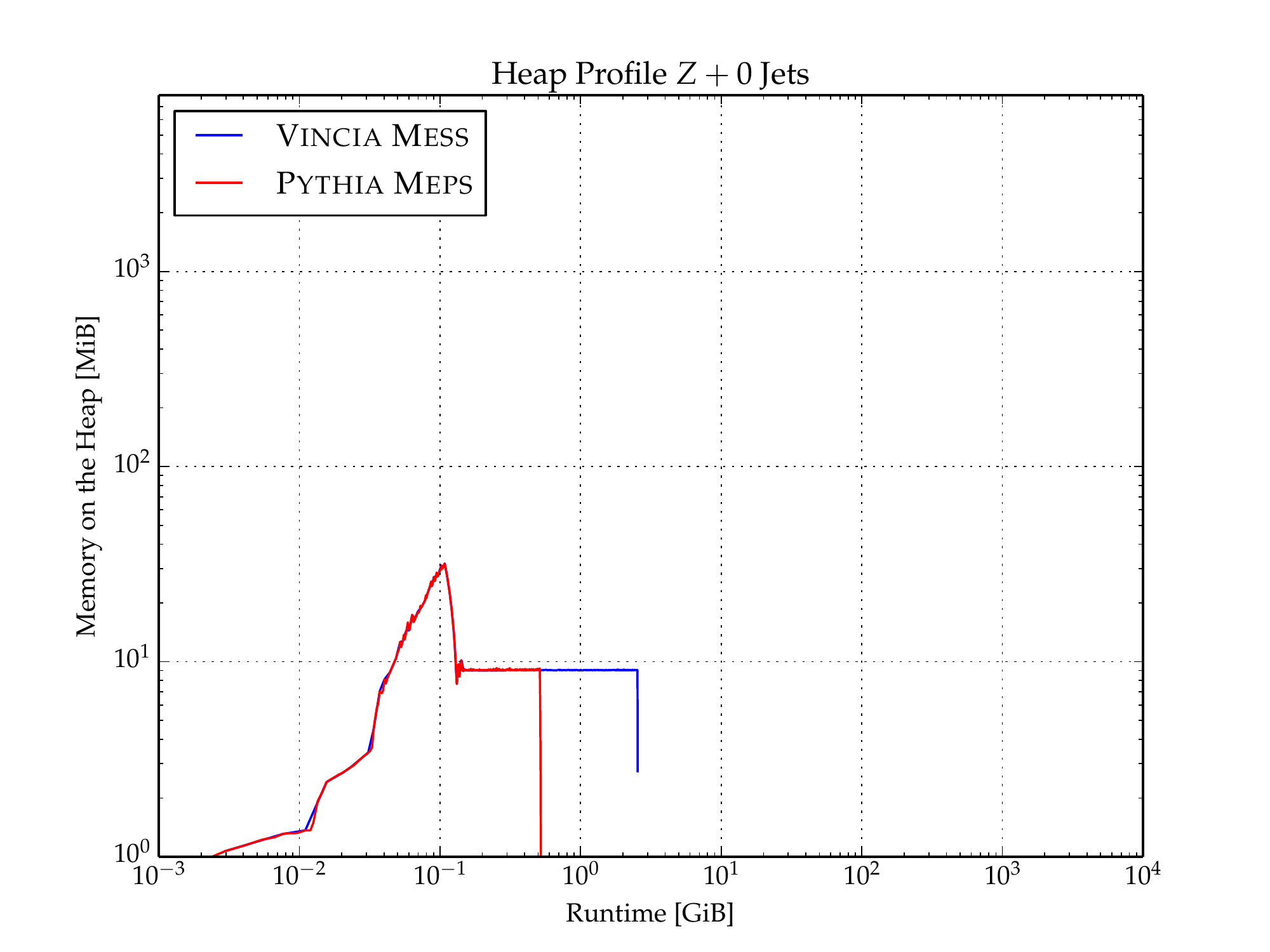}
    \includegraphics[width=0.3\textwidth]{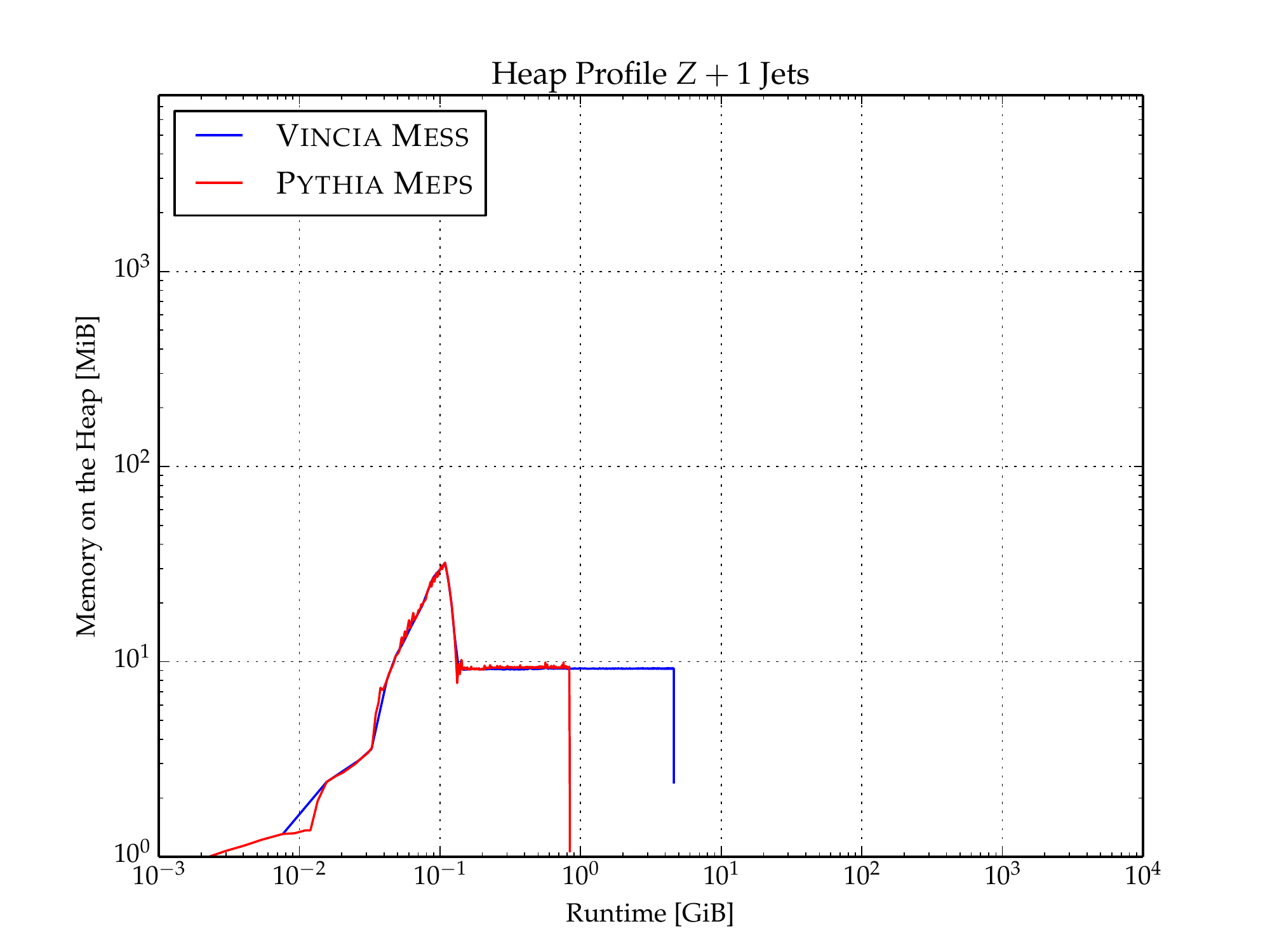}\\
    \includegraphics[width=0.3\textwidth]{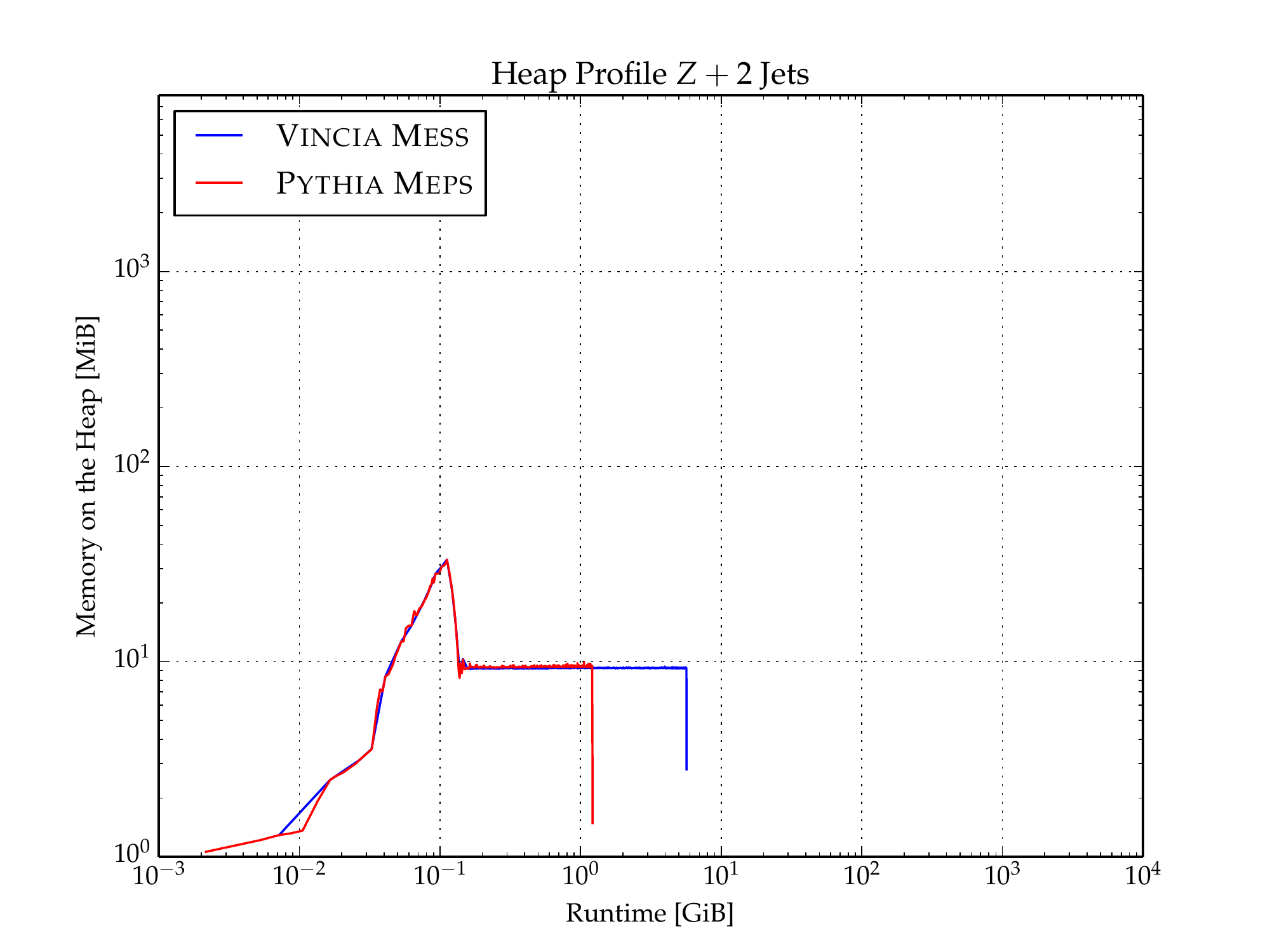}
    \includegraphics[width=0.3\textwidth]{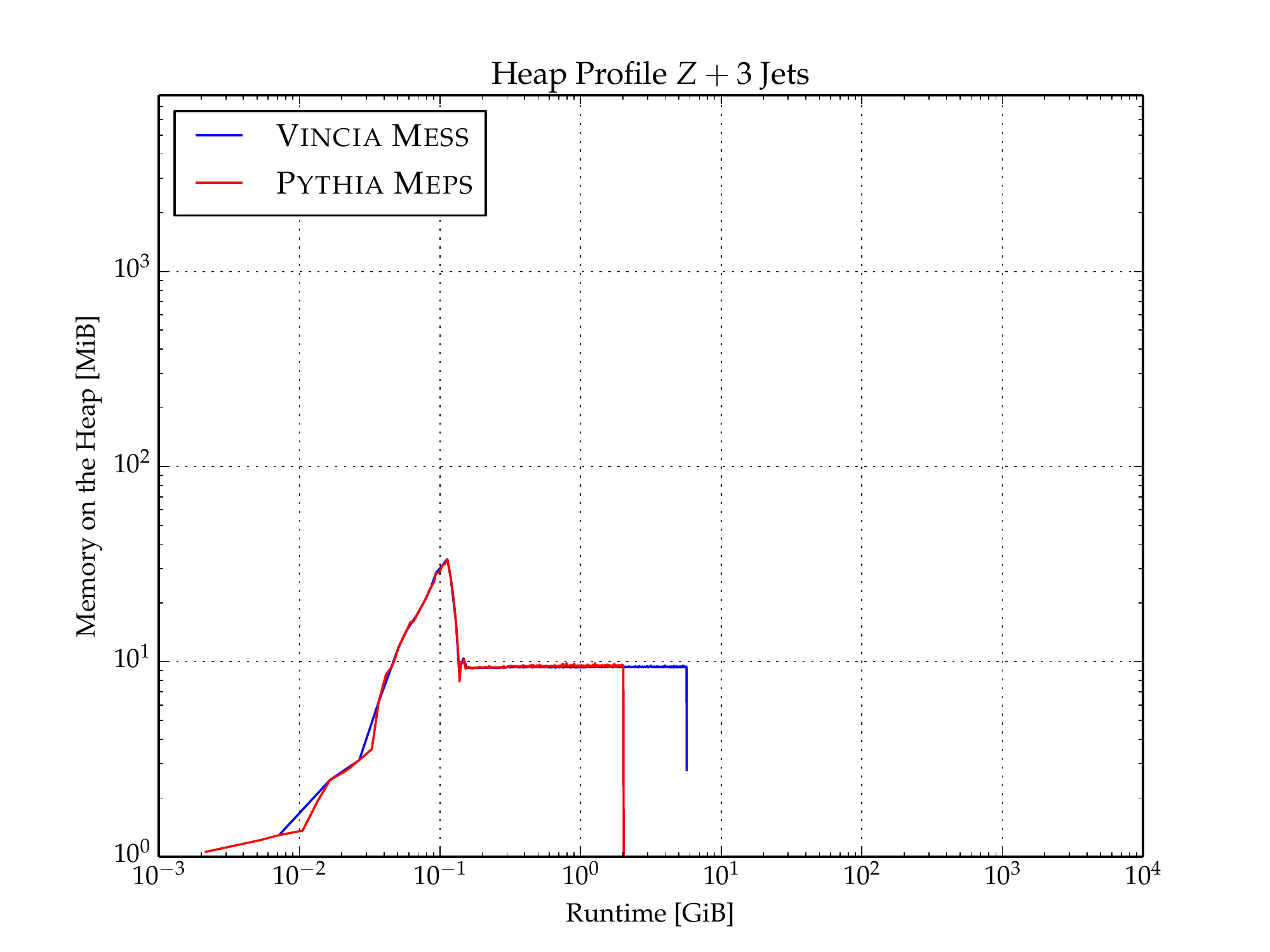}\\
    \includegraphics[width=0.3\textwidth]{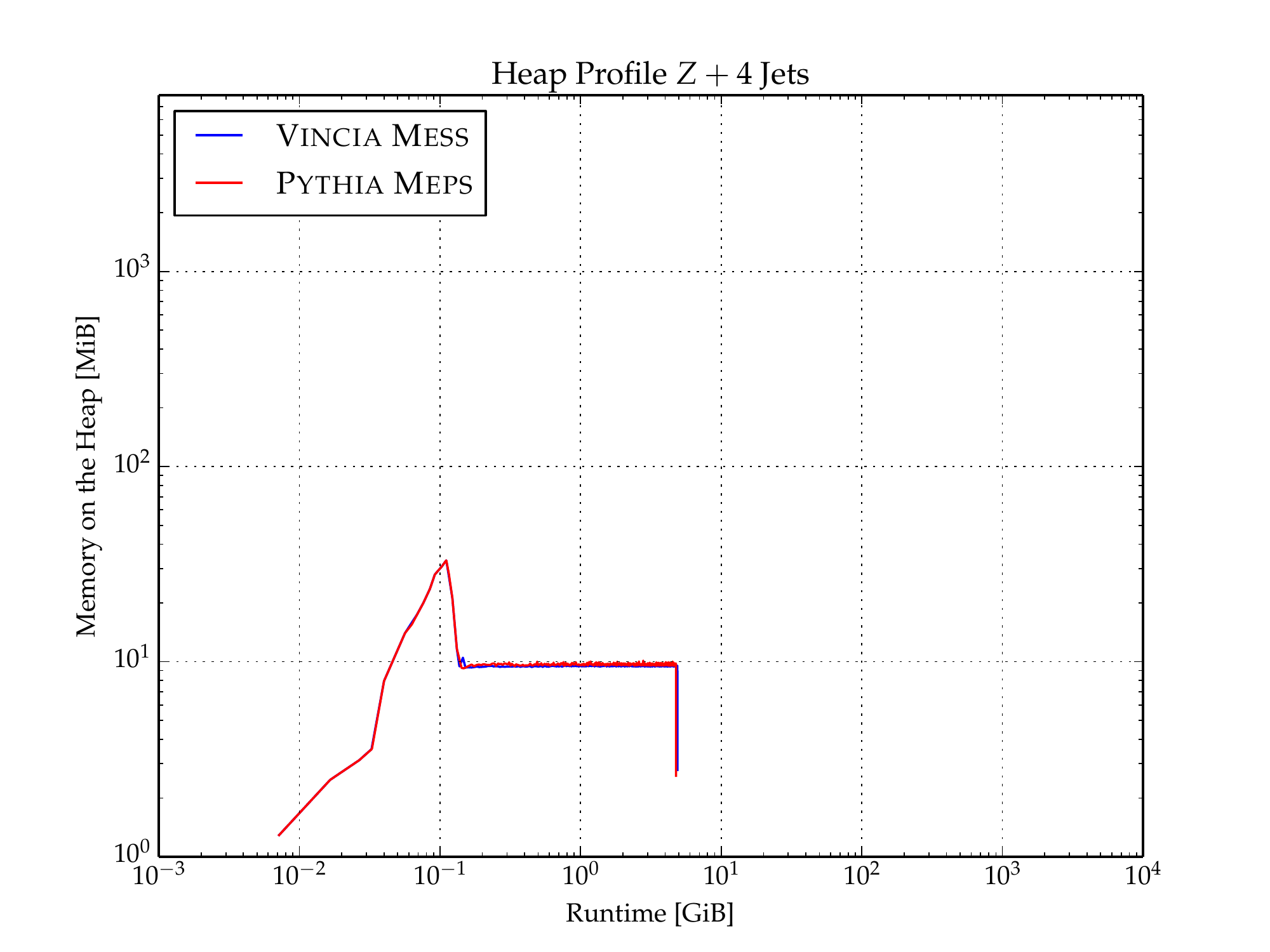}
    \includegraphics[width=0.3\textwidth]{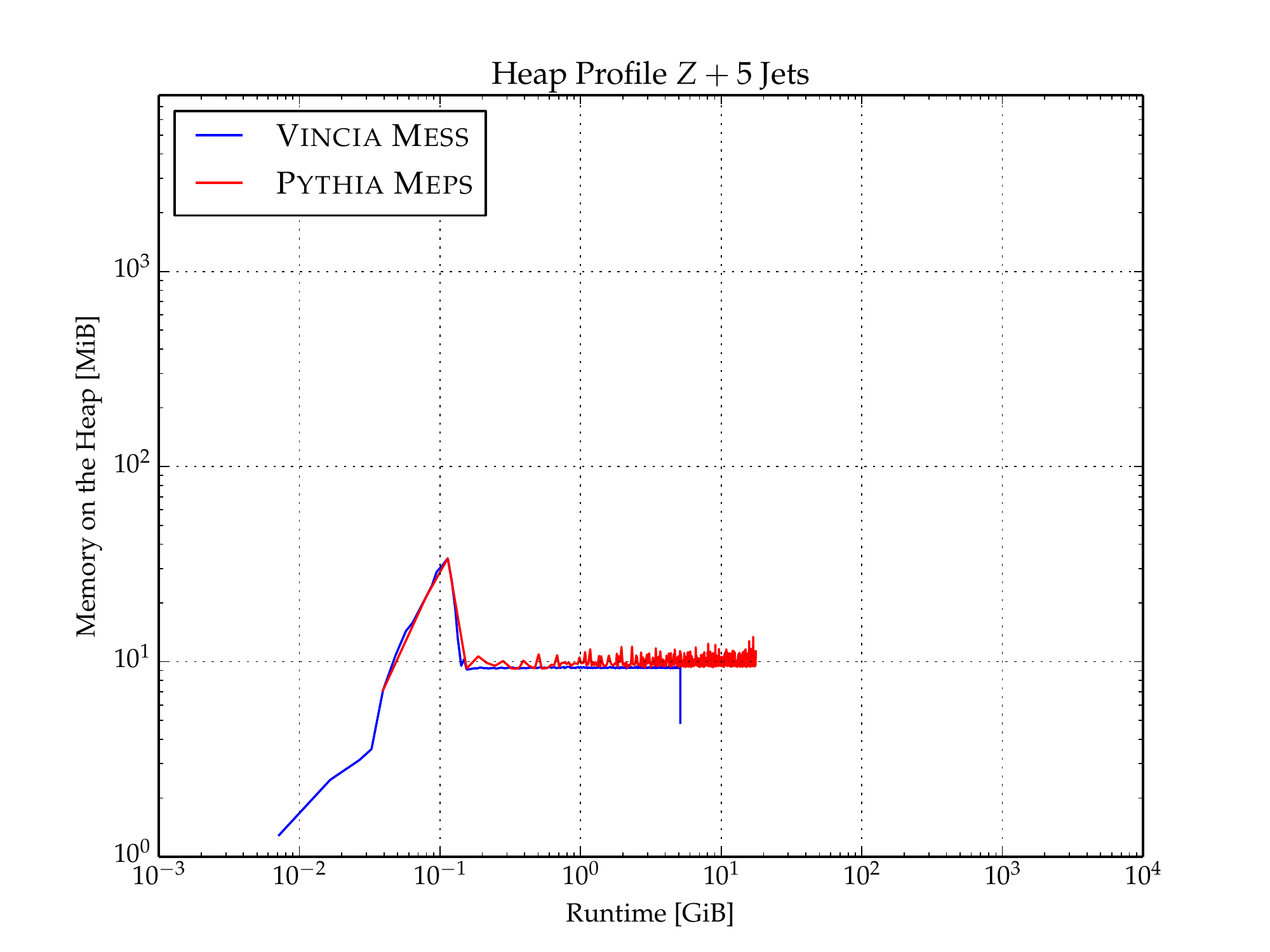}\\
    \includegraphics[width=0.3\textwidth]{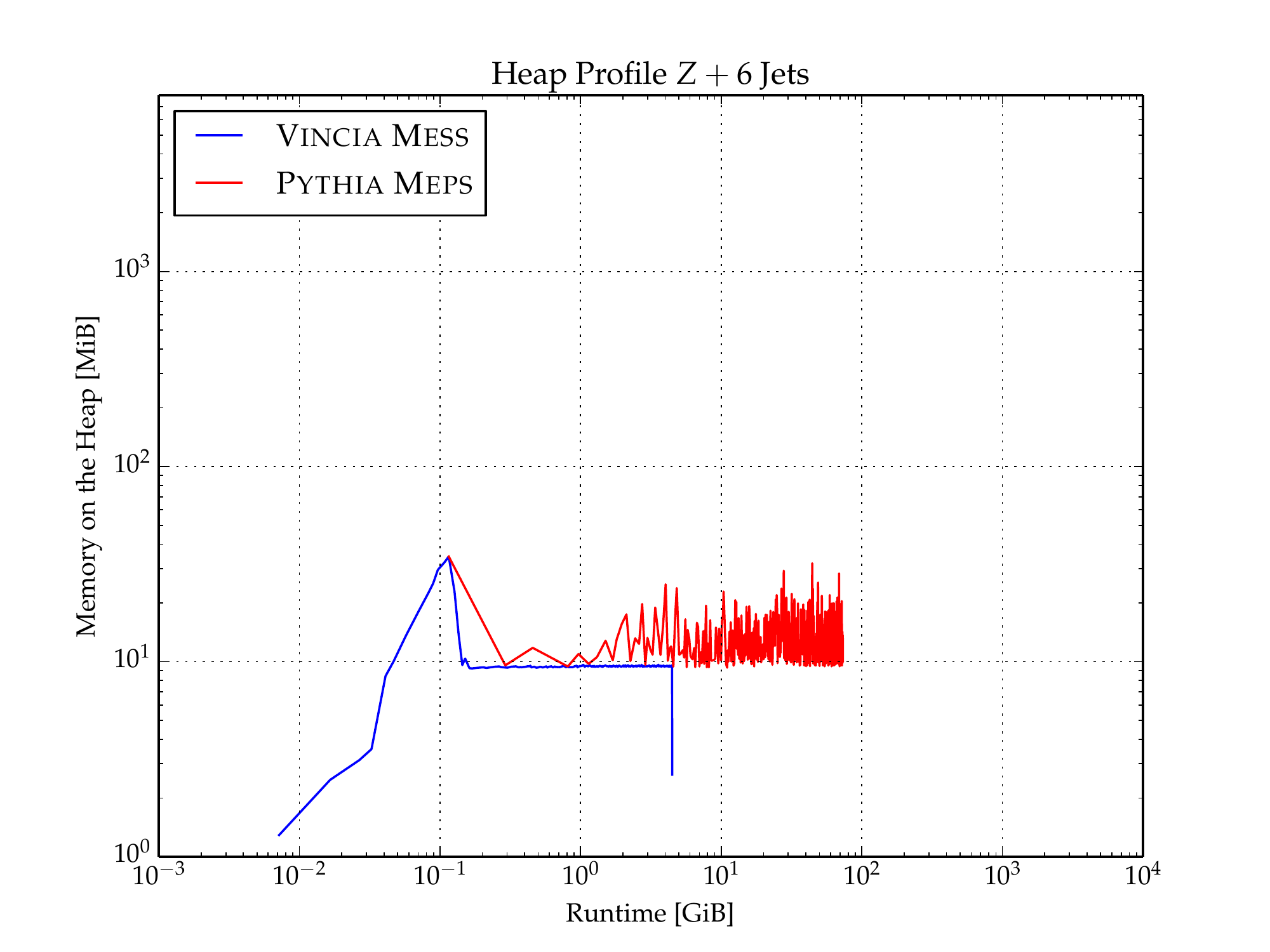}
    \includegraphics[width=0.3\textwidth]{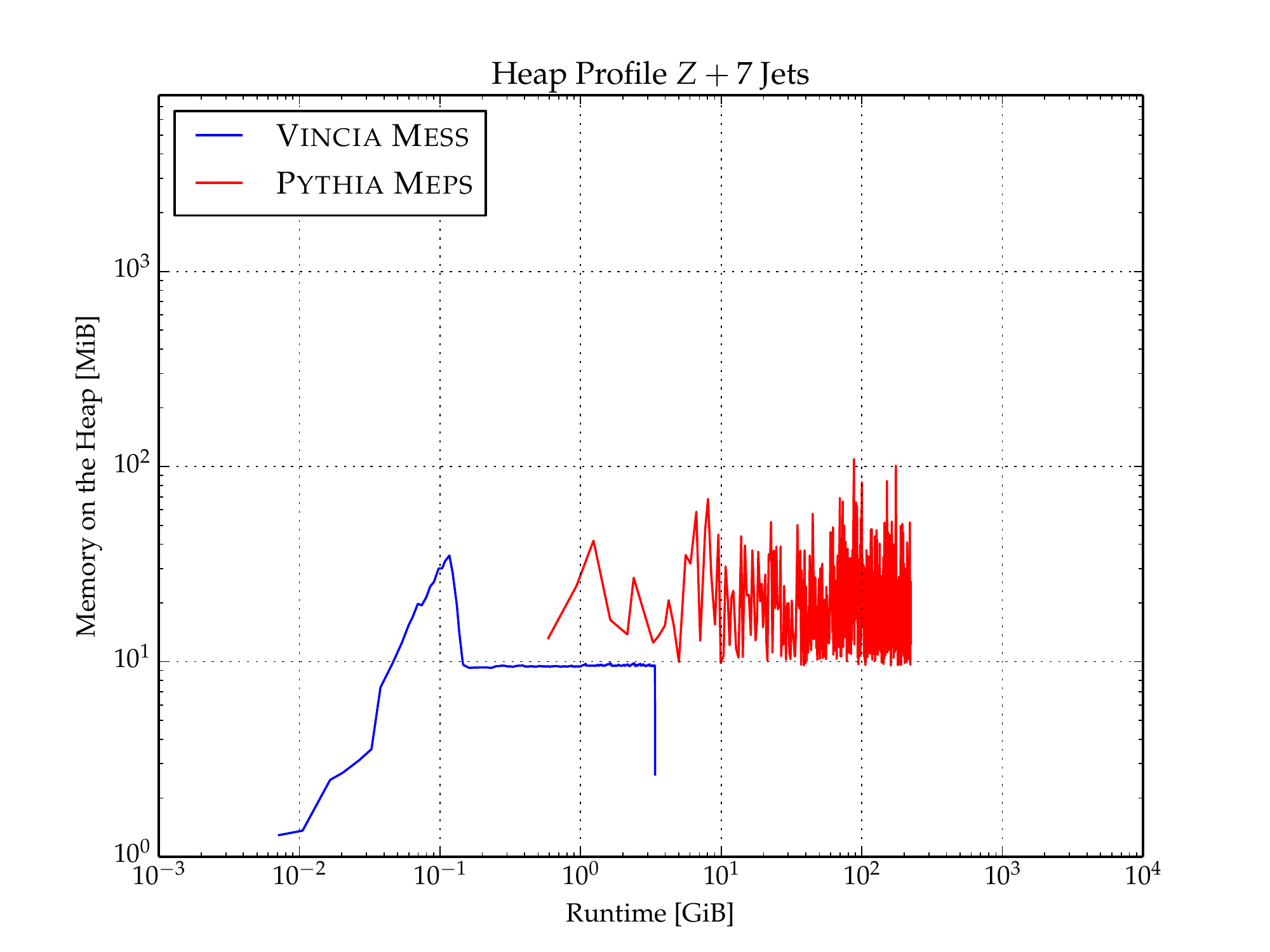}\\
    \includegraphics[width=0.3\textwidth]{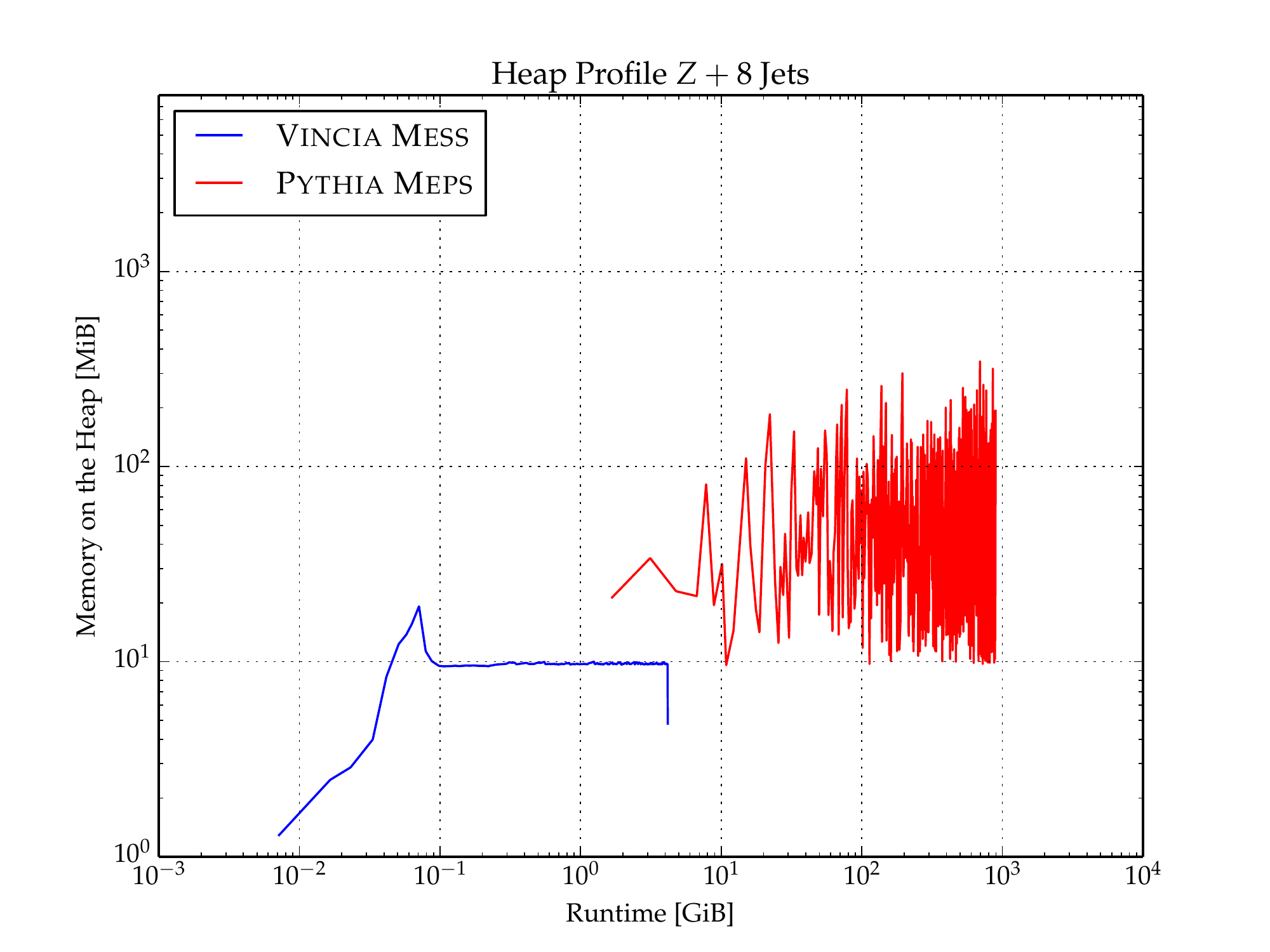}
    \includegraphics[width=0.3\textwidth]{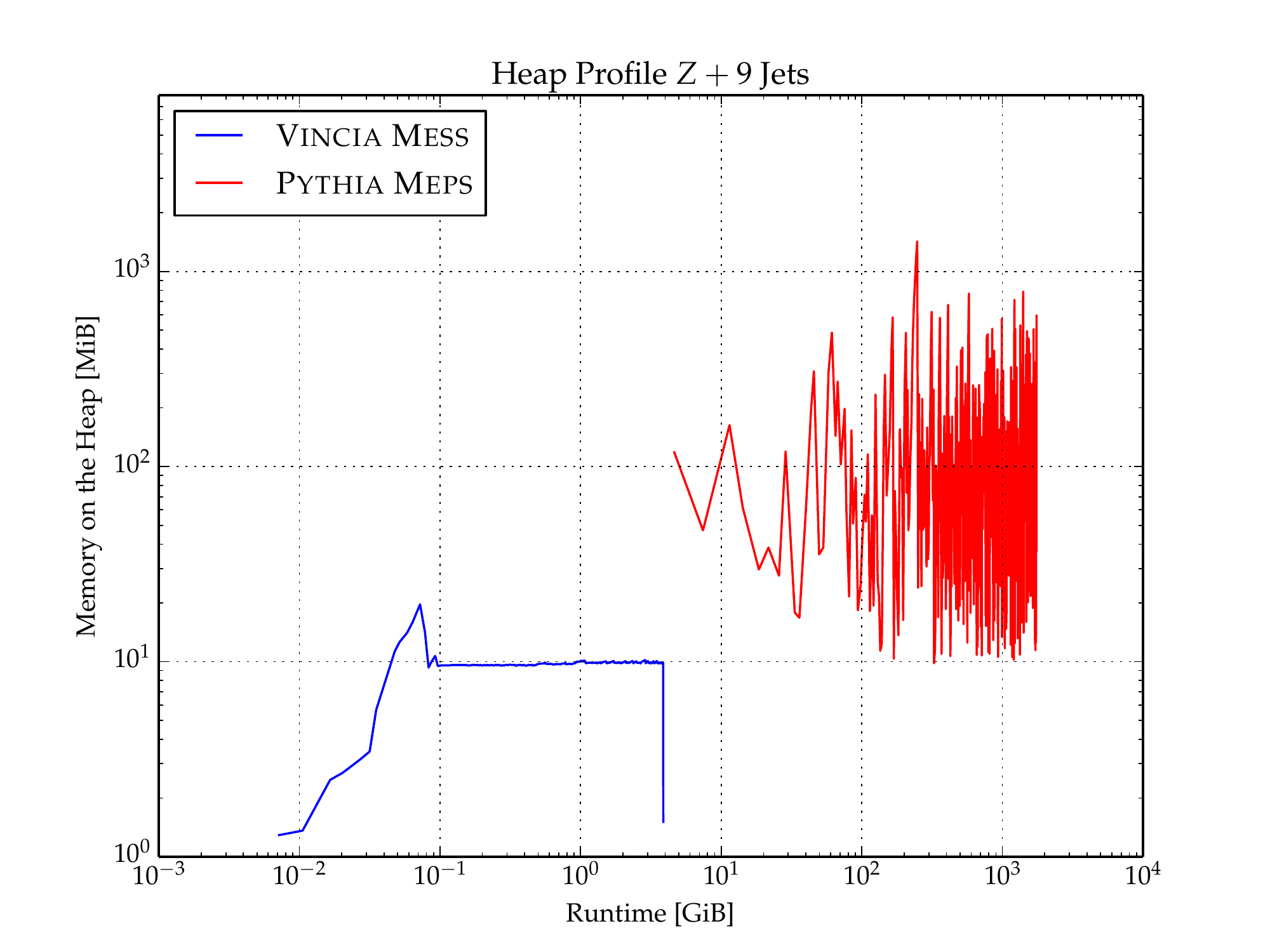}
    \caption{\Pythia\ and \Vincia\ memory usage profiles in $pp \to Z + \mathrm{jets}$ merging at $\sqrt{s} = 14~\tera e\volt$.}
    \label{fig:memoryProfiles}
\end{figure}

In \cref{fig:memoryProfiles}, the individual heap profiles of all event samples from the $pp \to Z$ to the $pp\to Z+9~\mathrm{jets}$ sample are shown. For samples with more than six jets, we only show a representative memory profile of one group. Additional profiles are collected in \ref{sec:memProfiles}.
The peak on the left-hand side of the plots corresponds to the read-in of the HDF5 event sample, which (for high-multiplicity runs) is not recorded by \texttt{Valgrind} for the default CKKW-L implementation in \Pythia\ anymore, in favour of higher allocation peaks later in the run. For merging with less than 4 additional jets, the heap profiles of \Pythia\ and \Vincia\ are very similar: after the high peak when reading the event file, only a number of very small peaks are recorded. For these runs, the \Pythia\ merging implementation has a shorter \textquotedblleft runtime\textquotedblright\ in terms of total allocated/deallocated memory, which, however, continuously increases with the number of additional jets. That \Vincia\ allocates more memory than \Pythia\ during these runs can be traced back to the differences in the shower implementations, which, as alluded to above, is more complicated for the sector shower. For every trial, a tentative post-branching state has to be constructed to evaluate the sector veto on. Although this does not amount to large peaks in memory allocation, it adds to the total allocated memory, i.e., the \textquotedblleft run time\textquotedblright.
Beginning with the $Z+5~\mathrm{jets}$ sample, sizeable effects become visible in the \Pythia\ memory profiles. The peak heights continuously grow for \Pythia\ and eventually outgrow the first file-reading peak. The \Vincia\ memory profile remains mostly constant and becomes negligible in comparison to \Pythia's profile for the 8- and 9-jet samples. \Pythia's history-construction technique directly translates to the memory profiles; after a history has been chosen, the memory allocation returns to the baseline value at which \Vincia\ remains throughout.

\clearpage

\FloatBarrier
\begin{figure}[ht]
    \centering
    \includegraphics[width=0.6\textwidth]{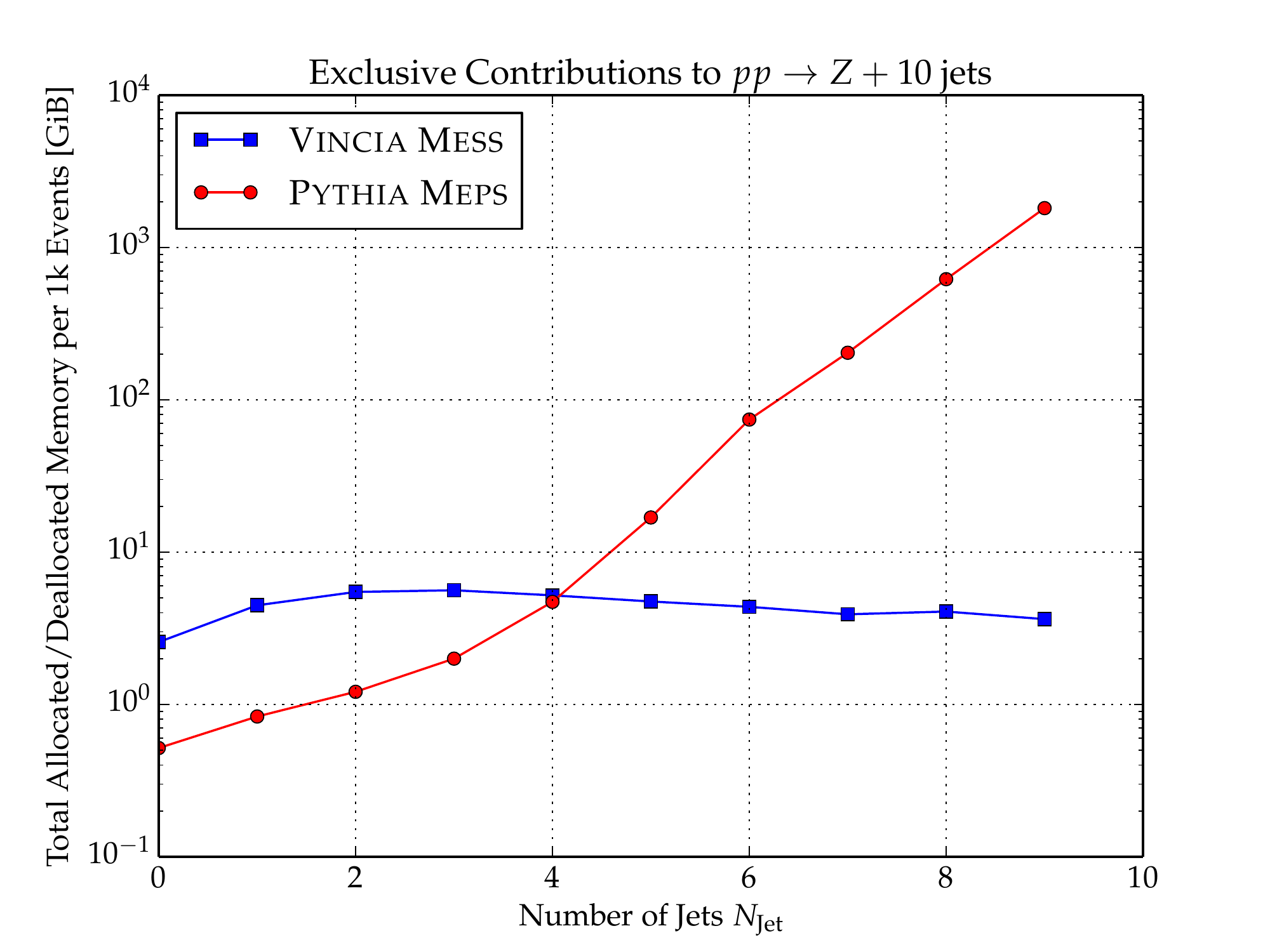}
    \caption{\Pythia\ and \Vincia\ memory usage scaling in $pp \to Z + \mathrm{jets}$ merging at $\sqrt{s} = 14~\tera e\volt$.}
    \label{fig:memoryScaling}
\end{figure}

As a gauge of the scaling behaviour of the memory usage in both merging implementations, we plot the total allocated/deallocated memory per 1k events in \cref{fig:memoryScaling}. For each multiplicity, we average over statistically independent runs and from 7 jets on, we also average over the different groupings.
While \Pythia\ shows a rather dramatic scaling, with allocating and deallocating a total of $1~\mathrm{TiB}$ of data for $Z+9~\mathrm{jets}$, the \Vincia\ curve remains almost flat, with only a small peak around 3 additional jets. The latter can be understood by considering that the sector shower has a comparable memory footprint as the merging and that in the latter maximally two histories are stored concurrently, cf.~\cref{sec:SectorHistories}. At high multiplicities, most of the events get vetoed during the trial showers and the sector shower is never started off these events. For samples with 1 -- 3 additional jets, on the other hand, a fair number of events are accepted and further processed by the sector shower, explaining the small increase in memory usage there.

\FloatBarrier

\section{Conclusions}\label{sec:Conclusions}
We here presented the first-ever implementation of the CKKW-L merging approach with sector showers, which alleviates the bottlenecks of conventional implementations while accurately calculating the Sudakov factors as generated by the shower. The merging scheme was implemented for the \Vincia\ antenna shower in the \Pythia~8.3 event generator; this implementation is mostly independent from the default CKKW-L one, and has been made public in the \Pythia 8.304 release.

We have validated the implementation for processes of immediate phenomenological interest and studied the scaling behaviour of the method in multi-jet merging in vector boson production at high multiplicities. While the time to construct sector shower histories scales approximately linearly with the number of hard jets, the overall event generation time as well as the memory usage stays approximately constant. Both provides a significant improvement over the exponential scaling of the default merging implementation in \Pythia. As a consequence, including merging hard jets with the sector shower in fact becomes easier with increasing multiplicity.
We gained a first estimate of renormalisation scale uncertainties arising at high merged multiplicities and compared preliminary results to \Pythia's CKKW-L implementation.

While we have here focused on the computational improvements, a dedicated physics study with the MESS framework is yet to follow. In such a study, the default \Vincia\ tune should be reviewed to achieve an accurate description on the hadron level when including higher-order matrix elements.
Moreover, immediate future work can be done on improving sampling methods for sector antennae in their respective sectors, which is currently only inefficiently achieved by means of multiple global antennae over their full phase spaces and rejecting branchings outside of appropriate sectors. This results in the overall slower shower algorithm, as seen above.

As the sector merging approach developed here is an adaption of the CKKW-L technique, existing refinements and extensions of it can readily be applied.
As such, it can be extended to retain unitarity by the methods presented in \cite{Lonnblad:2011xx,Bellm:2017ktr} or to include NLO corrections by the methods of \cite{Lavesson:2008ah,Lonnblad:2012ng,Platzer:2012bs,Gehrmann:2012yg,Hoeche:2012yf} in the future. Even extensions towards NNLO are feasible along the lines of the method presented in \cite{Hoeche:2014aia,Hoche:2014dla,Hoche:2018gti}.
For these, only the construction of the shower histories needs to be adapted to the sector case. To this end, the NL3 \cite{Lavesson:2008ah} and UNLOPS \cite{Lonnblad:2012ng} schemes are particularly well suited for the generalisation to NLO merging, as their adaption to sector showers would follow the exact same steps as in the CKKW-L case outlined in \cref{sec:SectorMerging}. This means that only the probabilistic construction of shower histories would be replaced by the sectorised history construction, while all other steps in NL3 and UNLOPS stay the same.

In a less straightforward way, it might also be possible to extend the LO merging technique presented here to higher orders via other schemes, such as the MINLO \cite{Hamilton:2012np} and MINNLO$_\text{PS}$ \cite{Monni:2019whf} or FxFx \cite{Frederix:2012ps} techniques. These extensions do, however, differ in the way histories are constructed and scales are associated to intermediate states. In addition to an adaptation of the history construction to sector showers, an implementation of these schemes would, for instance, also require the implementation of analytical Sudakov factors.

In the light of the observed scale uncertainties at high multiplicities, both unitarity-improved schemes and extensions to the NLO (or even higher orders) provide interesting and sensible avenues for future work.
It is worth pointing out that in both cases, event files are usually processed twice to generate counter terms. As the algorithm presented here will always yield a unique history for an input configuration, it bears the potential to make both unitary and NLO merging even more efficient, as event files might only have to be processed once. On the one hand, this reduces the overall run time and on the other hand, this might alleviate problems regarding negative-weight events.
As the implementation in the \Vincia\ shower furthermore implies a dedicated interleaved resonance shower framework \cite{Brooks:2019xso,Skands:2020lkd,Kleiss:2020rcg}, it may be worthwhile to explore merging in coloured as well as electroweak resonance systems in the future.

Although the discussion here was tailored to the implementation in \Vincia, it demonstrates the efficiency gains that can be obtained through a relatively straightforward adaption of the shower model.
We have sketched how our method could be adopted in other shower models and exemplified it for the case of Catani-Seymour dipole showers in \cref{sec:SectorShowers}.

We close by noting that with the merging scheme presented here, shower-plus-matrix-element calculations with more than 9 hard jets are readily possible on the shower side. The main bottlenecks of merged calculations remain entirely on the fixed-order side and generating large numbers of high-multiplicity configurations still remains a time- and resource-consuming endeavour.

\section*{Acknowledgements}
We would like to thank Peter Skands for many helpful and fruitful discussions and comments on the manuscript. CTP thanks Stefan H\"oche for help with the HDF5 event files and Stefan Prestel for help with the HDF5 data format and clarifications on the CKKW-L implementation in \Pythia. 
CTP also thanks Steffen Schumann and Enrico Bothmann for helpful comments on commensurate sector scales.
We thank Cody B Duncan for proofreading the manuscript.
We acknowledge support from the Monash eResearch Centre and
eSolutions-Research Support Services through the MonARCH HPC
Cluster. 
CTP is supported by the Monash Graduate Scholarship, the Monash International Postgraduate Research Scholarship, and the J.~L.~William Scholarship.
HB received funding from the Australian Research Council via Discovery
Project DP170100708 -- \textquotedblleft Emergent Phenomena in Quantum
Chromodynamics\textquotedblright. 
This work was also supported in part
by the European Union’s Horizon 2020 research and innovation programme
under the Marie Sklodowska-Curie grant agreement No 722104 --
MCnetITN3.

\clearpage
\appendix
\section{Perturbative Tune Parameters}\label{sec:tuneParms}
We use the preliminary default \Vincia\ tune of the perturbative parameters, cf.~\cite{Brooks:2020upa}.
We include two-loop running-coupling effects with an effective value of $\alpha_s$ chosen according to the CMW scheme~\cite{Catani:1990rr},
\begin{equation}
    \alphaS^\text{CMW} = \alphaS^{\overline{\text{MS}}}
    \left(1 + \frac{\alphaS^{\overline{\text{MS}}}}{2\pi} \left[
    C_A \left(\frac{67}{18} - \frac{\pi^2}{6}\right) - \frac{5 n_f}{9}
    \right]\right) \, , \quad \alphaS^{\overline{\mathrm{MS}}}(M_Z) = 0.118 \, ,
\end{equation}
supplemented by renormalisation-scale prefactors $k_R$, which modify the evolution scale,  $p_\perp$, in the argument of the running coupling,
\begin{equation}
    \alphaS^\mathrm{Vincia}(p_\perp^2) = \alphaS^\mathrm{CMW}(k_R p_\perp^2) \, .
\end{equation}
The default values for these additional scale prefactors are chosen based on preliminary studies of LEP event shapes and Drell-Yan $p_\perp$-spectra as
\begin{align}
    k_\text{R,Emit}^\text{F} &= 0.66 \, , \quad 
    k_\text{R,Split}^\text{F} = 0.8 \, , \\
    k_\text{R,Emit}^\text{I} &= 0.66 \, , \quad
    k_\text{R,Split}^\text{I} = 0.5 \, , \quad 
    k_\text{R,Conv}^\text{I} = 0.5 \, .
\end{align}

\clearpage
\section{Memory Profiles}\label{sec:memProfiles}
We here collect additional memory profiles for event samples with more than 6 jets, for which the event files are grouped according to similar process structures.
\begin{figure}[ht]
    \centering
    \includegraphics[width=0.3\textwidth]{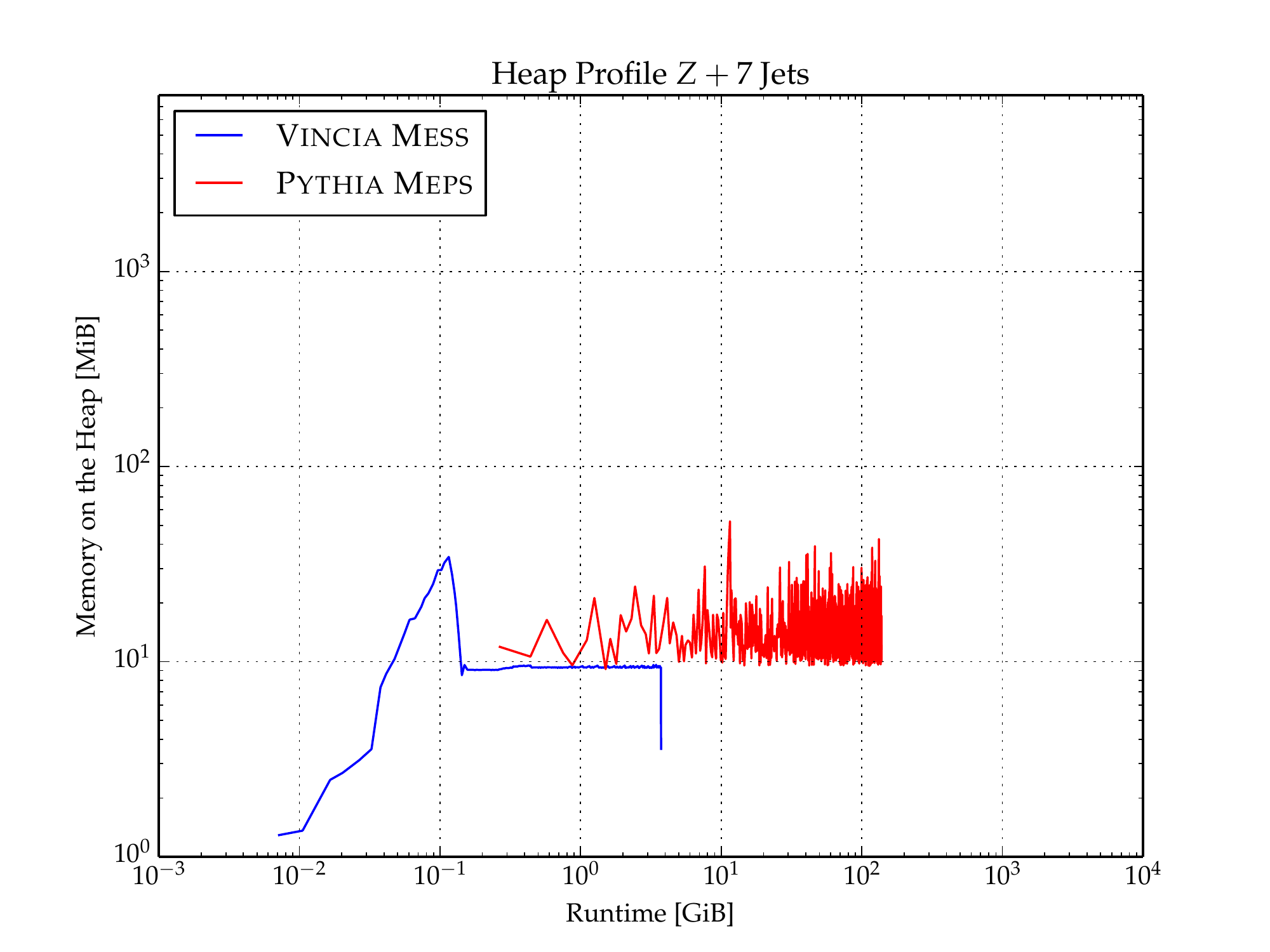}
    \includegraphics[width=0.3\textwidth]{memprofile-zj7-g2.pdf}
    \includegraphics[width=0.3\textwidth]{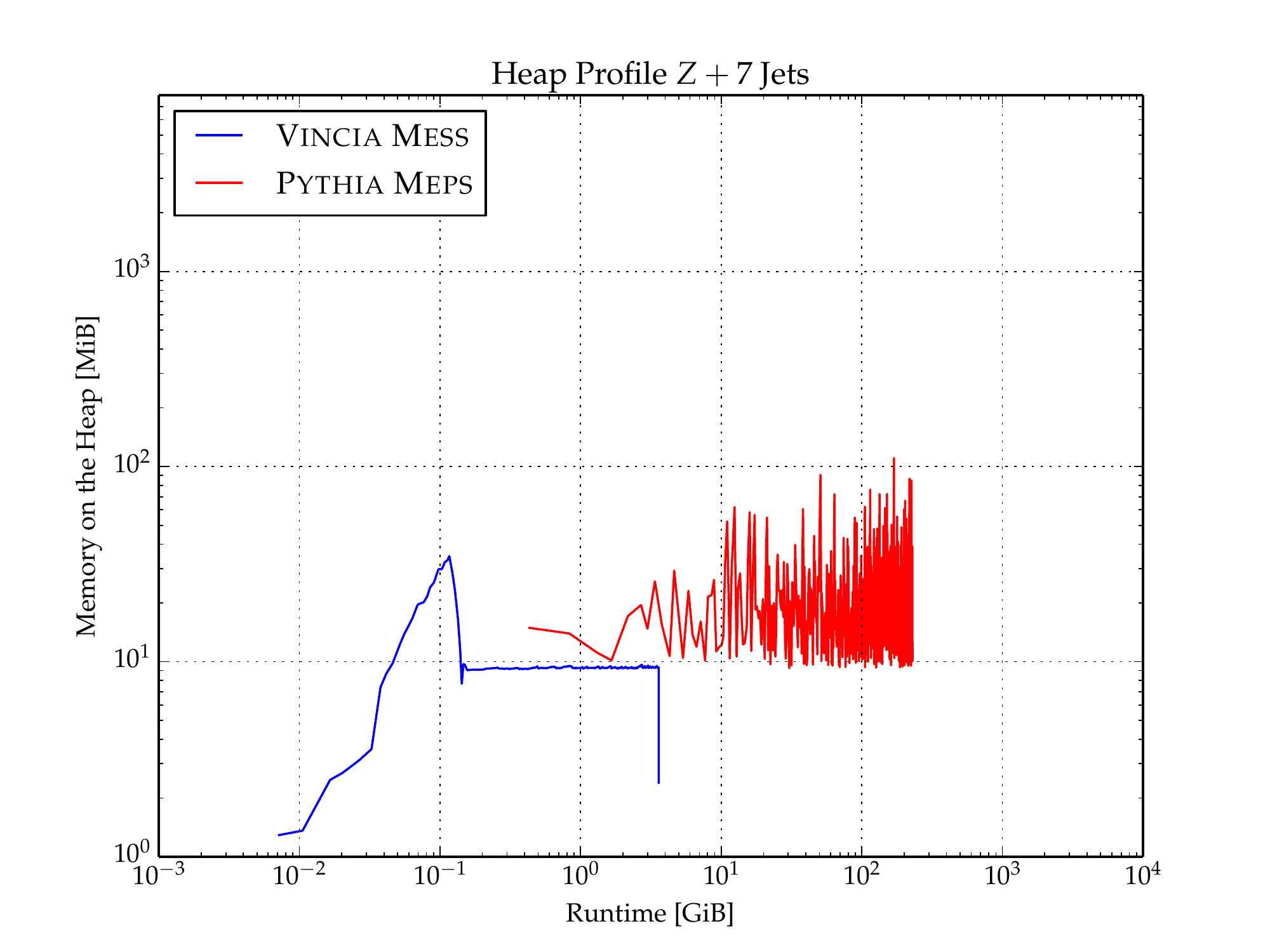}
    \includegraphics[width=0.3\textwidth]{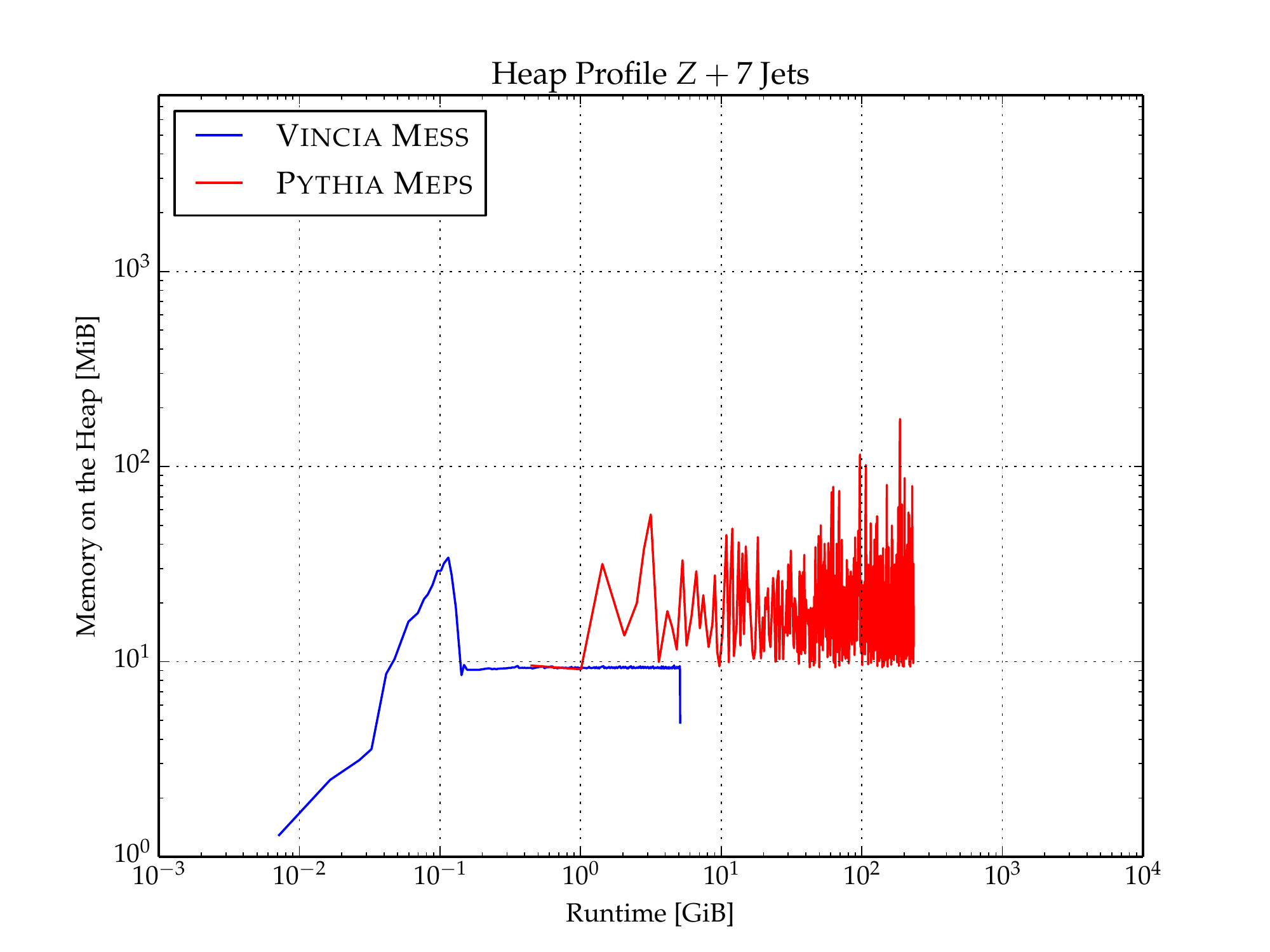}
    \includegraphics[width=0.3\textwidth]{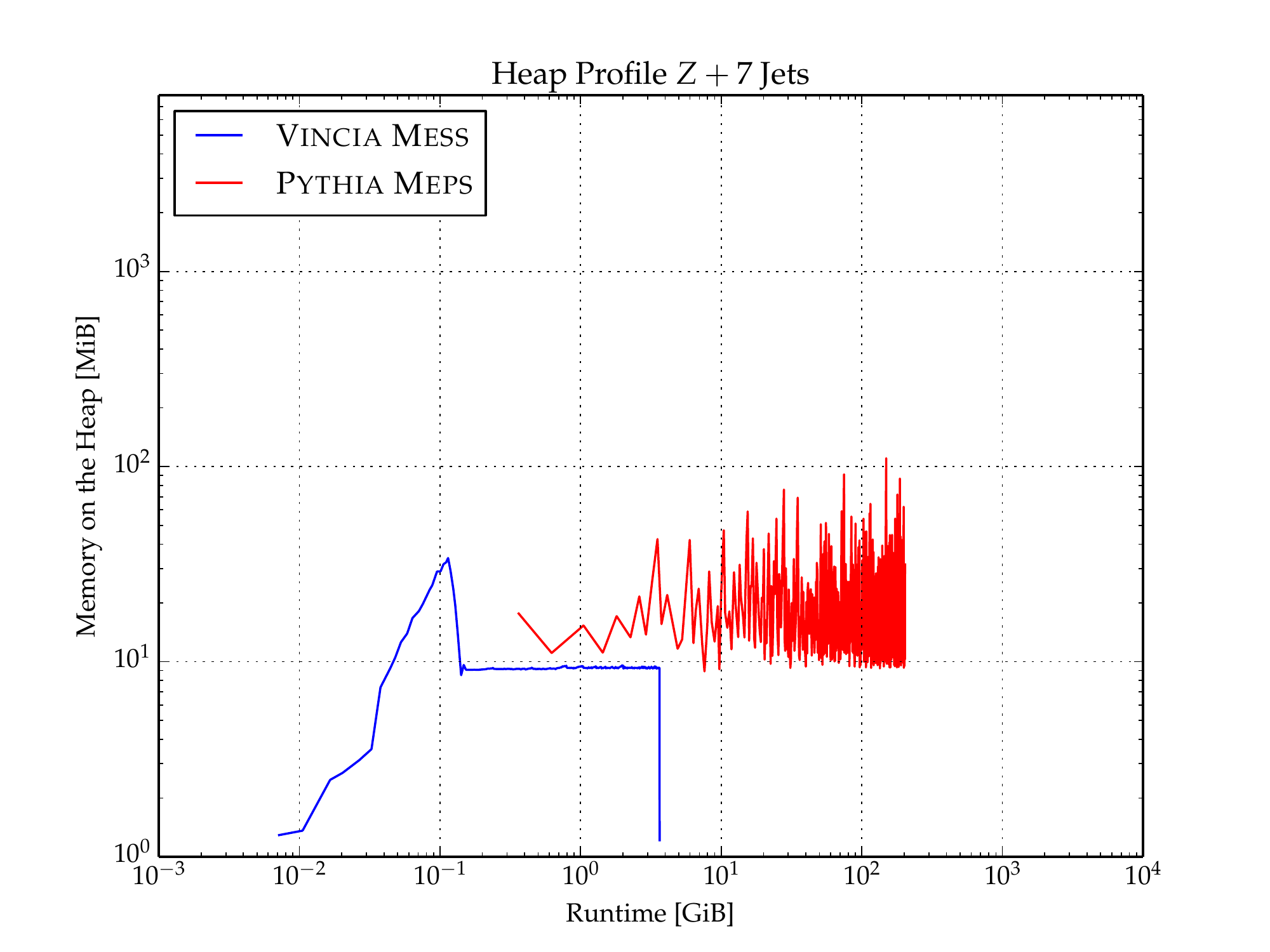}
    \includegraphics[width=0.3\textwidth]{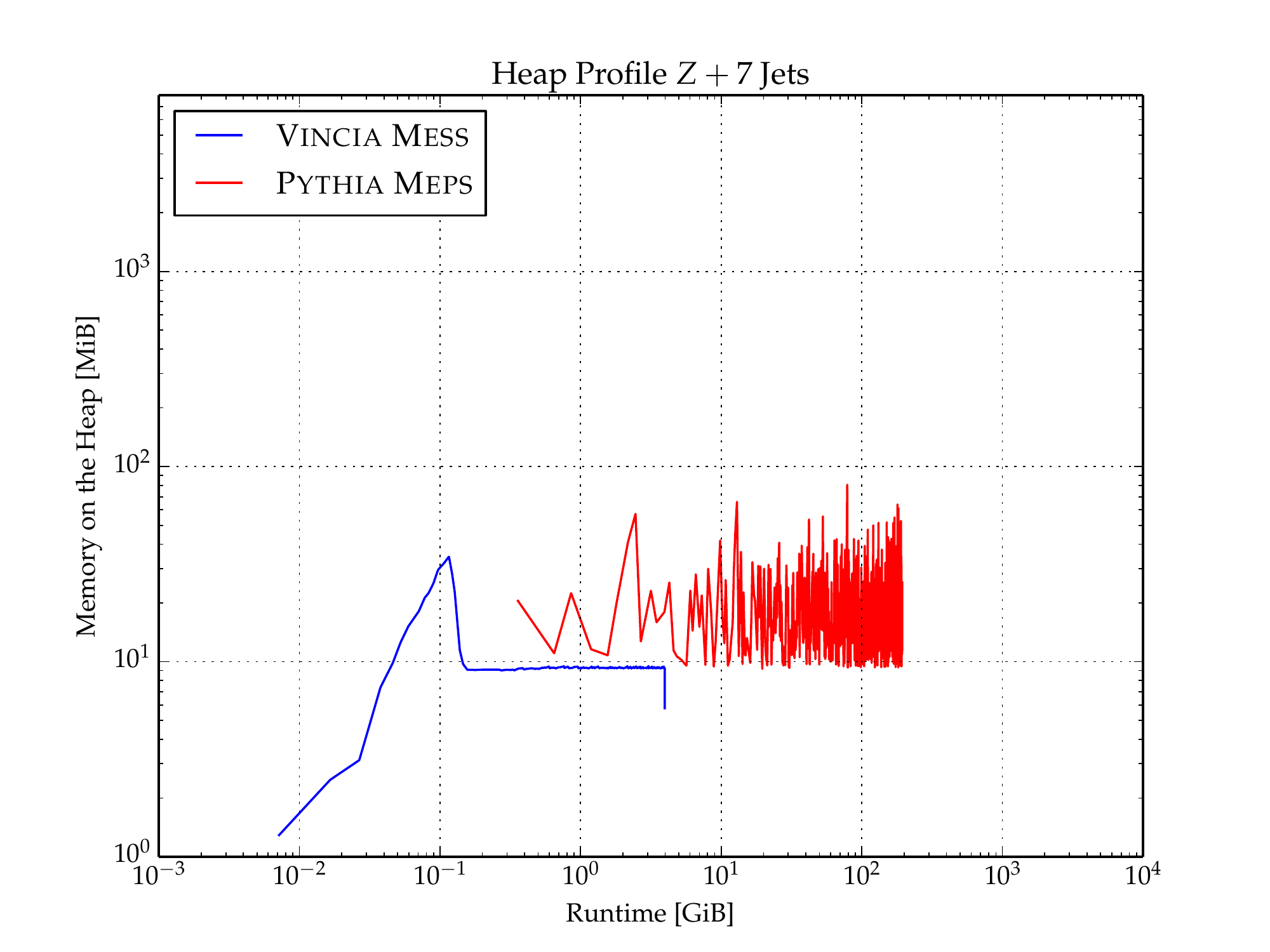}
    \caption{\Pythia\ and \Vincia\ memory usage profiles for different process groupings in $pp \to Z + 7~\mathrm{jets}$ samples at $\sqrt{s} = 14~\tera e\volt$.}
    \label{fig:memoryProfilesZj7}
\end{figure}
\begin{figure}[ht]
    \centering
    \includegraphics[width=0.3\textwidth]{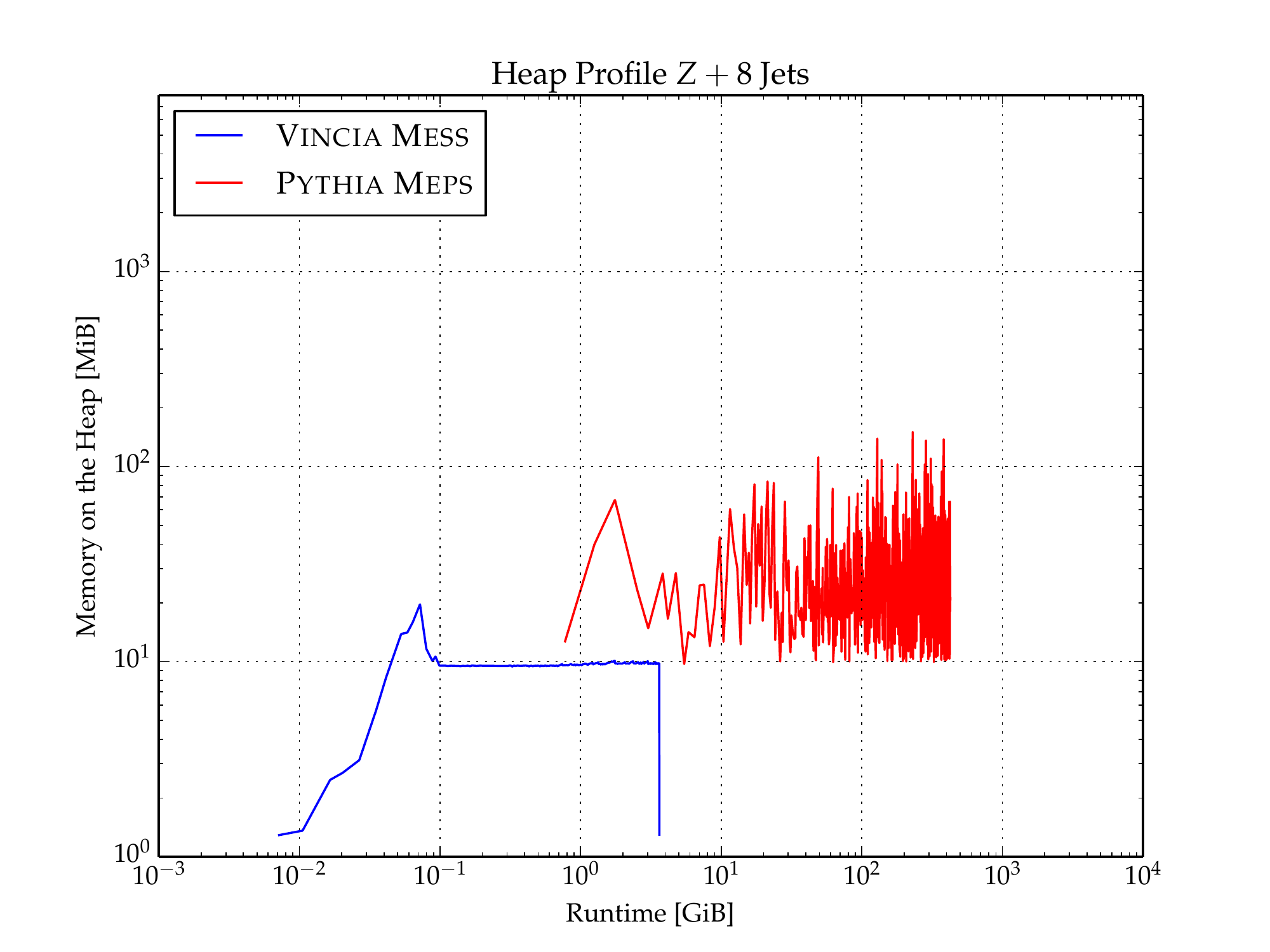}
    \includegraphics[width=0.3\textwidth]{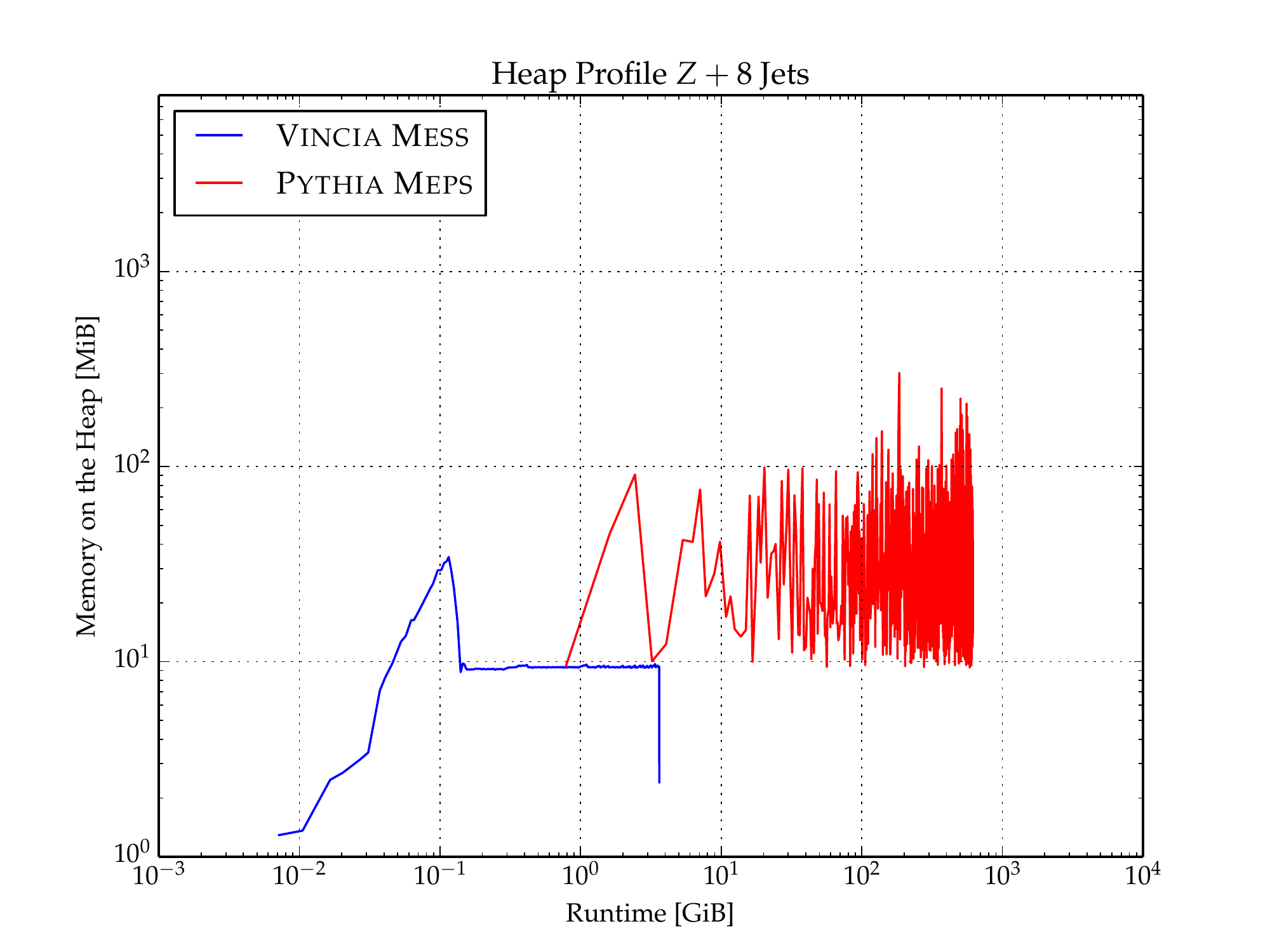}
    \includegraphics[width=0.3\textwidth]{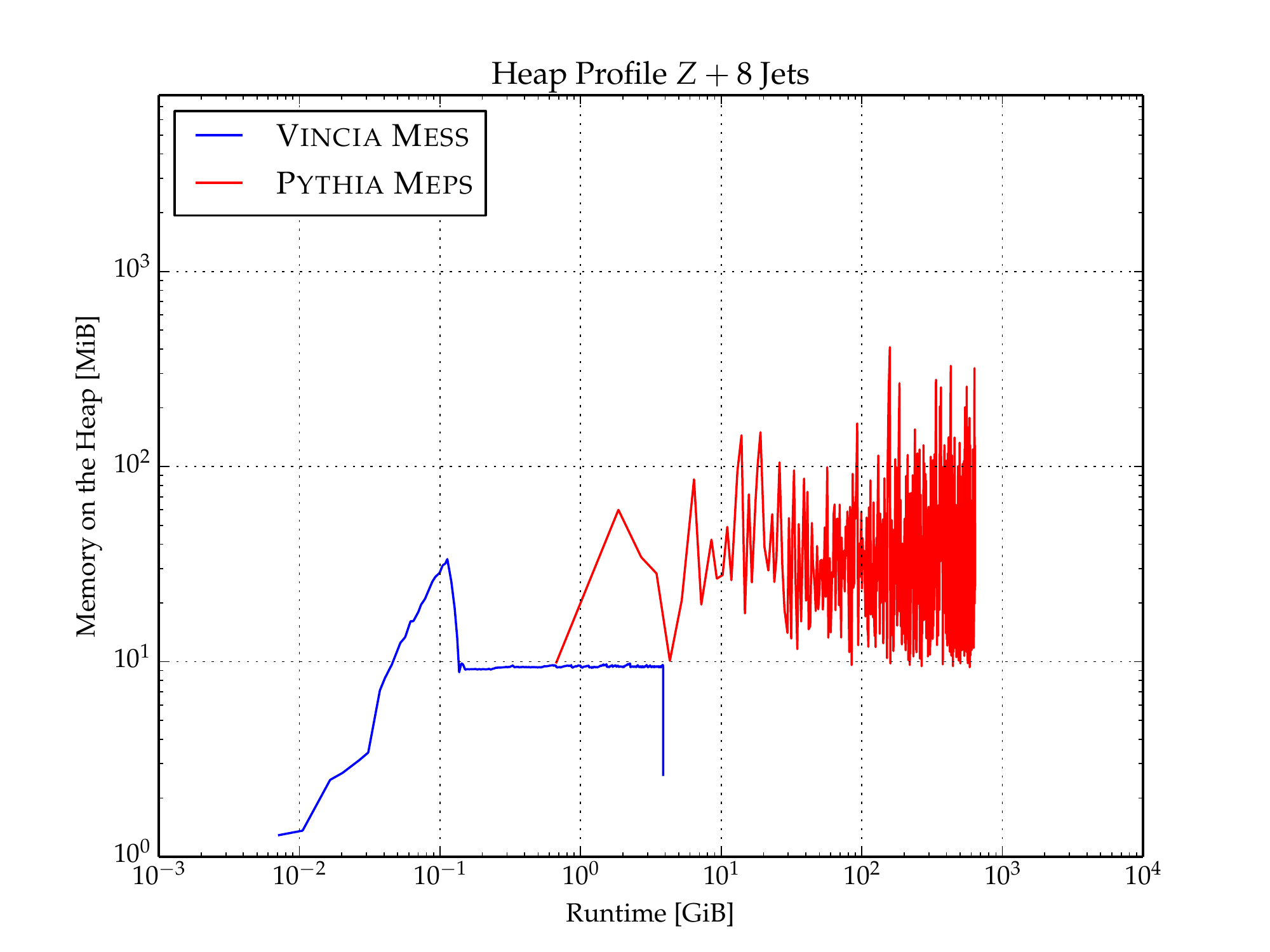}
    \includegraphics[width=0.3\textwidth]{memprofile-zj8-g4.pdf}
    \includegraphics[width=0.3\textwidth]{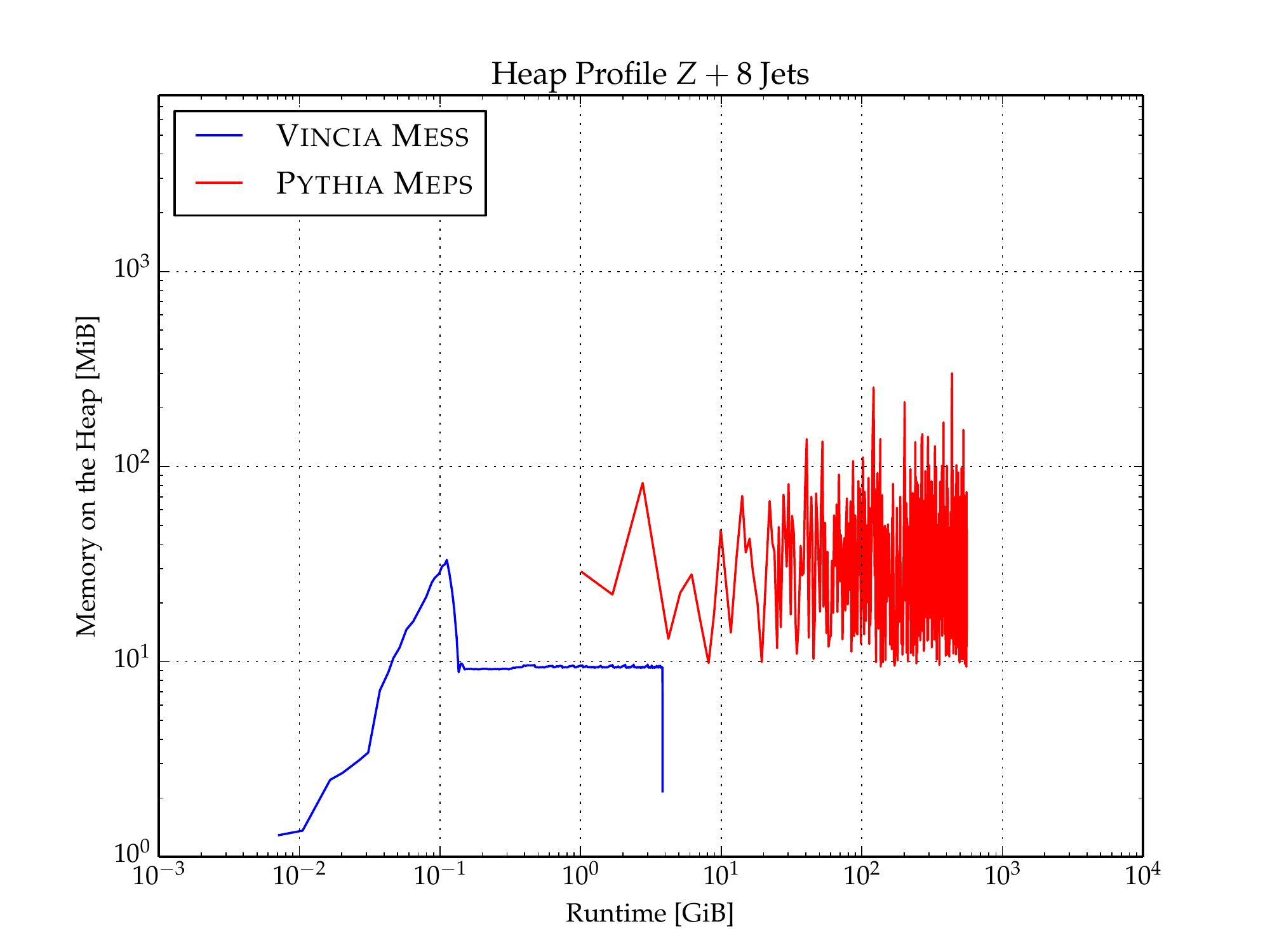}
    \includegraphics[width=0.3\textwidth]{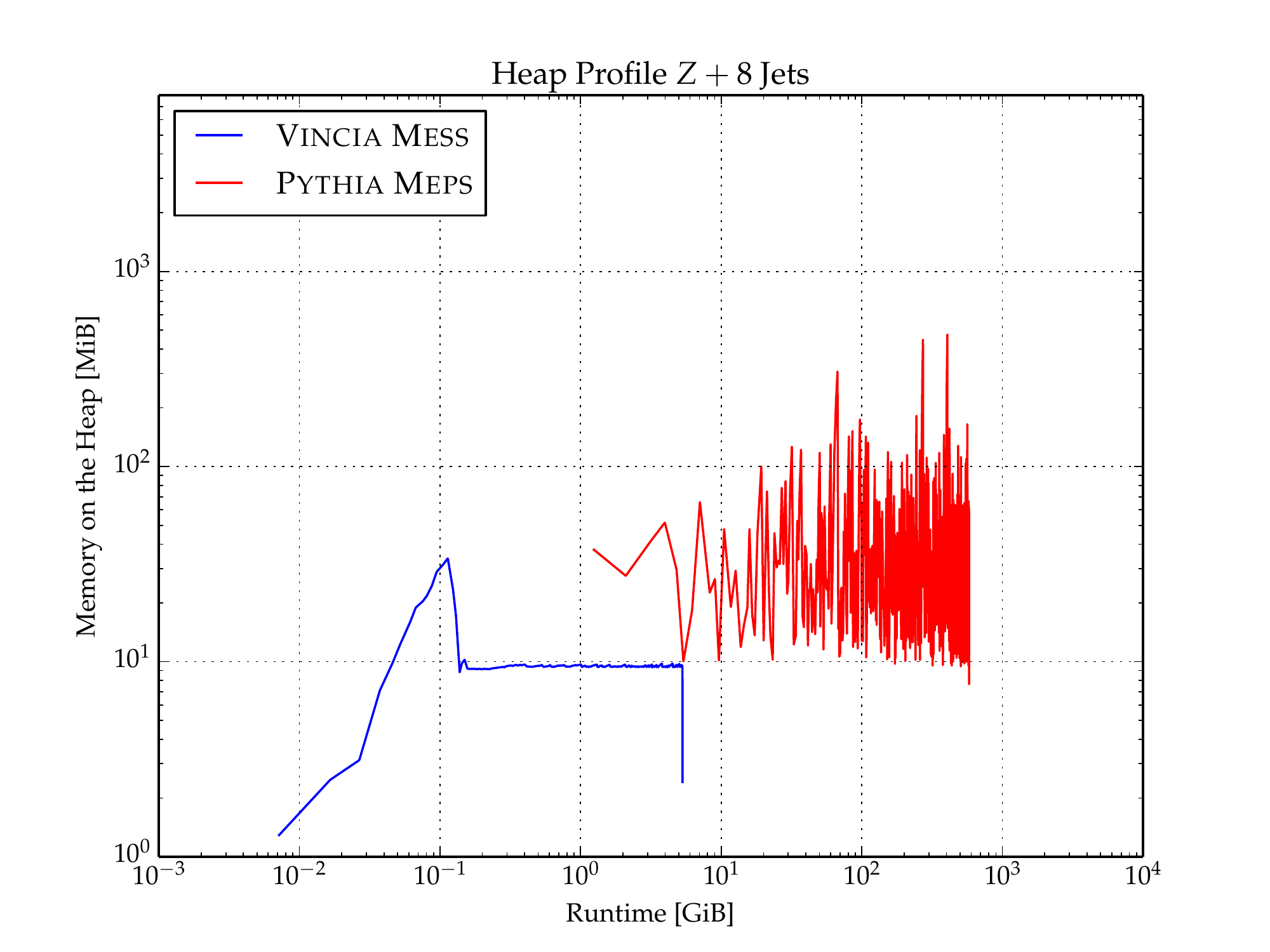}
    \caption{\Pythia\ and \Vincia\ memory usage profiles for different process groupings in $pp \to Z + 8~\mathrm{jets}$ samples at $\sqrt{s} = 14~\tera e\volt$.}
    \label{fig:memoryProfilesZj8}
\end{figure}
\clearpage
\begin{figure}[ht]
    \centering
    \includegraphics[width=0.3\textwidth]{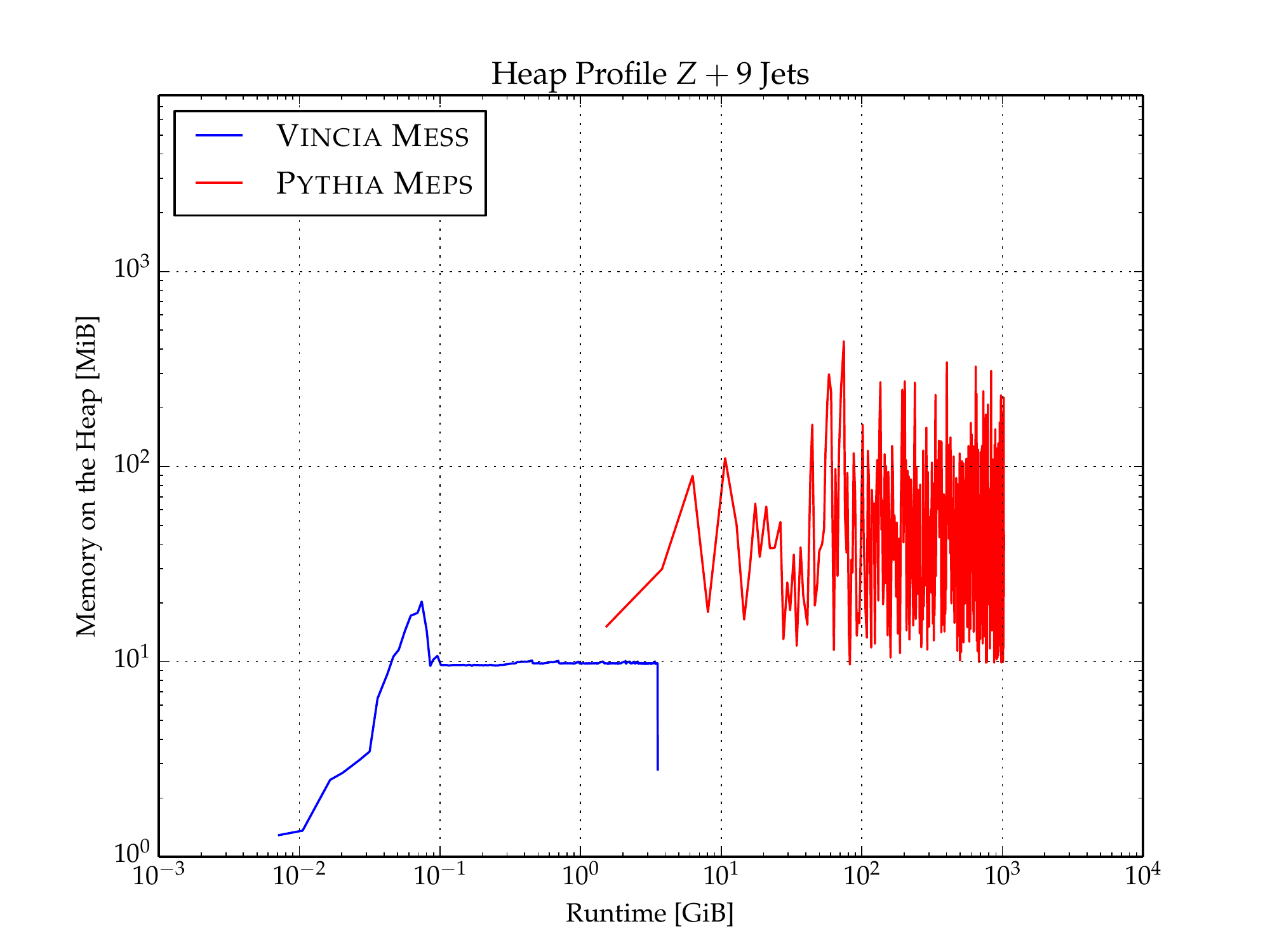}
    \includegraphics[width=0.3\textwidth]{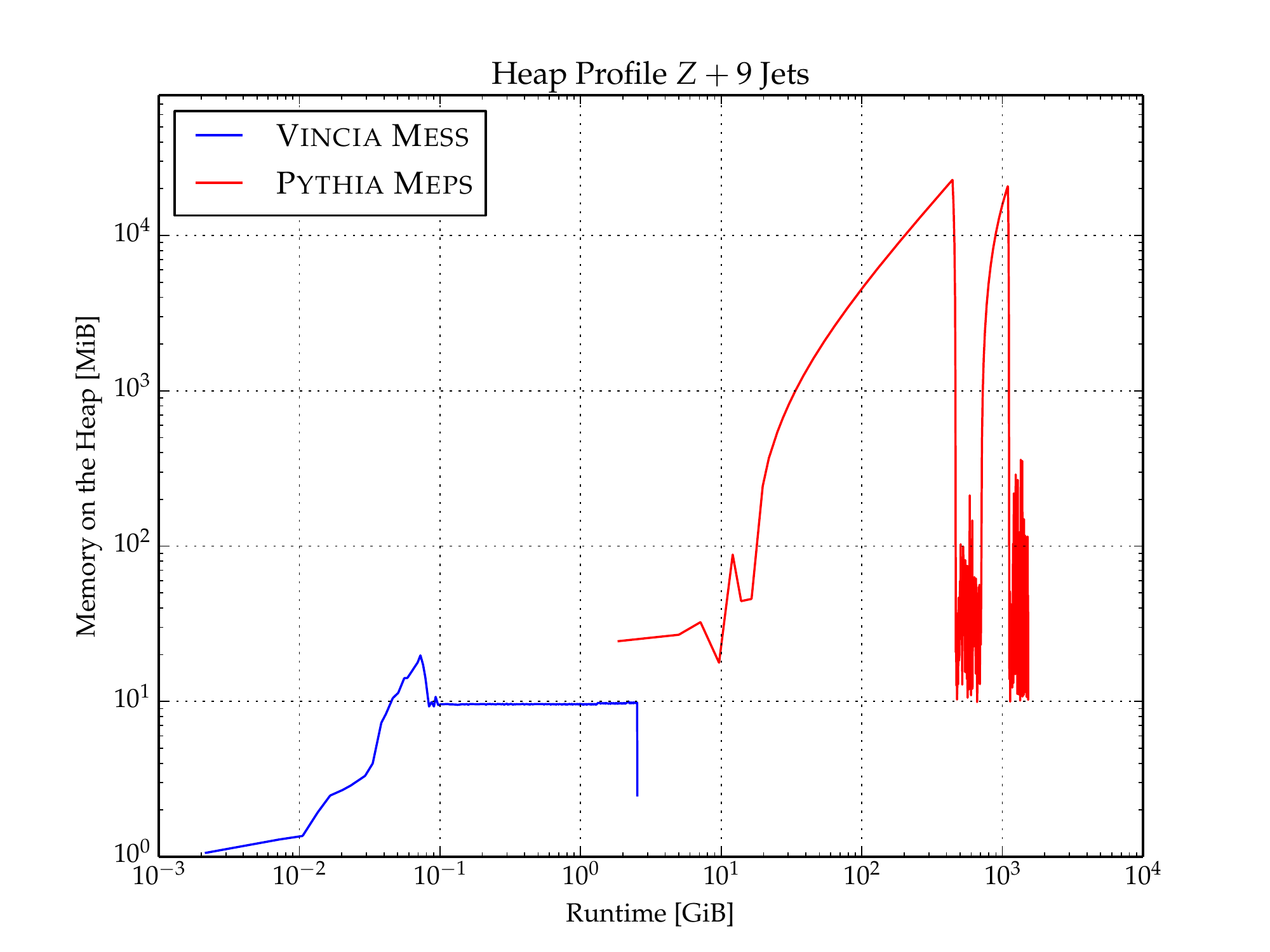}
    \includegraphics[width=0.3\textwidth]{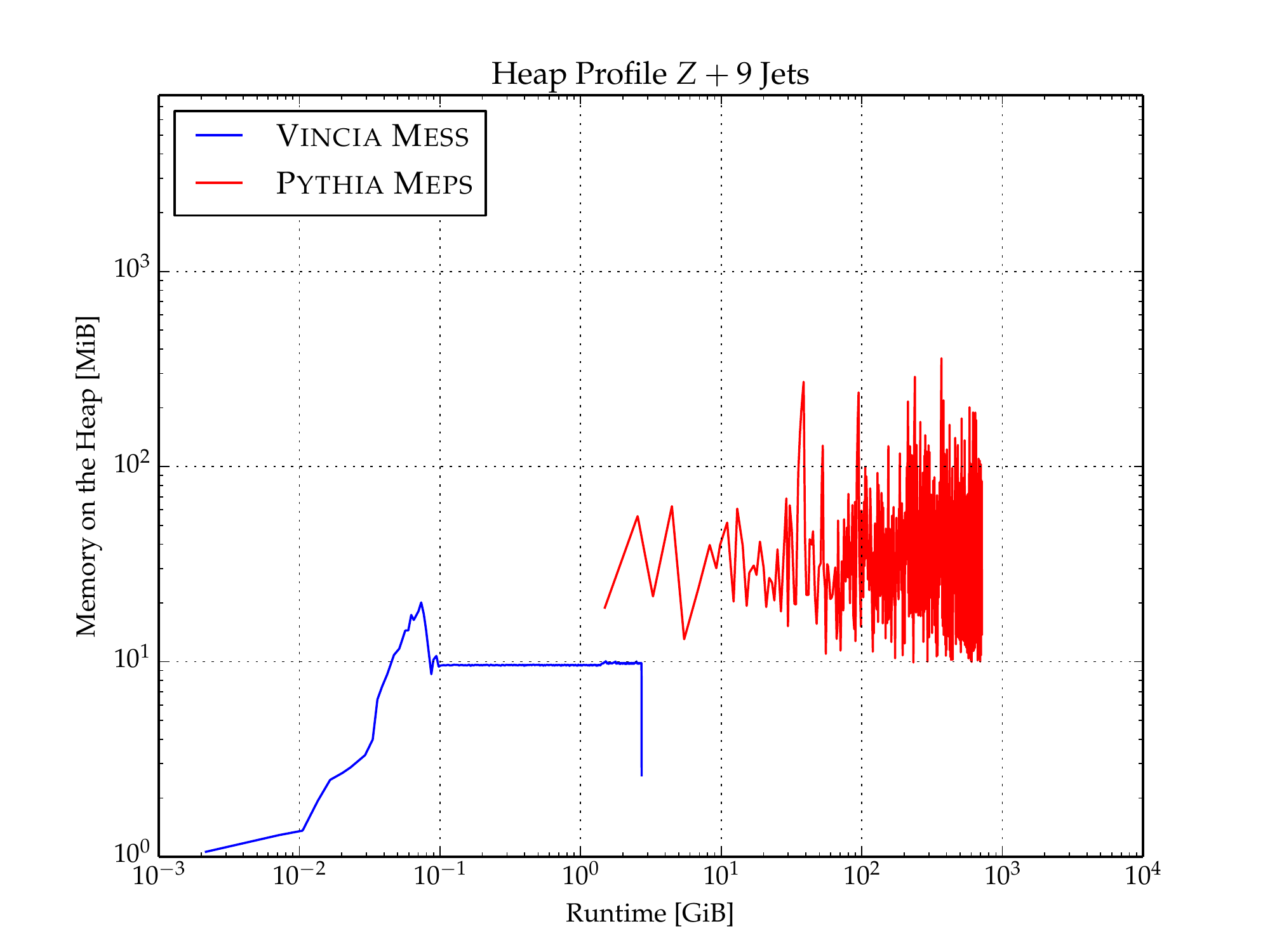}
    \includegraphics[width=0.3\textwidth]{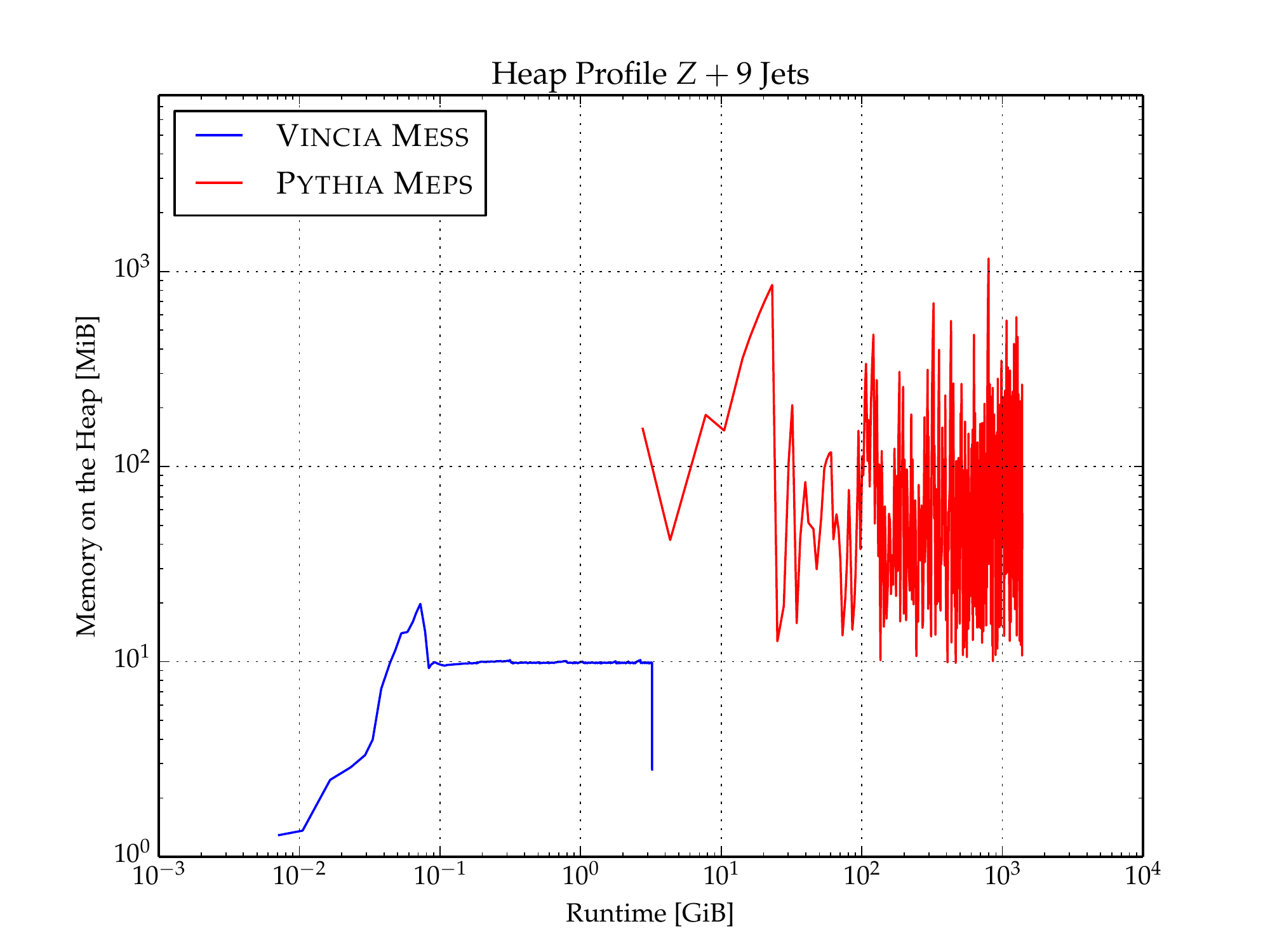}
    \includegraphics[width=0.3\textwidth]{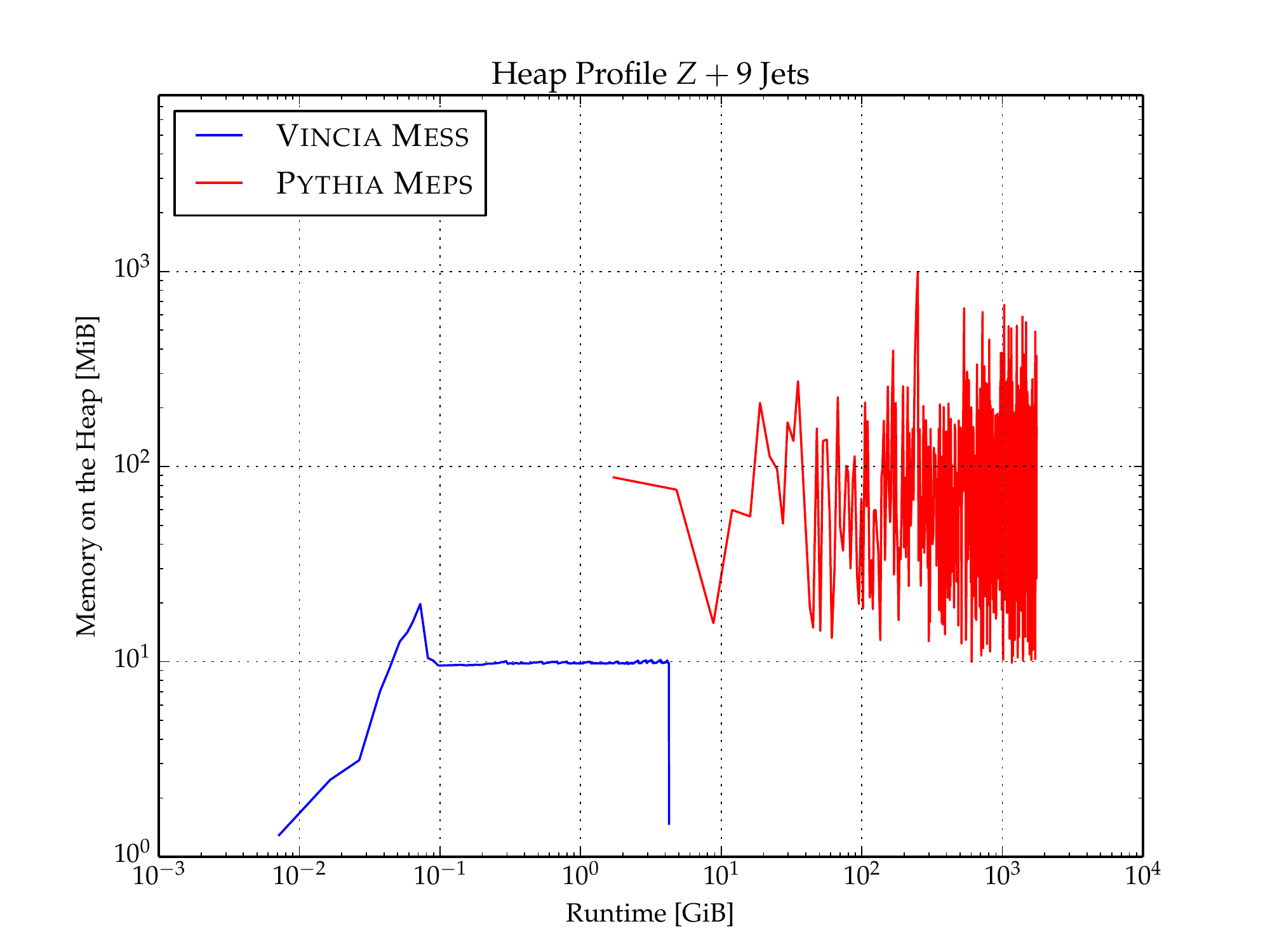}
    \includegraphics[width=0.3\textwidth]{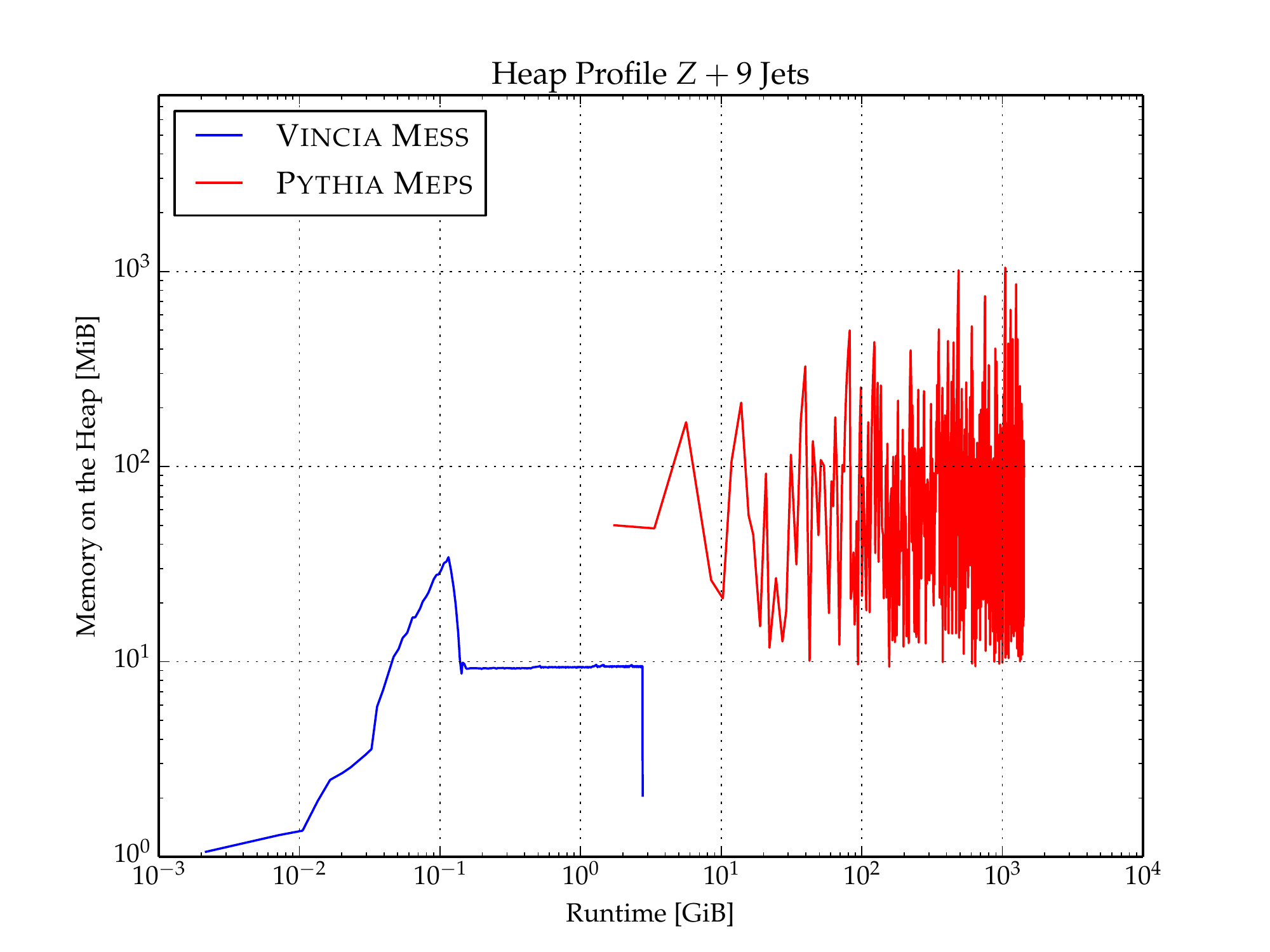}
    \includegraphics[width=0.3\textwidth]{memprofile-zj9-g7.pdf}
    \includegraphics[width=0.3\textwidth]{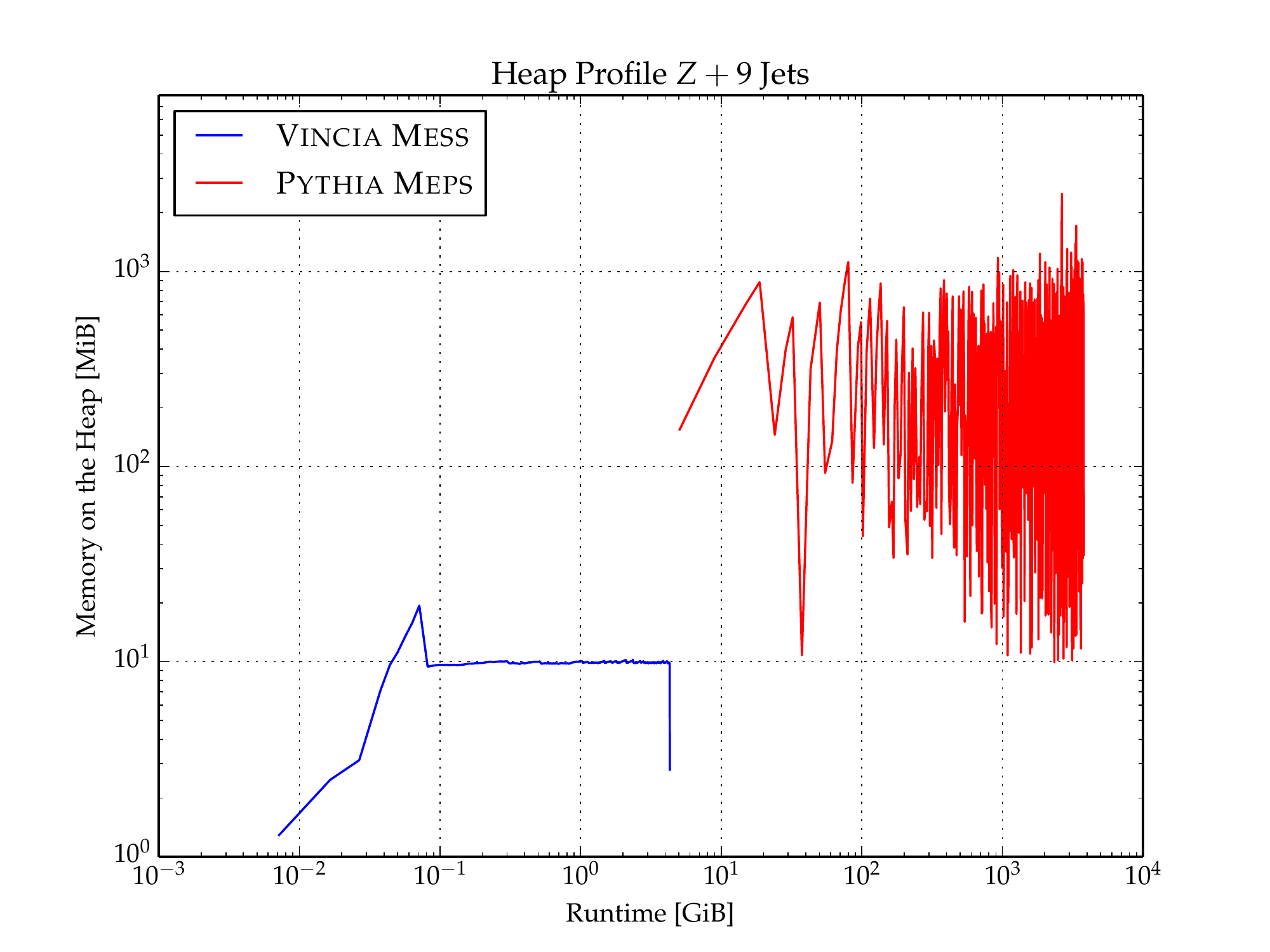}
    \includegraphics[width=0.3\textwidth]{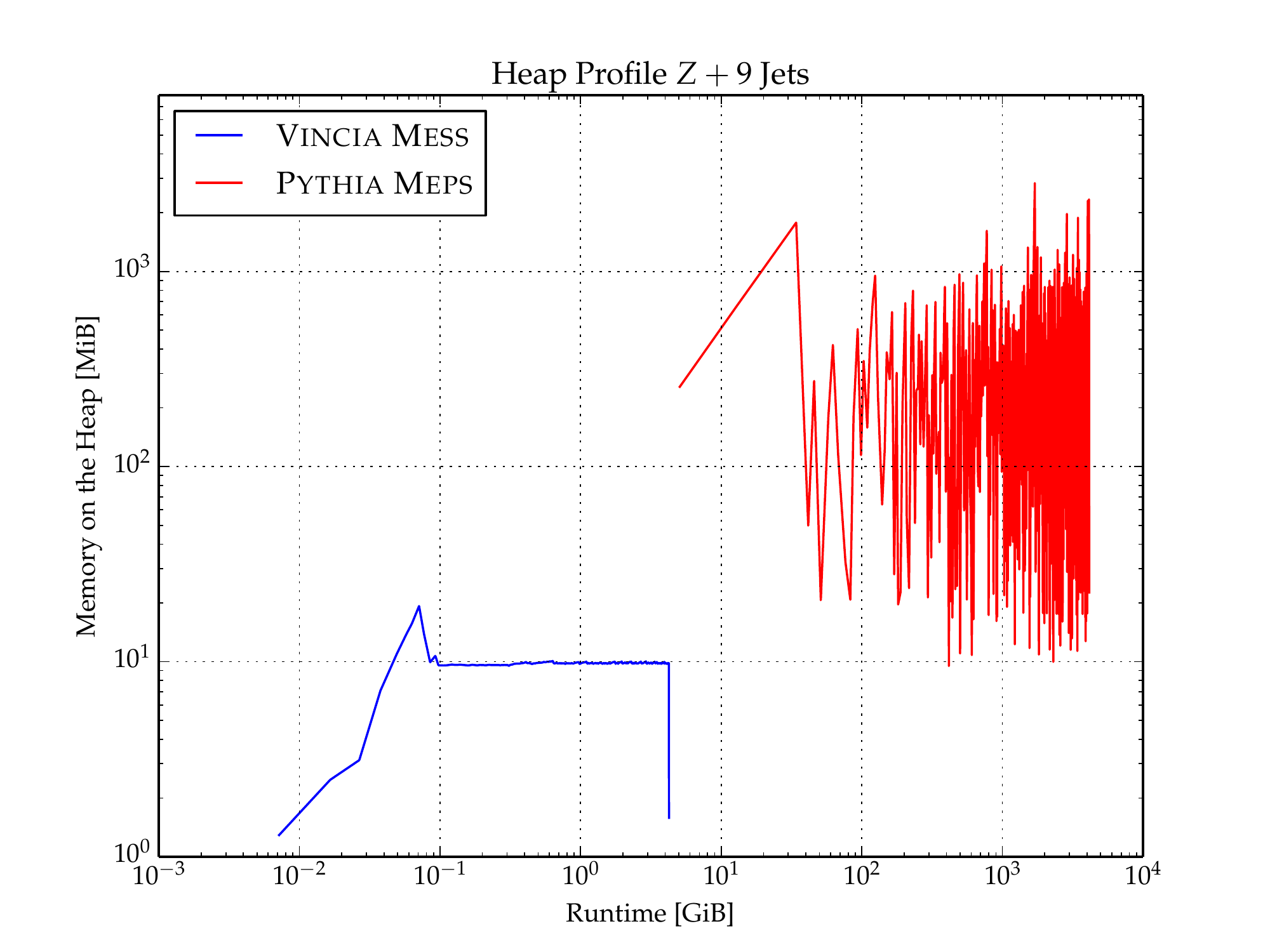}
    \includegraphics[width=0.3\textwidth]{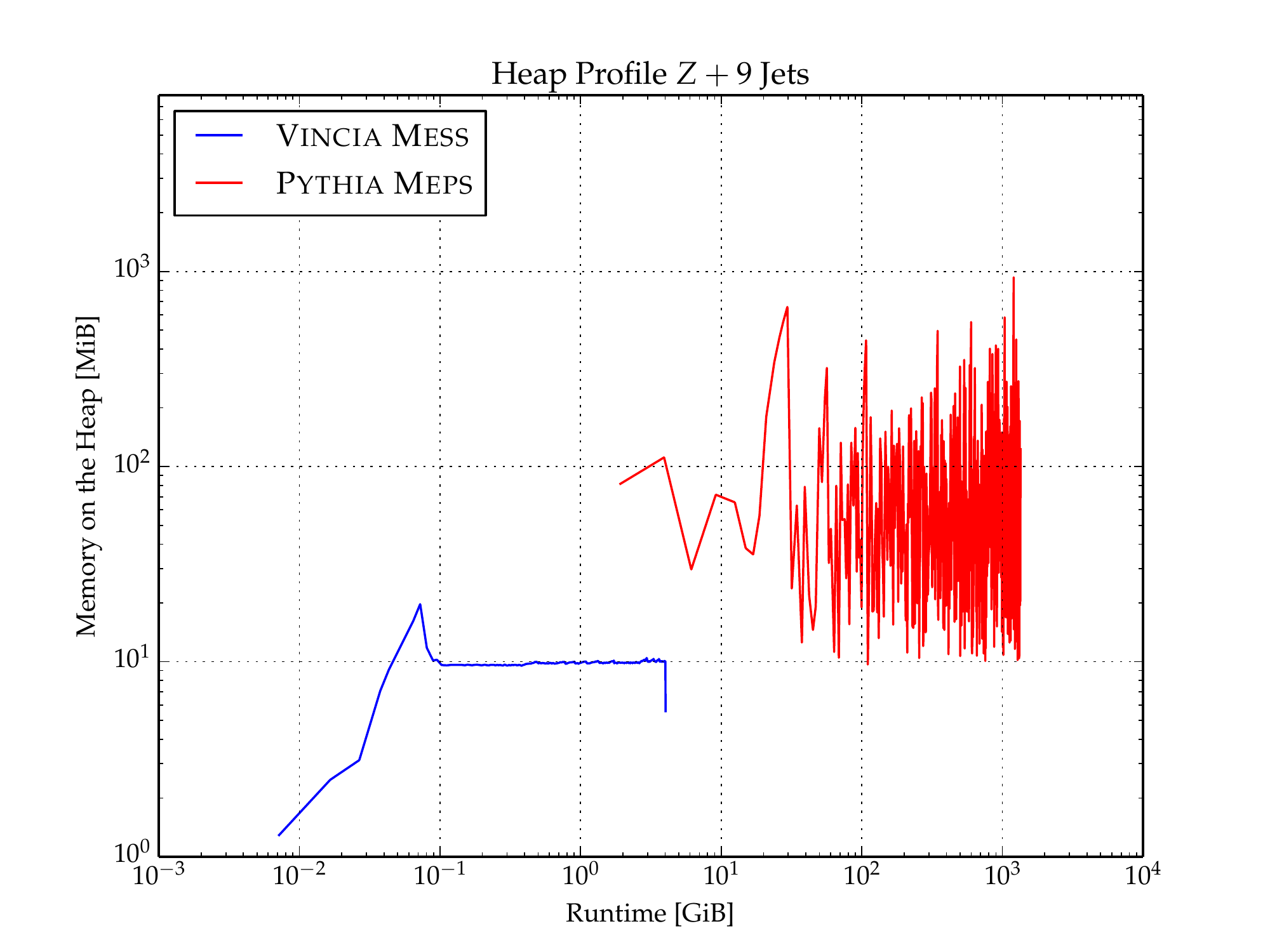}
    \includegraphics[width=0.3\textwidth]{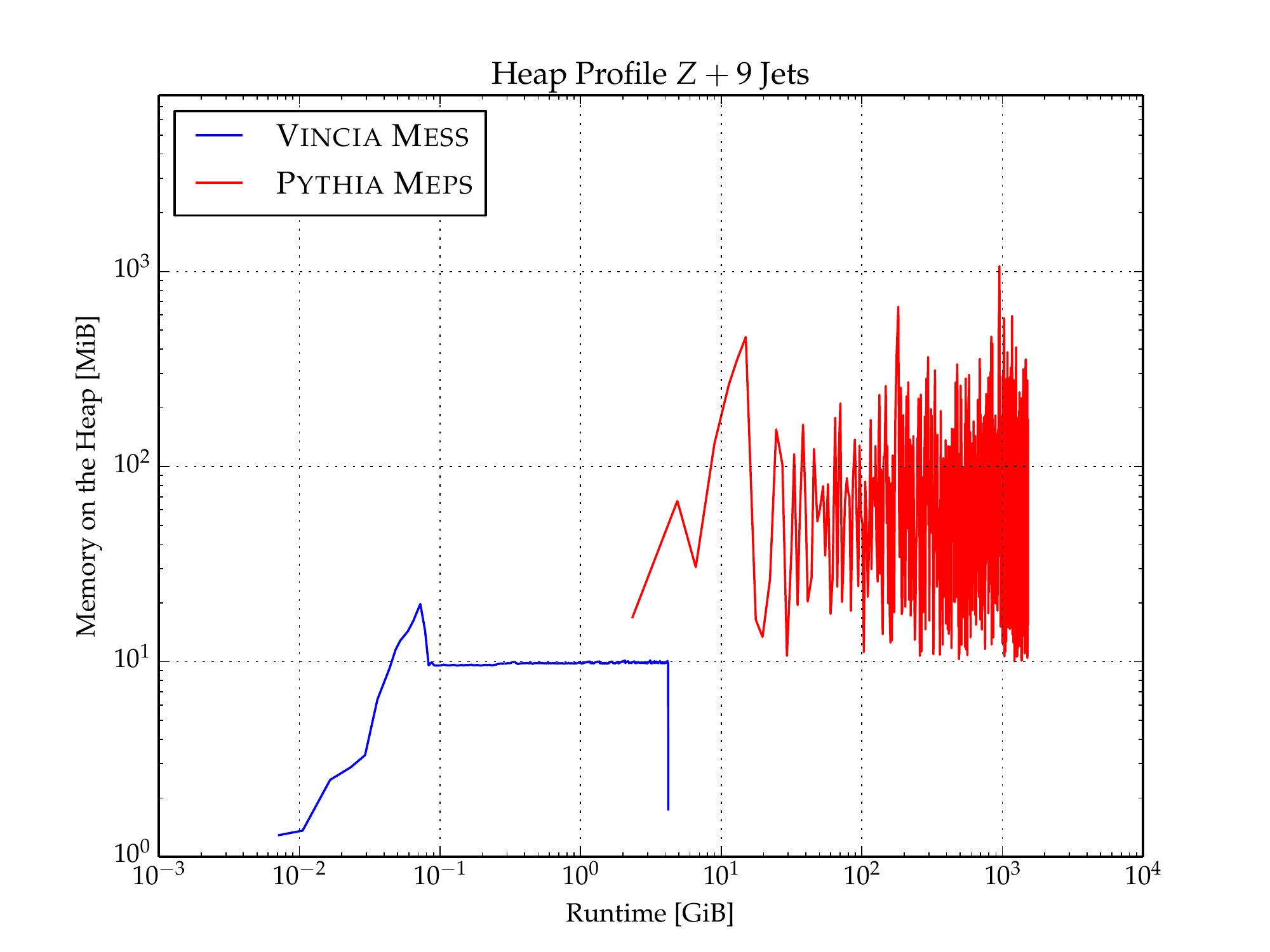}
    \includegraphics[width=0.3\textwidth]{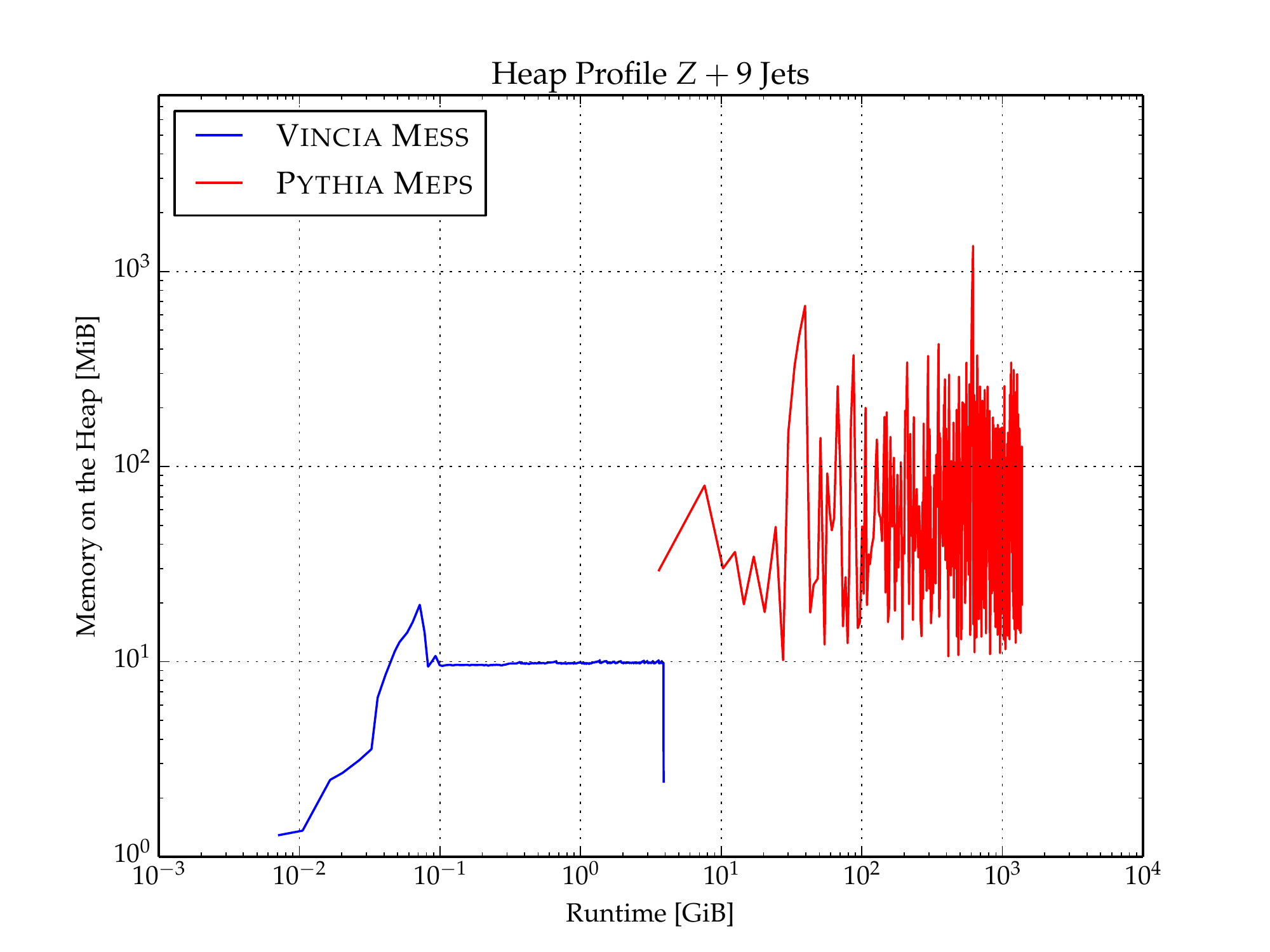}
    \caption{\Pythia\ and \Vincia\ memory usage profiles for different process groupings in $pp \to Z + 9~\mathrm{jets}$ samples at $\sqrt{s} = 14~\tera e\volt$.}
    \label{fig:memoryProfilesZj9}
\end{figure}

\bibliographystyle{elsarticle-num} 
\bibliography{refs}

\end{document}